\documentclass{article}

\usepackage{tocloft} 
\usepackage{microtype}
\usepackage{booktabs} 

\usepackage{hyperref}
\usepackage{url}
\usepackage[utf8]{inputenc} 
\usepackage[T1]{fontenc}    
\usepackage{hyperref}       
\usepackage{url}            
\usepackage{amsfonts}       
\usepackage{nicefrac}       
\usepackage{xcolor}   
\usepackage{graphicx}
\usepackage{subcaption}
\usepackage{colortbl}
\usepackage{tabularx}
\usepackage{bbm}
\usepackage{multirow}
\usepackage{adjustbox}
\usepackage{algorithm}
\usepackage{algpseudocode}
\usepackage[noEnd=false]{algpseudocodex}
\usepackage[toc,page]{appendix} 
\usepackage{titlesec}
\usepackage{booktabs}
\usepackage{threeparttable}
\usepackage{enumitem}
\setlist[enumerate]{left=10pt}
\usepackage[font=small]{caption}
\usepackage{wrapfig}

\usepackage{hyperref}

\usepackage[accepted]{icml2025}

\usepackage{amsmath}
\usepackage{amssymb}
\usepackage{mathtools}
\usepackage{amsthm}

\usepackage[capitalize,noabbrev]{cleveref}

\theoremstyle{plain}
\newtheorem{theorem}{Theorem}[section]
\newtheorem{proposition}[theorem]{Proposition}

\theoremstyle{definition}

\theoremstyle{remark}

\usepackage[textsize=tiny]{todonotes}

\icmltitlerunning{\textit{De Novo} Generation of Therapeutic Peptides with Multi-Objective-Guided Discrete Diffusion}

\begin{document}

\twocolumn[
\icmltitle{PepTune: \textit{De Novo} Generation of Therapeutic Peptides with Multi-Objective-Guided Discrete Diffusion}



\icmlsetsymbol{equal}{*}

\begin{icmlauthorlist}
\icmlauthor{Sophia Tang}{equal,1}
\icmlauthor{Yinuo Zhang}{equal,2}
\icmlauthor{Pranam Chatterjee}{1,3}
\end{icmlauthorlist}

\icmlaffiliation{1}{Department of Computer and Information Science, University of Pennsylvania}
\icmlaffiliation{2}{Center of Computational Biology, Duke-NUS Medical School}
\icmlaffiliation{3}{Department of Bioengineering, University of Pennsylvania}

\icmlcorrespondingauthor{Pranam Chatterjee}{pranam@seas.upenn.edu}

\icmlkeywords{Machine Learning, ICML}

\vskip 0.3in
]



\printAffiliationsAndNotice{\icmlEqualContribution} 

\begin{abstract}
We present \textbf{PepTune}, a multi-objective discrete diffusion model for simultaneous generation and optimization of therapeutic peptide SMILES. Built on the Masked Discrete Language Model (MDLM) framework, PepTune ensures valid peptide structures with a novel bond-dependent masking schedule and invalid loss function. To guide the diffusion process, we introduce \textbf{Monte Carlo Tree Guidance (MCTG)}, an inference-time multi-objective guidance algorithm that balances exploration and exploitation to iteratively refine Pareto-optimal sequences. MCTG integrates classifier-based rewards with search-tree expansion, overcoming gradient estimation challenges and data sparsity. Using PepTune, we generate diverse, chemically-modified peptides simultaneously optimized for multiple therapeutic properties, including target binding affinity, membrane permeability, solubility, hemolysis, and non-fouling for various disease-relevant targets. In total, our results demonstrate that MCTG for masked discrete diffusion is a powerful and modular approach for multi-objective sequence design in discrete state spaces.
\end{abstract}

\section{Introduction}

Peptides possess unique advantages as a therapeutic modality, including their low cytotoxicity and structural flexibility to bind to a diverse set of binding motifs without requiring stable binding pockets, making them ideal for targeting structurally diverse protein surfaces \cite{Dang2017, Wang2022}. However, peptides containing only the 20 wild-type amino acids have limitations, including susceptibility to enzymatic degradation and low membrane permeability \cite{Wang2022}. To overcome these limitations, non-natural amino acids (nAAs) containing diverse chemical modifications to the peptide backbone and side chains have been integrated into peptides to enhance their therapeutic properties. Despite this progress in peptide drug development, searching for the vast space of chemically modified peptides remains a major limitation \cite{Muttenthaler2021, Vinogradov2019}. This motivates the development of generative deep learning models that can effectively learn the space of clinically relevant peptides and sample \textit{de novo} peptides conditioned with various therapeutic properties.

Generative structure-based models are considered state-of-the-art for \textit{de novo} binder design, but they often rely on stable tertiary structures of target proteins \cite{Rettie2024, Bryant2023, Pacesa2024, Li2024, Watson2023}, precluding the design of peptide binders to disordered and dynamic targets, which drive a sizable portion of diseases \cite{Uversky2008}. Generative peptide design language models that depend only on the target sequence \cite{Bhat2023, Chen2024pepmlm} have demonstrated robust success on disordered and structurally diverse targets, but their use of only 20 wild-type amino acids limits these models from sampling from the space of chemically-modified or cyclic peptides. Furthermore, discrete generative models still face significant limitations in multi-objective-guided generation and optimization \cite{Austin2021, Lou2024, Shi2024, Sahoo2024, Gat2024, DDPP, Peng2025p2, fishflow, tang2025gumbelh}. Classifier-based and classifier-free guidance strategies have been explored to steer discrete diffusion objectives toward specific properties \cite{DPLM, DDPP, Stark2024, Nisinoff2024}, yet these approaches often struggle with conflicting objectives, gradient estimation, and the sparsity of quality data. 

In this work, we introduce \textbf{PepTune}, the first multi-objective-guided discrete diffusion model for \textit{de novo} peptide SMILES generation. Our key contributions include:
\begin{enumerate}
    \item \textbf{Masked Diffusion Language Model for Peptide SMILES.} We introduce the first discrete diffusion model for generating peptide SMILES with non-canonical amino acids and cyclic modifications.
    \item \textbf{Bond-Dependent Masking Schedule.} We derive the NELBO and reverse-posterior for a bond-dependent masking schedule that increases the structural validity of our generated peptide SMILES.
    \item \textbf{Global Sequence Invalid Loss.} We introduce a novel invalid loss based on our peptide SMILES filter that efficiently propagates penalties from a discrete sequence to continuous probability distributions.
    \item \textbf{Monte Carlo Tree Guidance.} We develop a robust framework for classifier-based multi-objective guidance for discrete diffusion by iteratively refining Pareto-optimal peptide sequences across therapeutic properties without gradient estimation or re-training.
    \item \textbf{Property Prediction Toolkit.} We train a set of classifiers and regressors for binding affinity, membrane permeability, solubility, hemolysis, and non-fouling.
\end{enumerate}

\section{Bond-Dependent Masked Discrete Diffusion}
\label{sec:Bond-Dependent Masked Discrete Diffusion}
\begin{figure*}
    \centering
    \includegraphics[width=0.8\linewidth]{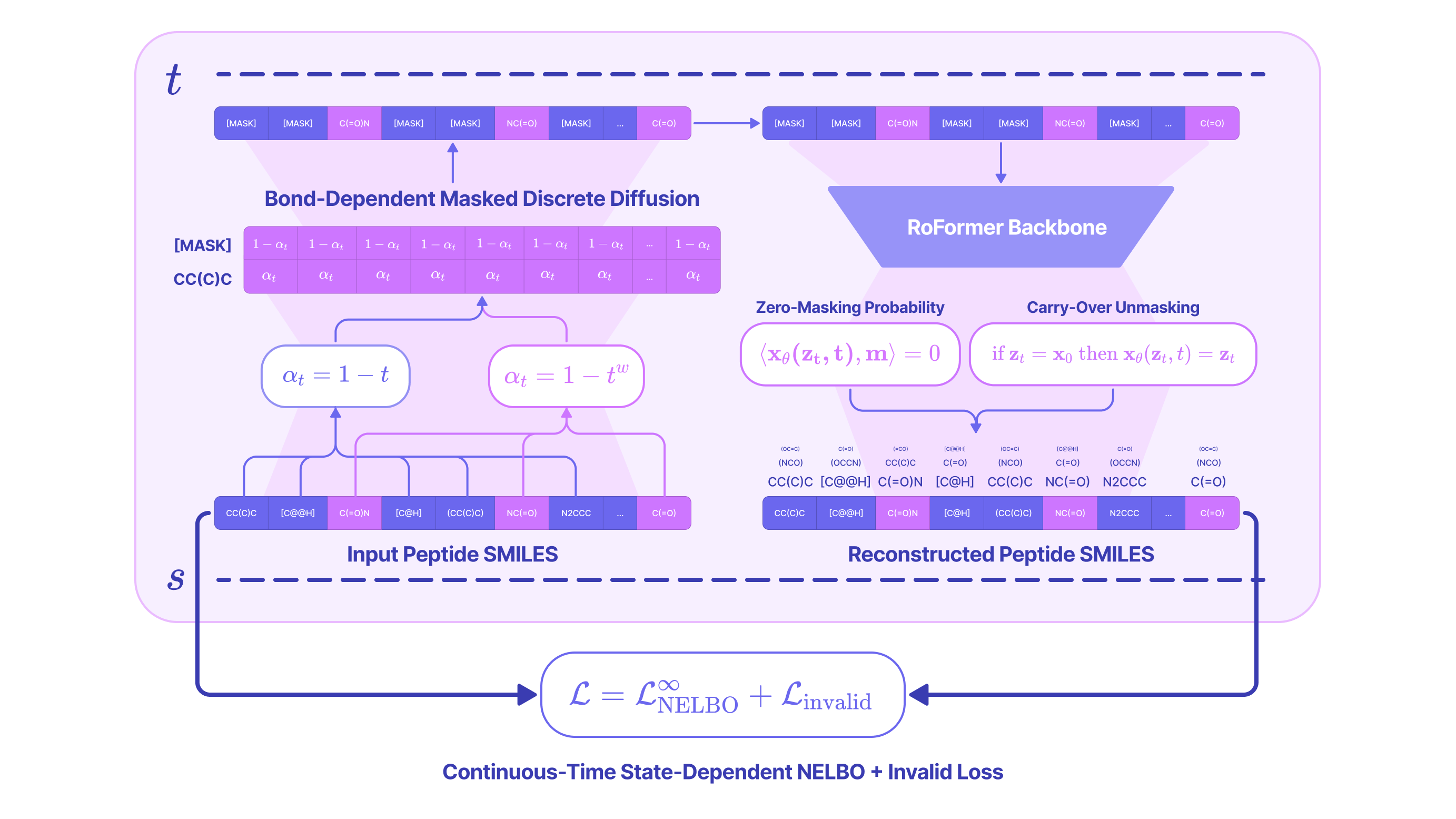}
    \caption{\textbf{PepMDLM}. PepMDLM is a discrete masked diffusion model for unconditional \textit{de novo} generation of peptide SMILES representations.}
    \label{fig:mdlm}
\end{figure*}
\subsection{Masked Discrete Diffusion Model} 
We based our unconditional generator on the Masked Diffusion Language Model (MDLM) framework, which learns to reconstruct clean sequences from sequences corrupted with [MASK] tokens (Figure \ref{fig:mdlm}) \cite{Sahoo2024, Shi2024, ou2024, Zheng2024}. The backbone model used to generate the predicted probabilities, denoted $\mathbf{x}_{\theta}(\mathbf{z}_t, t):\mathcal{V}^L\times[0,1]\to \Delta^{|\mathcal{V}|}$, of transitioning from a masked state to any token in the vocabulary $\mathcal{V}$ is predicted by a backbone RoFormer architecture (See Appendix \ref{appendix:D.2}). RoFormer leverages rotary positional embeddings (RoPE) \cite{Roformer}, which effectively captures the relative inter-token interactions in peptide SMILES, especially for cyclic peptides.

\subsection{Bond-Dependent Masking Schedule} Since all peptides follow a distinct SMILES structure consisting of un-modified or modified peptide bonds before and after each central carbon atom with an amino acid side chain, we hypothesized that applying bond-dependent masking and unmasking schedules would allow the reverse diffusion process to learn to unmask the crucial structural components of a peptide SMILES that are common across all peptides before filling in the segments in-between with diverse amino acid side-chains. 

Extending previous work in state-dependent masking \cite{Shi2024}, we devised a masking schedule where the probability of masking a token within a peptide bond increases at a slower rate in earlier times $t$ in the masking process compared to non-peptide bond tokens. To achieve this, we define the discrete-time log-linear masking schedule $\sigma(t)=-\log(1-t)$ for non-peptide bond tokens and the log-polynomial masking schedule $\sigma(t)=-\log(1-t^w)$ for peptide-bond tokens. We show in Appendix \ref{appendix:Bond-Dependent Masking Schedule} that the continuous-time probability of remaining unmasked at time $t$ in the forward diffusion process is given by the function $\alpha_t(\mathbf{x}_0): \mathbb{R}^{|\mathcal{V}|}\to \mathbb{R}$ that takes the vector encoding the token $\mathbf{x}_0$ and returns a probability 
\begin{align}
    \alpha_t(\mathbf{x}_0)=\begin{cases}1-t^w&\mathbf{x}_0=\mathbf{b}\\
                                1-t&\mathbf{x}_0\neq \mathbf{b}\end{cases}
\end{align}
where $\mathbf{b}$ is the vector with ones at indices of peptide bond tokens and zeroes in remaining indices. The tokens corresponding to peptide bonds are identified with our \textsc{BondMask} function (Algorithm \ref{alg:6}). Since the probability of transitioning to a [MASK] token at time $t$ is given by $1-\alpha_t(\mathbf{x}_0)$, there is a lower probability $t^w$ for $t\in (0, 1]$ of masking a peptide bond token than the probability $t$ of masking a non-peptide bond token, especially in earlier time steps for smaller $t$ (Figure \ref{fig:masking}A). As $t\to 1$, the probability of remaining unmasked approaches 0 ($\alpha_t(\mathbf{x}_0)\to 0$) and the probability of masking for both peptide and non-peptide bond tokens approaches 1, ensuring that the model can learn to reconstruct the full token sequence during the reverse diffusion process. 

With our bond-dependent masking rate $\alpha_t(\mathbf{x}_0)$, we define the forward transition matrix as
\begin{align}
    q(\mathbf{z}_t|\mathbf{x}_0)=\text{Cat}(\mathbf{z}_t;\alpha_t(\mathbf{x}_0)\mathbf{x}_0+(1-\alpha_t(\mathbf{x}_0))\mathbf{m})
\end{align}

\begin{proposition}[Bond-Dependent Reverse Posterior]\label{prop:bond-dependent reverse posterior}
The reverse posterior defining the probability distribution of the token $\mathbf{z}_s$ at time $s=t-\Delta t$ given the token $\mathbf{z}_t$ at time $t$ with our bond-dependent forward masking schedule is defined as 
\begin{align}
    &q(\mathbf{z}_s|\mathbf{z}_t, \mathbf{x}_0)=\nonumber
    \\&\begin{cases}
        \left\langle\left(\frac{s}{t}-\frac{s^w}{t^w}\right)\mathbf{b}+\frac{t-s}{t}\mathbf{1}, \mathbf{x}_0\right\rangle \mathbf{x}_0+
        \\\hspace{30pt}\left\langle \left(\frac{s^w}{t^w}-\frac{s}{t}\right)\mathbf{b}+\frac{s}{t}\mathbf{1}, \mathbf{x}_0\right\rangle \mathbf{m}&\mathbf{z}_t=\mathbf{m}\\
        \mathbf{z}_t&\mathbf{z}_t\neq \mathbf{m}
    \end{cases}
\end{align}
When the clean token is a peptide bond token (i.e. $\mathbf{x}_0=\mathbf{b}$), the transition distribution for a masked token $\mathbf{z}_s=\mathbf{m}$ reduces to $q(\mathbf{z}_s|\mathbf{z}_t=\mathbf{m}, \mathbf{x}_0=\mathbf{b})=\left(1-\frac{s^w}{t^w}\right)\mathbf{x}_0+\left(\frac{s^w}{t^w}\right)\mathbf{m}$. When the clean token is not a peptide bond token (i.e. $\mathbf{x}_0\neq \mathbf{b}$), the transition distribution for a masked token $\mathbf{z}_s=\mathbf{m}$ reduces to $q(\mathbf{z}_s|\mathbf{z}_t=\mathbf{m}, \mathbf{x}_0\neq \mathbf{b})=\left(1-\frac{s}{t}\right)\mathbf{x}_0+\left(\frac{s}{t}\right)\mathbf{m}$. If the token is already unmasked, it remains unmasked at the same token with probability 1.
\end{proposition}

The derivation is provided in Appendix \ref{appendix:G.2}. To estimate the reverse posterior, we define a parameterized RoFormer model $\mathbf{x}_{\theta}(\mathbf{z}_t, t):\mathcal{V}^L\times[0, 1]\to \Delta ^{|\mathcal{V}|}$ that takes the partially masked sequence at time $t$ and predicts a vector of token probabilities over the $|\mathcal{V}|$-dimensional simplex for each position in the sequence. By substituting $\mathbf{x}_0\approx\mathbf{x}_{\theta}(\mathbf{z}_t, t)$ into the true reverse transition, we get the predicted reverse transition distribution.
\begin{align}
    &p_{\theta}(\mathbf{z}_s|\mathbf{z}_t)=\nonumber
    \\&\begin{cases}
        \left\langle\left(\frac{s}{t}-\frac{s^w}{t^w}\right)\mathbf{b}+\frac{t-s}{t}\mathbf{1}, \mathbf{x}_{\theta}(\mathbf{z}_t,t)\right\rangle\mathbf{z}_s+
        \\\hspace{10pt}\left\langle\left(\frac{s^w}{t^w}-\frac{s}{t}\right)\mathbf{b}+\frac{s}{t}\mathbf{1}, \mathbf{x}_{\theta}(\mathbf{z}_t,t)\right\rangle \mathbf{m}&\mathbf{z}_t=\mathbf{m}\\
        \mathbf{z}_t&\mathbf{z}_t\neq \mathbf{m}\label{eq:8}
    \end{cases}
\end{align}
For larger $w$, peptide bonds are masked at later timesteps, encouraging earlier unmasking in the reverse diffusion process. However, setting $w$ too large can result in the model over-fitting to the dataset \cite{Shi2024}. Empirically, we found that $w=3$ increased peptide validity while maintaining diversity across generated samples. 

\subsection{Loss Functions}
\label{sec:Loss Functions}
\paragraph{Bond-Dependent Continuous-Time Diffusion Loss} To optimize the parameters $\theta$ of the reverse diffusion model, we maximize the evidence lower bound (ELBO) of the distribution $\log p(\mathbf{x}_0)$, which is the log-probability distribution of generating the peptide sequences $\mathbf{x}_0$ present in the dataset. Therefore, we define our loss function as the negative ELBO (NELBO) (Appendix \ref{appendix:B.2}).

Training on samples masked for continuous values of $t\sim \text{Uniform}(0,1)$ yields a tighter lower bound compared to discrete values of $t$ \cite{Kingma2021}. When the predicted probability distribution $\mathbf{x}_{\theta}(\mathbf{z}_t, t)$ is exactly the one-hot encoding vector $\mathbf{x}_0$ for each position $\ell$ in the true sequence, the loss reduces to 0, which supports our objective.

\begin{proposition}[Bond-Dependent NELBO]\label{prop:bond-dependent NELBO}
The bond-dependent continuous-time NELBO decomposes into the sum of the negative log-losses (NLLs) for all non-peptide bond tokens that follow a log-linear masking schedule and the sum of the NLLs for all peptide bond tokens that follow a log-polynomial schedule. 
\begin{align}
    \mathcal{L}^{\infty}_{\text{NELBO}}&=\mathbb{E}_{t, q(\mathbf{z}_t|\mathbf{x}_0)}\bigg[-\sum_{\ell:\mathbf{x}^{(\ell)}_0=\mathbf{b}}\frac{w}{t}\log\langle\mathbf{x}^{(\ell)}_0,\mathbf{x}^{(\ell)}_{\theta}(\mathbf{z}_t, t)\rangle\nonumber\\
    &-\sum_{\ell:\mathbf{x}^{(\ell)}_0\neq\mathbf{b}}\frac{1}{t}\log\langle\mathbf{x}^{(\ell)}_0,\mathbf{x}^{(\ell)}_{\theta}(\mathbf{z}_t, t)\rangle\bigg]\label{eq:13}
\end{align}
where $\ell\in \{1, \dots , L\}$ denotes the position in the sequence.
\end{proposition} 

The derivation is provided in Appendix \ref{appendix:G.3}. Since the NLL term is minimized when the predicted probability of the ground truth token is close to 1, we show that applying the log-polynomial masking schedule for an exponent $w> 1$ scales the diffusion loss NELBO by a factor of $w$ from the log-linear schedule. However, for earlier timesteps as $t\to 0$, both NLL weights increase to $\infty$, ensuring high precision in the final unmasking steps (Appendix Figure \ref{fig:masking}). 

Given that peptide bonds form the fundamental backbone structure of a peptide, our bond-dependent masking strategy for peptide bonds acts as a peptide bond loss that introduces a higher penalty when the token predictions at positions of peptide bonds are inconsistent from the ground truth tokens during training, forcing the model to learn the specific structure of peptide SMILES strings in a vast space of SMILES strings that are not valid peptides.

\paragraph{Invalid Peptide Loss} To further discourage the generation from predicting token logits that produce invalid peptide SMILES, we incorporate a loss to penalize sampling of invalid peptide SMILES during training by taking the \texttt{argmax} of the predicted logits and assigning a penalty based on our peptide validity filter (Appendix \ref{alg:7}). Given the clean peptide sequence $\tilde{\mathbf{x}}_0\in \mathcal{V}^L$ generated from the \texttt{argmax} tokens with the highest probability from the predicted logits $\mathbf{x}_{\theta}(\mathbf{z}_t, t)$, we minimize a penalty determined by our validity filter $\mathbf{1}[\tilde{\mathbf{x}}_0\text{ is Invalid}]$ which returns 0 when the sequence is a valid peptide SMILES and 1 when the sequence either contains invalid SMILES notation or cannot be decoded into a peptide sequence. Since the \texttt{argmax} function is not differentiable, we use the \texttt{softmax} probability of the sampled tokens to scale the penalty score for each token in the loss function. 
\begin{align}
    \mathcal{L}_{\text{invalid}}&= \sum_{\ell=1}^L \tilde{\mathbf{x}}_{0}^{(\ell)\top}\text{SM}\big(\mathbf{x}_{\theta}^{(\ell)}(\mathbf{z}_t, t)\big)\cdot \mathbf{1}[\tilde{\mathbf{x}}_0\text{ is Invalid}] \nonumber\\
    &=\sum_{\ell=1}^L\frac{\exp(x^{(\ell)}_{\theta, k})}{\sum_{j=1}^K \exp(x_{\theta,j}^{(\ell)})}\cdot \mathbf{1}[\tilde{\mathbf{x}}_0\text{ is Invalid}]\label{eq:invalid-loss}
\end{align}
where $k=\arg\max_j(\mathbf{x}^{(\ell)}_{\theta}\big(\mathbf{z}_t,t)\big)$ is the token with the highest predicted probability at position $\ell$ of the sequence. 

\begin{proposition}[Gradient Flow of Invalid Loss]\label{prop:invalid loss}
    By differentiating the invalid loss with respect to the probability vector $\mathbf{x}_{\theta}^{(\ell)}(\mathbf{z}_t, t)$ for position $\ell$, the gradient with respect to the predicted probability of the sampled token $j=k$ and all other tokens in the vocabulary $j\neq k$ is given by
    \begin{align}
        \nabla\mathcal{L}_{\text{invalid}}=\begin{cases}
            \text{SM}(x^{(\ell)}_{\theta, k})\left(1-\text{SM}(x^{(\ell)}_{\theta. k})\right)&j=k\\
            -\text{SM}(x^{(\ell)}_{\theta, j})\text{SM}(x^{(\ell)}_{\theta, k})&j\neq k
        \end{cases}
    \end{align}
\end{proposition}

The derivation is provided in Appendix \ref{appendix:G.4}. Minimizing this objective function updates the parameters to lower the predicted probabilities for tokens that result in invalid peptide SMILES and increase the probabilities of the remaining tokens proportional to their original distribution, such that the relative probability distribution of all other tokens $j\neq k$ is maintained.  

\paragraph{Training} To train the MDLM to accurately approximate the true reverse transition distribution $q(\mathbf{z}_s|\mathbf{z}_t, \mathbf{x}_0)$ of a training sample $\mathbf{x}_0$ for all continuous timesteps $t=1\to 0$, we train a parameterized model that takes the partially masked sequence $\mathbf{z}_t$ and returns a probability distribution $\mathbf{x}_{\theta}(\mathbf{z}_t, t)$ that approximates the clean one-hot vector $\mathbf{x}_0$ (Algorithm \ref{alg:1}). For each dynamic training batch $B$, we randomly sample $|B|$ values $t\in \text{Uniform}(0, 1)$ and off-set each time $t$ by $\delta= \left[0, \frac{1}{|B|},\frac{2}{|B|}, \dots, \frac{|B|-1}{|B|}, 1\right]$ to get a vector $\mathbf{t}=(\mathbf{t}+\delta) \text{ mod } \mathbf{1}  $ of evenly distributed time steps to ensure the model learns to regenerate the clean sample $\mathbf{z}_t$ for a continuous range of time steps. After applying bond-dependent masking to each training sequence, obtaining the predicted probabilities $\mathbf{x}_{\theta}(\mathbf{z}_t, t)$ and sampling the discrete sequence $\tilde{\mathbf{x}}_0$ from greedy \texttt{argmax} sampling, we minimize the total loss function $\mathcal{L}$ given by 
\begin{align}
    \mathcal{L}&=\mathcal{L}_{\text{NELBO}}^{\infty}+\mathcal{L}_{\text{invalid}}
\end{align}
By increasing batch size and applying dynamic batching (Appendix \ref{appendix:C.2}), we obtain a tighter ELBO of the true distribution $\log p(\mathbf{x}_0)$. The model used to generate the validation results in this manuscript is trained on our in-house 8$\times$A6000 Nvidia GPUs (50G memory) for 1600 GPU hours using the AdamW optimizer with a learning rate of 0.0003 and weight decay of 0.075. After training for 8 epochs with 11 million peptide SMILES (Appendix \ref{appendix:C.1}), we achieved a train loss of 0.832 and a validation loss of 0.880. 

\paragraph{Sampling} 
To sample from the unconditional PepMDLM model, we start with a sequence $\mathbf{x}_0= \text{[MASK]}^L$ of length $L$ of only [MASK] tokens. We first compute the diffusion time steps $t\in \{\frac{1}{T},\frac{2}{T}, \dots, 1\}$ where $T$ is the number of denoising steps ($T=128$). From the predicted token probabilities $\mathbf{x}_{\theta}(\mathbf{z}_t, t)$ generated by feeding $\mathbf{z}_t$ through the trained RoFormer backbone, we compute the reverse transition token distribution $p_{\theta}(\mathbf{z}_s|\mathbf{z}_t)$ following Equation (\ref{eq:8}) and perform Gumbel-max sampling to get the next token $\mathbf{z}_s$.
\begin{align}
    \mathbf{z}_s&\sim \arg \max \left(\log p_{\theta}(\mathbf{z}_s|\mathbf{z}_t)+\mathbf{G}\right)
\end{align}
$G_j=-\log(-\log(u_j+\epsilon)+\epsilon)$ is the i.i.d. sampled Gumbel noise applied to the $j$th token probability, $u_j\sim \text{Uniform}(0,1)$, and $\epsilon =1e-10$. Then, we return the newly sampled tokens only when $\mathbf{z}_t=\mathbf{m}$, while keeping all unmasked tokens unchanged. After $T$ timesteps, we obtain a fully unmasked sequence $\mathbf{x}$.

\section{Multi-Objective Guided Discrete Diffusion}
\label{sec:Multi-Objective Guidance for Discrete Diffusion}
In this section, we describe the concept of Pareto dominance and non-dominance for multiple objectives and introduce \textbf{Monte Carlo Tree Guidance (MCTG)}, a novel algorithm that reformulates the Monte Carlo Tree Search (MCTS) framework \cite{Coulom2007} for multi-objective-guided discrete diffusion. 

\begin{figure*}
    \centering
    \includegraphics[width=\textwidth]{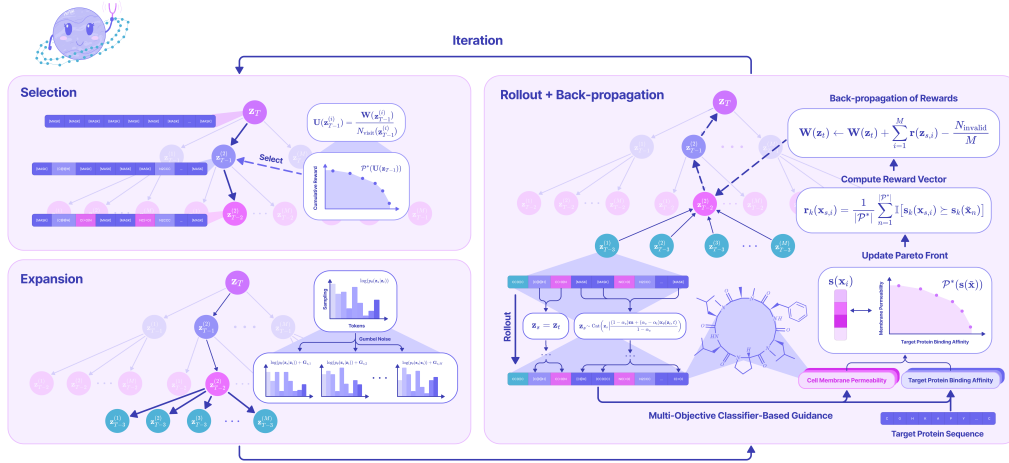}
    \caption{\textbf{PepTune}. PepTune is a multi-objective discrete diffusion model with  Monte Carlo Tree Guidance (MCTG). The full algorithm is detailed in Algorithm \ref{alg:2}.}
    \label{fig:mcts}
\end{figure*}

\paragraph{Pareto Optimization} When optimizing sequences for multiple objectives (e.g., affinity to multiple protein targets, membrane permeability, solubility, etc.), there is likely no single best sequence that achieves the highest score across all objectives. Optimizing one objective often leads to sacrificing performance on another objective. 

Therefore, we focus on finding a set of Pareto optimal sequences that minimize the trade-offs between objectives to achieve overall optimal performance across all objectives. Formally, Pareto-optimal sequences (or non-dominated sequences) cannot be further optimized in any single objective without sacrificing performance in another objective. 

Let $\mathbf{s}(\mathbf{x})=[s_1(\mathbf{x}), \dots, s_K(\mathbf{x})]\in \mathbb{R}^K$ be a vector of scores that measures the performance of a sequence $\mathbf{x}$ in $K$ different objectives, with higher scores indicating better performance. A sequence $\mathbf{x}^*$ is said to dominate another sequence $\mathbf{x}$ (denoted as $\mathbf{s}(\mathbf{x}^*)\succ\mathbf{s}(\mathbf{x})$) if and only if it satisfies the following property. For all objectives $k\in \{1,\dots , K\}$, the score for the $k$th objective for $\mathbf{x}^*$ is greater than or equal to the score for the $k$th objective for $\mathbf{x}$, and for at least one objective $k’$, the score for $\mathbf{x}^*$ is strictly greater than the score for $\mathbf{x}$. 

A Pareto-optimal sequence $\mathbf{x}$ is a sequence where there does not exist another sequence $\mathbf{x}^*$ in the current Pareto-optimal set $\mathcal{P}^*$ that dominates it. Since there are trade-offs between objectives, this does not mean that $\mathbf{x}$ is dominant over all other sequences.
\begin{align}
    \underbrace{\nexists \mathbf{x}^*\in \mathcal{P}^*\;\;\text{s.t.}\;\;\mathbf{s}(\mathbf{x}^*)\succ \mathbf{s}(\mathbf{x})}_{\mathbf{x} \text{ is non-dominated}}
\end{align}
We define the Pareto front as the set of non-dominated sequences $\mathbf{x}$ and their $K$-dimensional objective score vectors.
\begin{align}
    \mathcal{P}^*=\left\{\big(\mathbf{x}, \mathbf{s}(\mathbf{x})\big)\;|\;\nexists \mathbf{x}^*\in \mathcal{P}^*\;\;\text{s.t.}\;\;\mathbf{s}(\mathbf{x}^*)\succ \mathbf{s}(\mathbf{x})\right\}
\end{align}
Since infinitely many trade-offs can exist between the $K$ objectives, there can be an infinite number of Pareto-optimal sequences. Therefore, multi-objective optimization aims to approximate a finite set of Pareto-optimal sequences with a reasonable number of iterations.

\paragraph{Notation} Let $\mathbf{z}_t$ denote the partially unmasked sequence at time $t$. $\mathbf{z}_t$ also corresponds to a node in the MCTS tree with a set of $M$ children nodes denoted as $\text{children}(\mathbf{z}_t)=\{\mathbf{z}_{s,1}, \dots, \mathbf{z}_{s, M}\}$. Each child node is itself a partially unmasked sequence at time $s$ derived from sampling the MDLM reverse posterior $p_{\theta}(\mathbf{z}_{s}|\mathbf{z}_{t})$. The children nodes at each iteration of MCTS are rolled out into a set of clean sequences denoted as $\{\mathbf{x}_{s,1}, \dots, \mathbf{x}_{s, M}\}$, for each of which we compute a score vector $\mathbf{s}(\mathbf{x}_{s,i})\in \mathbb{R}^K$ and a rewards vector $\mathbf{r}(\mathbf{x}_{s,i})\in \mathbb{R}^K$, where $K$ is the number of objectives guiding the MCTS search. 

Let $\mathcal{P}^*=\{\mathbf{x}^*_n\}$ be the set of $|\mathcal{P}^*|$ Pareto non-dominated sequences indexed $n\in \{1, \dots,  |\mathcal{P}^*|\}$, which is updated at each iteration. At each node $\mathbf{z}_t$, we store a cumulative rewards vector $\mathbf{W}(\mathbf{z}_t)$ and a counter for the number of times the node has been visited across all iterations $N_{\text{visit}}(\mathbf{z}_t)$. Finally, we denote the total number of search iterations with $N_{\text{iter}}$.

\paragraph{Initialization} We initialize a sequence $\mathbf{z}_{t(T)}= \text{[MASK]}^L$ of length $L$ of [MASK] tokens as the root node of the MCTS tree corresponding to time $t(T)$ and an empty set $\mathcal{P}^*$ that will maintain clean sequences with Pareto-optimal score vectors. We initialize a set of scoring functions $\mathbf{s}: \mathcal{V}^L\to \mathbb{R}^K$ that takes a clean sequence $\mathbf{x}_{s, i}\in \mathcal{V}^L$ generated from the partially masked sequence $\mathbf{z}_{s, i}$ and outputs a vector of real values $\mathbf{s}(\mathbf{x}_{s, i})\in \mathbb{R}^K$ that measures its performance in each of the $K$ objectives. We also set the hyperparameters, including the number of iterations $N_{\text{iter}}$ and the number of children $M$. 
 
At each iteration, four steps are performed to update the set of Pareto optimal solutions: traversing the tree by selecting a Pareto-optimal unmasking step until reaching a unexpanded leaf node (selection), expanding the leaf node into $M$ distinct partially unmasked sequences (expansion), fully unmasking each child node into a clean sequence and computing multi-objective score and reward vector (rollout), and finally back-propagating the total rewards to the predecessor nodes to guide the selection process at the next iteration (backpropagation). 

\paragraph{Selection} At each iteration, we traverse the tree starting at the root node (fully masked sequence) $\mathbf{z}_{t(T)}$ and selecting a child node based on the selection score vector $\mathbf{U}(\mathbf{z}_t, \mathbf{z}_{s, i})$ that balances child nodes that generate high reward sequences from previous iterations and unexplored unmasking actions that could lead to a larger pool of diverse sequences. 
\begin{align}\label{eq:define_U}
    \mathbf{U}(\mathbf{z}_t, \mathbf{z}_{s, i})=\frac{\mathbf{W}(\mathbf{z}_{s,i})}{N_{\text{visit}}(\mathbf{z}_{s,i})}+c\cdot p_{\theta}(\mathbf{z}_{s, i}|\mathbf{z}_t)\frac{\sqrt{N_{\text{visit}}(\mathbf{z}_t)}}{1+N_{\text{visit}}(\mathbf{z}_{s,i})}\nonumber
\end{align}
The first term is the cumulative reward vector $\mathbf{W}(\mathbf{z}_{s,i})$ normalized by the number of times the node was previously visited. This guides the selection process towards the unmasking step that has resulted in fully unmasked sequences with optimal properties without biasing towards highly visited nodes. The second term is a scalar added element-wise to the normalized rewards. The scalar probability of the unmasking step based on the unconditional reverse posterior $p_{\theta}(\mathbf{z}_{s, i}|\mathbf{z}_t)$ guides the selection towards the unmasking step with the highest probability to generate a valid peptide based on the pre-trained MDLM. When the number of times the parent node has been explored is high and the number of visits to a child node is low, the $\frac{\sqrt{N_{\text{visit}}(\mathbf{z}_t)}}{1+N_{\text{visit}}(\mathbf{z}_{s,i})}$ term encourages exploration of the unexplored unmasking scheme given that $p_{\theta}(\mathbf{z}_{s, i}|\mathbf{z}_t)$ is sufficiently high. However, as the number of visits to a child node increases, the impact of the second term decreases, and the cumulative rewards dominate the selection score vector. $c$ is a scalar hyperparameter that determines the degree of exploration compared to exploiting high-reward nodes, which is selected to be $c=0.1$. 

Then, we select uniformly at random from the pool of children nodes $\mathbf{z}_{s,i}\in \mathcal{P}^*_{\text{select}}$ whose selection score vectors are non-dominated, such that there does not exist another child node $\mathbf{z}_{s,j}$ where the selection score vector has a score strictly greater than that of $\mathbf{z}_{s,i}$ in at least one of the $K$ objectives and equal scores across all the remaining objectives.
\begin{align}
    \mathcal{P}^*_{\text{select}}=&\{\mathbf{z}_{s,i}\;|\;\nexists \mathbf{z}_{s, j}\in \text{children}(\mathbf{z}_t)\nonumber\\
    &\text{s.t.}\;\;\mathbf{U}(\mathbf{z}_t,\mathbf{z}_{s,j})\succ \mathbf{U}(\mathbf{z}_t,\mathbf{z}_{s,i})\}
\end{align}
If the selected node is a non-leaf node, the loop repeats with the selected node $\mathbf{z}_{s,i}$ as the new parent node. If a fully unmasked node with $t=0$ is reached, we restart the selection process from the root node. Once a leaf node is reached, the loop ends, and the next step executes. 

\paragraph{Expansion}  At the iteration at time $t$, we sample $M$ sequences from the reverse posterior $p_{\theta}(\mathbf{z}_{s}|\mathbf{z}_t)$ defined in Equation (\ref{eq:8}) to get a set of partially masked sequences which form the set of children nodes of $\mathbf{z}_t$: $\text{children}(\mathbf{z}_t)=\{\mathbf{z}_{s,1}, \dots, \mathbf{z}_{s, M}\}$. All the children nodes are added to the tree. 

To ensure that the expansion step results in $M$ distinct unmasking steps, we experimented with two different batched unmasking techniques from the single partially masked sequence at a parent node. For the first method, we repeated the array corresponding to the parent node tokens over $M$ dimensions and added independently sampled Gumbel noise values $G_{i,j}$, where $i$ denotes the sequence in the batch and $j$ denotes the token index. 
\begin{align}\label{eq:define_gumbel}
    \log\tilde{p}_{\theta, i}\big(\mathbf{z}_{s,i}|\mathbf{z}_t) &= \log p_{\theta}\big(\mathbf{z}_{s,i}|\mathbf{z}_t\big)+\mathbf{G}_i
\end{align}
where $\mathbf{G}_i\sim  \text{Gumbel}(0, 1)$ and $\tilde{p}_{\theta, i}$ denotes the $i$th perturbed reverse transition distribution after applying Gumbel noise independently sampled for each sequence $i$ in the batch, where $i\in\{1,\dots, M\}$. Then, we sample $M$ distinct child sequences from each of the distributions $\mathbf{z}_{s,i}\sim \tilde{p}_{\theta, i}\big(\mathbf{z}_{s,i}|\mathbf{z}_t\big)$.

The second method involves taking the \texttt{softmax} (denoted as SM) across the top $k$ probabilities after applying Gumbel noise and drawing random samples from the re-normalized softmax distribution over only the top $k$ most probable tokens.
\begin{align}
    \tilde{p}_{\theta, i}\big(\mathbf{z}^{(\ell)}_{s,i}|\mathbf{z}^{(\ell)}_t) &= \text{SM}\bigg(\text{top}k\big\{\log p_{\theta}\big(\mathbf{z}^{(\ell)}_{s,i}|\mathbf{z}^{(\ell)}_t\big)+\mathbf{G}_i^{(\ell)}\big\}\bigg)\nonumber
\end{align}
After empirical experimentation, we found that the first method results in higher diversity across sequences, whereas the second method prevents unlikely tokens. Since the reward generated by a sequence ultimately determines whether it is selected in subsequent iterations, we chose the first method to allow for greater exploration during the expansion step.

\paragraph{Rollout} From each child node generated at time $s$, we completely unmask the sequence by greedily sampling the \texttt{argmax} tokens from the predicted reverse transition distribution ${p}_{\theta,i}(\mathbf{z}_{s'}|\mathbf{z}_s)$ for all remaining time steps $s\to 0$ and $s'=s-\frac{1}{T}$ to get a set of clean sequences $\{\mathbf{x}_{s,1}, \dots, \mathbf{x}_{s, M}\}$ of SMILES tokens. We feed each clean sequence $\mathbf{x}_{s, i}$ as input to the scoring functions for all of the $K$ objectives to generate the score vector $\mathbf{s}(\mathbf{x}_{s,i})=\big[s_1(\mathbf{x}_{s,i}), \dots , s_K(\mathbf{x}_{s,i})\big]\in \mathbb{R}^K$. Then, we use the score vector to compute a vector of rewards $\mathbf{r}(\mathbf{x}_{s,i})=\big[r_1(\mathbf{x}_{s,i}), \dots, r_K(\mathbf{x}_{s,i})\big]\in \mathbb{R}^K$. To generate the property scores given an input peptide SMILES, we train regression models for target-binding affinity and cell membrane permeability and binary classification models for solubility, hemolysis, and non-fouling specifically on peptide SMILES data (Appendix \ref{appendix:E}). 

The reward of a child node sequence for the $k$th objective is the fraction of the sequences $\mathbf{x}^*_n$ in the current set of Pareto-optimal sequences $\mathcal{P}^*$ where the child node has a higher classifier score in that objective. Specifically, the reward for the $i$th child node $\mathbf{z}_{s, i}$ and the resulting unmasked sequence $\mathbf{x}_{s, i}$ for the $k$th objective is given by
\begin{align}
    r_k(\mathbf{x}_{s, i})=\frac{1}{|\mathcal{P}^*|}\sum_{n=1}^{|\mathcal{P}^*|}\mathbf{1}\big[s_k(\mathbf{x}_{s, i})\geq s_k(\mathbf{x}^*_n)\big]
\end{align}
where $\mathbf{1}$ is an indicator function that returns 1 if the score for the $k$th objective of the $i$th child node is greater than or equal to the score of the $n$th sequence in the Pareto-optimal set $\mathcal{P}^*$. In parallel to computing the reward, we add all non-dominated children sequences to the set of Pareto optimal sequences $\mathcal{P}^*$ and remove all dominated sequences (Algorithm \ref{alg:5}). 
\begin{align}
    \mathcal{P}'^*&=\mathcal{P}^*\cup \big\{(\mathbf{z}_{s,i}, \mathbf{s}(\mathbf{x}_{s, i}))\;|\;\forall \tilde{\mathbf{x}}\in \mathcal{P}^*\;\;\mathbf{s}(\mathbf{x}_{s, i})\succeq\mathbf{s}(\tilde{\mathbf{x}})\big\}\nonumber\\
    \mathcal{P}'^*&=\mathcal{P}^*\setminus \big\{\tilde{\mathbf{x}}\;|\;\exists \mathbf{x}_{s, i}\in \text{children}(\mathbf{z}_t)\;\text{s.t.}\;\mathbf{s}(\mathbf{x}_{s, i})\succ\mathbf{s}(\tilde{\mathbf{x}})\big\}\nonumber
\end{align}
In Appendix \ref{appendix:I.1}, we show a proof-of-concept for a time-dependent multi-objective guidance strategy where the update to the Pareto-optimal set $\mathcal{P}^*$ depends on the rewards for only a subset of the $K$ objectives that varies depending on the current iteration, enabling the prioritization of properties with larger influence on peptide structure and function in earlier iterations and fine-tuning on additional properties in later iterations. 

\paragraph{Back-propagation}   At each child node $\mathbf{z}_{s, i}$, the reward vector $\mathbf{r}(\mathbf{x}_{s,i})$ is used to initialize the cumulative reward vector $\mathbf{W}(\mathbf{z}_{s,i})$, and the number of visits $N_{\text{visits}}(\mathbf{z}_{s,i})$ is initialized to 1.
\begin{align}
    \mathbf{W}(\mathbf{z}_{s,i})\gets \mathbf{r}(\mathbf{x}_{s,i}), \;N_{\text{visit}}(\mathbf{z}_{s,i})\gets 1
\end{align}
Then, we backtrack through the predecessor nodes of $\mathbf{z}_{s, i}$ up to the root node $\mathbf{z}_{t(T)}$, adding the child reward vector to the cumulative reward vector and incrementing the number of visits for each node in the path. For all nodes from $\mathbf{z}_{t}= \text{parent}(\mathbf{z}_{s,i})$ to $\mathbf{z}_{t}=\mathbf{z}_{t(T)}$ we apply the following update
\begin{small}
    \begin{align}
    \mathbf{W}(\mathbf{z}_{t})&\gets \mathbf{W}(\mathbf{z}_{t})+\sum_{i=1}^M\mathbf{r}(\mathbf{x}_{s, i})\\
    N_{\text{visit}}(\mathbf{z}_{t})&\gets N(\mathbf{z}_{t})+1
\end{align}
\end{small}
These updated scores are used to guide the selection process in the next iteration, such that the unmasking paths that result in the highest reward sequences have a greater chance of being selected and explored further. 


\begin{table*}
\caption{\textbf{Evaluation metrics for generative quality of peptide SMILES sequences of max token length set to 200.}}
\label{table:data_composition_evaluation}
\vskip 0.05in
\begin{center}
\begin{small}
\resizebox{0.8\textwidth}{!}{
\begin{tabular}{@{}lcccccc@{}}
\toprule
\textbf{Model} & \textbf{Validity ($\uparrow$)} & \textbf{Uniqueness} ($\uparrow$)& \textbf{Diversity} ($\uparrow$)& \textbf{SNN} ($\downarrow$)& \textbf{Randomness} ($\uparrow$)& \textbf{KL-Divergence} ($\uparrow$)\\
\midrule
Data & 1.000 & 1.000 & 0.885 & 1.000 & 4.55 & 0 (Reference)\\
PepMDLM & 0.450 & 1.000 & 0.705 & 0.513 & 4.11 & 0.174 \\
PepTune & 1.000 & 1.000 & 0.677 & 0.486 & 4.12 & 0.173 \\ 
\bottomrule
\end{tabular}
}
\end{small}
\end{center}
\end{table*}

\paragraph{Output} The output after $N_{\text{iter}}$ iterations is the set $\mathcal{P}^*$ of Pareto-optimal sequences across the $K$ objectives. Our strategy simultaneously guides the unmasking process towards optimality across multiple objectives directly in the discrete state space while exploring the diverse space of peptide sequences using the trained unconditional MDLM generator. Furthermore, we generate a set of Pareto-optimal sequences from a single run through the MCTS-search algorithm, which are non-dominated from the total of $N_{\text{iter}}\cdot M$ total sequences sampled across all iterations.

\section{Therapeutic Property Classifiers}
While several classifiers exist for predicting properties of small-molecule SMILES sequences and amino-acid representations of peptides, there exists a gap in high-quality property models trained specifically on peptide SMILES data. To fill this gap, we train regression models for target-binding affinity and cell membrane permeability (Appendix Table \ref{table:affinity-predictor}; Figure \ref{fig:binding_affinity and permeability}) and binary classification models for solubility, hemolysis, and non-fouling specifically on peptide SMILES data. Our prediction models achieve significantly enhanced performance in peptide property prediction compared to the state-of-the-art PeptideBERT \cite{Guntuboina2023} baseline (Table \ref{table:solubility_hemolysis_non_fouling}). 

\begin{table}[H]
    \caption{\textbf{Benchmarks of solubility, hemolysis, and non-fouling prediction against PeptideBERT \cite{Guntuboina2023}.} We leveraged PeptideCLM embedding representations of the SMILES tokens and trained XGBoost models for binary classification.}
    \vskip 0.05in
    \begin{small}
    \resizebox{\linewidth}{!}{
    \begin{tabular}{@{}lccccccc@{}}
    \toprule
    \multicolumn{1}{c}{} & \multicolumn{2}{c}{\textbf{Solubility}} & \multicolumn{2}{c}{\textbf{Hemolysis}} & \multicolumn{2}{c}{\textbf{Non-fouling}} \\ 
    \cmidrule(lr){2-3} \cmidrule(lr){4-5} \cmidrule(lr){6-7}
    Metric & Ours & PeptideBERT & Ours & PeptideBERT & Ours & PeptideBERT\\ 
    \midrule
    \textbf{F1} & \textbf{0.660} & 0.597 & \textbf{0.846} & 0.483 & \textbf{0.768} & 0.699\\ 
    \textbf{Accuracy} & \textbf{0.661} & 0.651 & \textbf{0.846} & 0.823 & 0.766 &  \textbf{0.873}\\ 
    \bottomrule
    \end{tabular}
    }
    \end{small}
    \label{table:solubility_hemolysis_non_fouling}
\end{table}

\section{Experiments}
\paragraph{PepMDLM generates diverse chemically-modified and cyclic peptides.} Our optimized unconditional MDLM (PepMDLM) shows increased uniqueness and diversity with lower SNN compared to the autoregressive generator of macrocyclic peptides, HELM-GPT \cite{Xu2024}, demonstrating our capability to comprehensively search the sub-space of valid peptide SMILES (Table \ref{table:benchmark}). Furthermore, our unconditional generator PepMDLM generates valid peptides with a higher average nAA frequency than experimentally-validated peptide SMILES for membrane permeability and binding affinity (Figure \ref{fig:nAA}), demonstrating our unique ability to design \textit{de novo} peptides with cyclic and nAA modifications and expanding the search space of therapeutic peptides well beyond any generative model trained on canonical amino acid representations.

Even though the multi-objective selection process favors high-reward unmasking steps, we show that the resulting pool of PepTune-generated peptides retains similar uniqueness and diversity scores to the peptides generated by PepMDLM and in the training dataset (Table \ref{table:data_composition_evaluation}). In addition, the fraction of valid peptides consistently reaches 100\% after only 20 iterations of the MCTS search algorithm, demonstrating the effectiveness of backpropagating the classifier-based rewards.

\begin{figure}
    \centering
    \includegraphics[width=\linewidth]{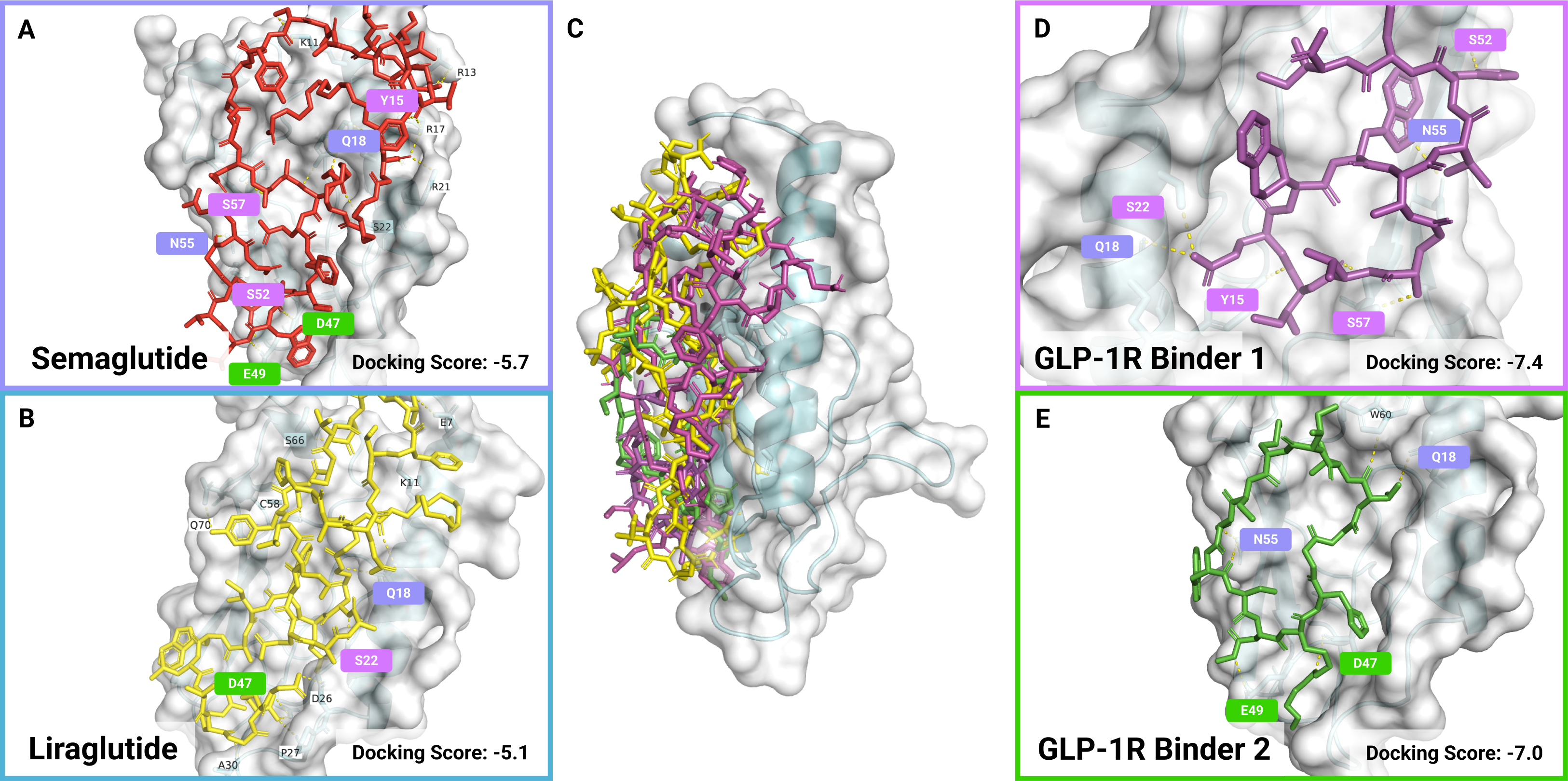}
    \caption{\textbf{Comparison of docked PepTune-generated peptides to existing GLP-1R agonists.} (\textbf{A, B}) Docking images of semaglutide (score: -5.7 kcal/mol) and liraglutide (score: -5.1 kcal/mol) binding to GLP-1R. \textbf{(C)} Full view of the positive control GLP-1R agonists and the PepTune-generated binders on GLP-1R. (\textbf{D, E)} Docking images of binder 1 (score: -7.4 kcal/mol) and 2 (score: -7.0 kcal/mol) were generated using PepTune, conditioned on predicted affinity to GLP-1R, solubility, hemolysis, and non-fouling. Shared polar contacts between binder 1 and either controls are highlighted in pink, shared polar contacts between binder 2 and either controls are highlighted in green, and the shared contacts across both binders are highlighted in purple.}
    \label{fig:glp-control}
\end{figure}

\paragraph{PepTune enables multi-objective generation of therapeutic peptide binders.}
\label{results:Case Study for Multi-Objective Generation of Peptide Binders}
Given the significant development of glucagon-like peptide-1 (GLP-1R) peptide agonists for the treatment of type-2 diabetes and obesity \cite{Alfaris2024}, we compared GLP-1R binding affinity-conditioned peptides generated using PepTune with recent blockbuster GLP-1R agonists: semaglutide and liraglutide. Both semaglutide and liraglutide are over 30 amino acids in length and act by mimicking the binding of natural GLP-1 by binding to the activation pocket of GLP-1R with high precision (Figure \ref{fig:glp-control}) \cite{Mahapatra2022, Nauck2022}. 

Shorter agonists or antagonists to GLP-1R would serve several benefits to the treatment of insulin-related disorders, including reduced cost and complexity of synthesis, lower immunogenicity, and faster tissue penetration. Therefore, we sought to generate shorter-chain peptides that are capable of binding to GLP-1R with comparable affinity to the existing agonists. We first generated a pool of peptide binders conditioned on binding affinity with the GLP-1R sequence, solubility, hemolysis, and non-fouling. After selecting the peptides with the highest predicted binding affinity scores from the Pareto non-dominated set, we performed docking and determined docking scores of -7.4 kcal/mol and -7.0 kcal/mol for the two best candidates. Our peptides show superior docking affinity to GLP-1R while interacting at overlapping binding motifs to semaglutide and liraglutide derived from the natural hormone ligand, GLP-1 (Figure \ref{fig:glp-control}). These results suggest that our PepTune-derived peptides can serve as potent agonists or antagonists of GLP-1R signaling. 

\paragraph{PepTune generates optimized dual-target-binding peptides.}\label{results:Case Study for Dual-Target Binding Peptides}
\begin{table*}
\centering
\caption{\textbf{Property metrics for PepTune-generated dual-target binders to TfR and GLAST.} The predicted binding affinity scores by our trained classifier are placed in brackets beside the docking score. Larger scores indicate stronger binding for our classifier.}
\label{table:dual TfR and GLAST}
\vskip 0.05in
\begin{small}
\resizebox{0.80\textwidth}{!}{%
\begin{tabular}{@{}lccccc@{}}
\toprule
\textbf{Binder ID}& \textbf{TfR Docking Score} (kcal/mol) ($\downarrow$) & \textbf{GLAST Docking Score} (kcal/mol) ($\downarrow$) & \textbf{Solubility} ($\uparrow$)& \textbf{Hemolysis} ($\uparrow$) & \textbf{Non-fouling} ($\uparrow$)\\ 
\midrule
Binder 1 & -8.8 (8.800) & -8.9 (7.775) & \textbf{0.975} & 0.743 & 0.118  \\
Binder 2 & -8.0 (7.599) & -7.9 (6.751) & 0.938 & 0.835 & \textbf{0.309}\\
Binder 3 & -8.3 (7.537) & -8.2 (6.662) & 0.972 & \textbf{0.914} & 0.214 \\
Binder 4 & -7.6 (7.748) & -7.5 (6.946) & 0.959 & 0.902 & 0.290 \\
Binder 5 & \textbf{-10.5} (8.714) & -8.5 (7.398) & 0.811 & 0.748 & 0.202 \\
Binder 6 & -8.4 (8.197) & -7.5 (7.076) & 0.971 & 0.855 & 0.165 \\
Binder 7 & -9.3 (8.321) & \textbf{-9.2} (7.190) & 0.881 & 0.860 & 0.212 \\
\bottomrule
\end{tabular}
}
\end{small}
\end{table*}
\begin{figure*}
    \label{figure: 3D-tfr-glast}
    \centering
    \includegraphics[width=0.84\linewidth]{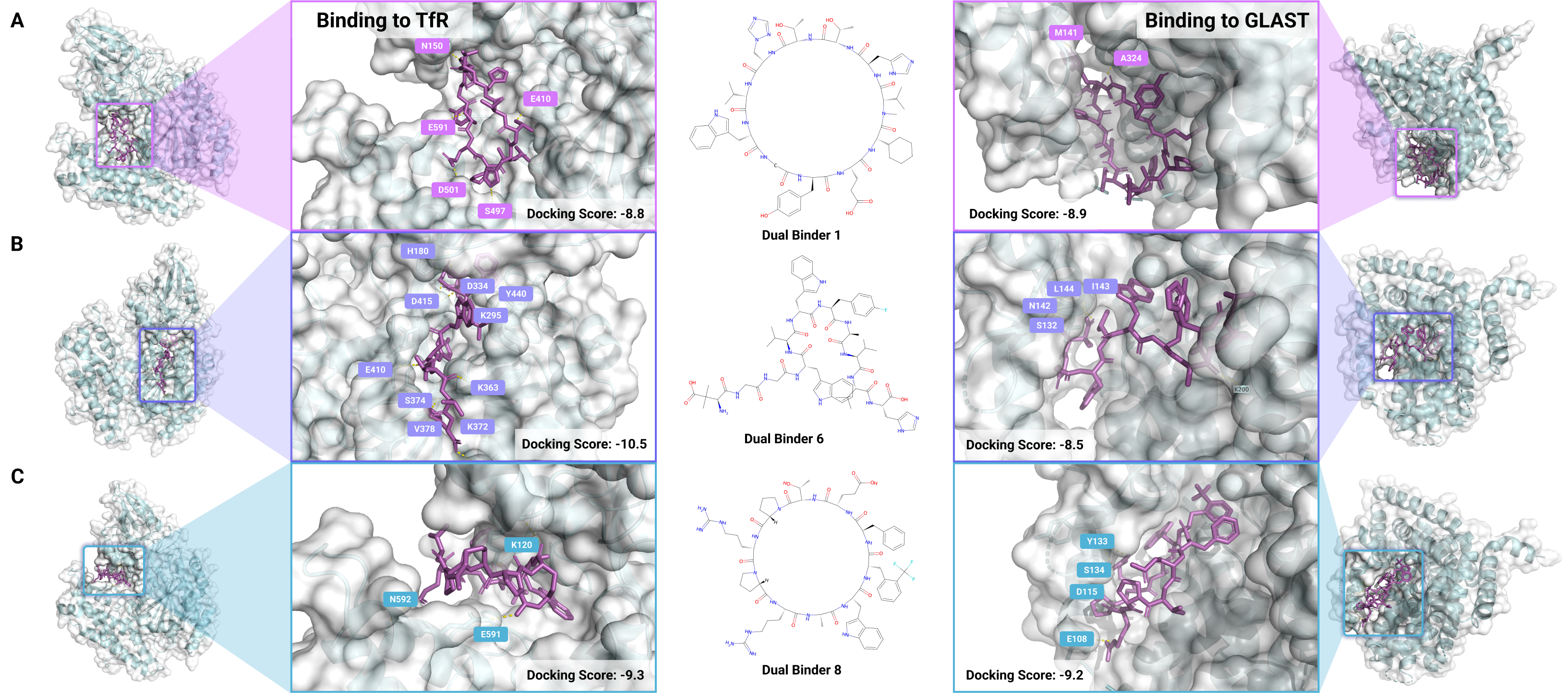}
    \caption{\textbf{PepTune-generated peptides to TfR and GLAST.} Full protein binding location and close-up binding position for (\textbf{A}) dual binder 1, (\textbf{B}) dual binder 6, and (\textbf{C}) dual binder 8 with TfR (left) and GLAST (right). Polar contacts within 3.5 \AA\ are highlighted.}
    \label{fig:3D-tfr-glast}
\end{figure*}
Multi-target drug discovery is of significant interest in various fields, including cancer therapeutics and drug delivery for neurological disorders, given their ability to perform multiple different functions such as binding to biological barriers like the blood-brain barrier, penetrating target cells, and inhibiting protein-protein interactions \cite{Tang2025rev, Li2023-2, Chan2016}. 

To evaluate PepTune's capabilities in multi-target guidance, we generate bi-specific peptide binders to TfR and glutamate-aspartate transporter (GLAST) protein abundant on the surface of astrocytes, a type of glial cell in the brain. Successfully generating these peptides can facilitate BBB-crossing via TfR binding and uptake in astrocytes via GLAST binding for intravenous delivery of therapeutics for a multitude of neurological disorders where astrocytes are involved, including Alexander disease \cite{Li2018}, Alzheimer's disease \cite{Habib2020}, Parkinson's disease \cite{Yun2018}, Huntington's disease \cite{Khakh2017}, multiple sclerosis \cite{Wheeler2020}, and several psychiatric disorders \cite{MartinFernandez2017}. 

We generated 100 peptide binders conditioned on five properties: predicted binding affinity to TfR, binding affinity to GLAST, solubility, hemolysis, and non-fouling. Remarkably, all property scores improved over iterations, with final solubility, hemolysis, and non-fouling scores surpassing binders conditioned solely on TfR binding affinity (Appendix Figure \ref{fig:dual-curves}). This highlights that our multi-target guidance strategy avoids significant property trade-offs.

To validate binding to both TfR and GLAST, we selected seven binders and docked them against each target. All seven achieved docking scores $\leq -7.5$ kcal/mol, with the top binder scoring $-10.5$ kcal/mol for TfR and $-9.2$ kcal/mol for GLAST (Table \ref{table:dual TfR and GLAST}). These top binders exhibited diverse secondary structures (Figure \ref{fig:3D-tfr-glast}), positive solubility, and low hemolysis probabilities (Table \ref{table:dual TfR and GLAST}). Binding positions and polar interactions varied, showing PepTune can discover diverse, high-affinity peptides without relying on specific motifs.

\section{Discussion}
We introduce \textbf{PepTune}, a multi-objective-guided discrete diffusion model for \textit{de novo} generation of peptide SMILES containing non-natural amino acids and cyclic modifications. We propose a bond-dependent masking schedule and invalid loss to ensure peptide validity. PepTune leverages \textbf{Monte Carlo Tree Guidance (MCTG)}, a novel multi-objective guidance framework for discrete diffusion, to identify peptide sequences optimized across multiple therapeutic properties. Unlike previous discrete guidance methods \cite{Nisinoff2024, Gruver2023}, MCTG operates strictly in the discrete state space and can be integrated at inference time with no additional training. By balancing exploration through batched unmasking with Gumbel noise and reward-based exploitation of optimal unmasking paths, MCTG enables the generation of a diverse set of Pareto-optimal sequences across an arbitrary number of objectives. Unlike recent binder design methods \cite{Rettie2024, Pacesa2024, Li2024, Watson2023}, PepTune requires no 3D target structures, enabling the design of peptides for conformationally diverse proteins—such as fusion oncoproteins \cite{Vincoff2025} and post-translationally modified isoforms \cite{Peng2025ptm}—while optimizing for properties beyond local geometric interactions.

\section*{Impact Statement}

We propose a robust framework for multi-objective guidance of diffusion that could be applied to various generative tasks requiring simultaneous optimization of multiple constraints. Our work focuses on designing therapeutically-viable peptides, potentially advancing treatments for many diseases by enabling the discovery of peptides optimized for binding, solubility, and other critical properties. However, like any powerful generative framework, there is a potential for misuse, such as generating peptides with harmful biological properties. We encourage careful oversight and ethical application of PepTune to ensure its use aligns with positive societal outcomes.

\section*{Declarations}
\paragraph{Acknowledgments} We thank Alexander Tong for reviewing the theoretical formulations of PepTune. We also thank Sophia Vincoff and Lauren Hong for their assistance with figure generation.

\paragraph{Author Contributions} S.T. devised and developed PepTune architecture and theoretical formulations, and trained and benchmarked generation, prediction, and sampling models. Y.Z. advised on model design and theoretical framework, trained classifier models, and performed molecular docking. S.T. drafted the manuscript and S.T. and Y.Z. designed the figures. P.C. conceived, designed, supervised, and directed the study, and reviewed and finalized the manuscript.

\paragraph{Data and Materials Availability} Our peptide filtering, analysis, and visualization tool, SMILES2PEPTIDE,
is freely available on HuggingFace: \url{https://huggingface.co/spaces/ChatterjeeLab/SMILES2PEPTIDE}. The PepTune codebase is freely accessible to the academic community via a non-commercial license at \url{https://huggingface.co/ChatterjeeLab/PepTune}. 

\paragraph{Funding Statement} This research was supported by NIH grant R35GM155282 as well as a gift from the EndAxD Foundation to the lab of P.C.

\paragraph{Competing Interests} P.C. is a co-founder of Gameto, Inc. and UbiquiTx, Inc. and advises companies involved in peptide therapeutics development. P.C., S.T., and Y.Z. have and are currently filing patent applications related to this work. P.C.’s interests are reviewed and managed by Duke University in accordance with their conflict-of-interest policies.

\nocite{langley00}

\bibliography{citation}
\bibliographystyle{icml2025}

\newpage
\appendix
\onecolumn
\section*{Overview of Appendix}
In Appendix \ref{appendix:A}, we discuss additional case studies demonstrating PepTune's ability to generate peptides with high binding affinity and enhanced therapeutic properties to several therapeutic targets, including receptors on the blood-brain barrier (\ref{appendix:A.1}), intracellular proteins with enhanced cell permeability (\ref{appendix:A.2}), targets without existing peptide binders (\ref{appendix:A.3}), and dual-targeting for target-protein degradation (\ref{appendix:A.4}). 

Appendix \ref{appendix:B} provides a background on continuous-time discrete diffusion (\ref{appendix:B.1}), the NELBO loss objective (\ref{appendix:B.2}), and guidance for diffusion models (\ref{appendix:B.3}). Appendix \ref{appendix:C} provides details on data curation and tokenization. Appendix \ref{appendix:D} provides additional implementation details and generation results of our unconditional bond-dependent masked discrete diffusion model, PepMDLM. Appendix \ref{appendix:E} provides details on the model architecture and training of our property prediction models for binding affinity (\ref{appendix:E.1}), membrane permeability (\ref{appendix:E.2}), solubility, hemolysis, and non-fouling (\ref{appendix:E.3}). Appendix \ref{appendix:F} contains details on our evaluation methods, including our SMILES2PEPTIDE filter (\ref{appendix:F.1}). 

In Appendix \ref{appendix:G}, we provide the theoretical basis for Section \ref{sec:Bond-Dependent Masked Discrete Diffusion}, including formal proofs for Proposition \ref{prop:bond-dependent NELBO} (\ref{appendix:G.2}), Proposition \ref{prop:bond-dependent reverse posterior} (\ref{appendix:G.3}), and Proposition \ref{prop:invalid loss} (\ref{appendix:G.4}). Appendix \ref{appendix:H} discusses the choices of hyperparameters. Appendix \ref{appendix:I} provides results of additional experiments, including one that integrates time-dependence into the MCTG algorithm (\ref{appendix:I.1}) and ablation studies investigating the impact of bond-dependent masking and the invalid loss on generation quality (\ref{appendix:I.2}). Finally, we provide pseudo code for all of our algorithms in Appendix \ref{appendix:J}.

\section{Additional Case Studies}
\label{appendix:A}
With our trained property classifiers, we conduct experiments for diverse, therapeutically relevant protein targets to evaluate our multi-objective MCTS guidance strategy. To demonstrate generalizability, we include targets with known peptide binders such as TfR, and proteins with no known binders, including GFAP, NCAM1, and AMHR2. We also design bi-specific binders for GFAP and RBX1 as a case study for Alexander disease therapeutics. These targets include both receptor proteins involved with active transport pathways, as well as intracellular targets where cell membrane permeability is crucial to achieving therapeutic effects. For each target, we condition the generation on the binding affinity score given the target protein sequence, along with solubility, hemolysis, non-fouling, and cell membrane permeability for intracellular targets. For external testing and validation, we use Autodock Vina \cite{eberhardt2021autodock} to compute \textit{in silico}  binding affinities of our generated binders (Appendix \ref{appendix:F.3}). 

\subsection{Targeting Receptors on the Blood-Brain Barrier} 
\label{appendix:A.1}
The Transferrin receptor (TfR) is a receptor protein abundant on the selectively permeable blood-brain barrier (BBB) that is responsible for transporting iron-binding transferrin (Tf) proteins into the brain parenchyma \cite{Johnsen2017}. Given its selective expression on brain endothelial cells and glioma cells and its ability to recycle back to the luminal surface after facilitating the internalization of cargo through the BBB \cite{Jefferies1984}, TfR has been extensively studied as a target for the intravenous delivery of various therapeutics and therapeutic nanocarriers through the BBB \cite{Gosk2004, Zhang2024}. 

\begin{figure}
    \centering
    \includegraphics[width=0.8\linewidth]{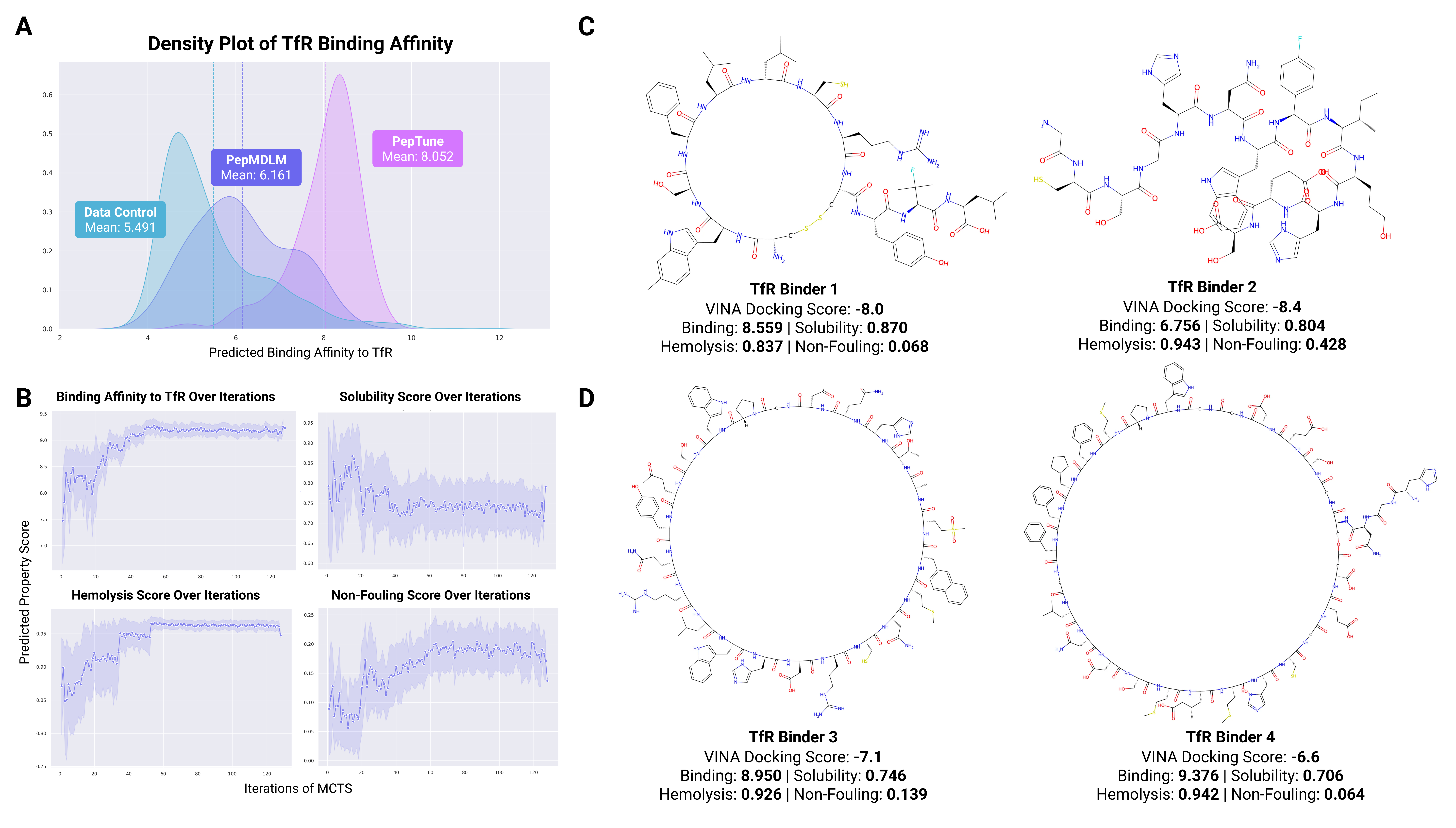}
    \caption{\textbf{PepTune-generated peptide binders to TfR.} (\textbf{A}) Density plot depicting the frequency of predicted binding affinity scores from our trained regression model for the sequences in the data used to train the regression model, the generated peptides from our unconditional PepMDLM model, and our PepTune model conditioned on TfR binding affinity, solubility, hemolysis, and non-fouling. (\textbf{B}) Plots depicting the mean scores for each property over the number of iterations or traversals of the MCTS algorithm for 128 iterations and a maximum token length of 200. The shaded region represents the standard deviation. (\textbf{C}) Two-dimensional visualization of generated binders with token length 100, their corresponding docking scores ($\downarrow$) computed using Vina docking, and predicted classifier scores ($\uparrow$) from the trained classifiers. (\textbf{D}) Visualizations of generated binders with token length 200, their docking scores, and predicted classifier scores.}
    \label{fig:tfr}
\end{figure}

To generate relevant binders for TfR, we condition PepTune on binding affinity with the TfR sequence, in addition to solubility, hemolysis, and non-fouling. At each iteration, we measured the mean of the property scores across all rolled-out sequences from the selected node to evaluate the effectiveness of the optimization strategy. We show that all properties, except solubility, exhibited an upward trend over iterations, with the average score of the binding affinity classifier exhibiting a significant increase in score to over 9.0 (Figure \ref{fig:tfr}B). After plotting the distribution of 100 peptides generated from a single run of PepTune with the minimum number of sequences set to 100, we confirm that our multi-objective MCTS algorithm shifted the distribution to a higher predicted binding affinity than the unconditionally generated peptides (PepMDLM) and the data used to train the binding regression model (Figure \ref{fig:tfr}A). Despite being conditioned on four distinct properties, PepTune is capable of generating higher-affinity binders than the unconditional model, supporting the effectiveness of our multi-objective guidance strategy.  

Encouraged by these results, we sampled the Pareto-optimal sequences from the generated peptides and used Vina docking to compute their optimized docking score. Notably, we observed that all of the generated binders that were selected for docking produced affinity scores below -6.0 kcal/mol, with our top-performing binder achieving a -8.4 kcal/mol binding affinity (Figure \ref{fig:tfr}C). From the docking scores, we took the two binders with the best docking scores and visualized their binding conformation with TfR, showing that they bind to distinct motifs on the protein surface (Figure \ref{fig:tfr}B, F, G). 

\begin{figure}
    \centering
    \includegraphics[width=0.8\linewidth]{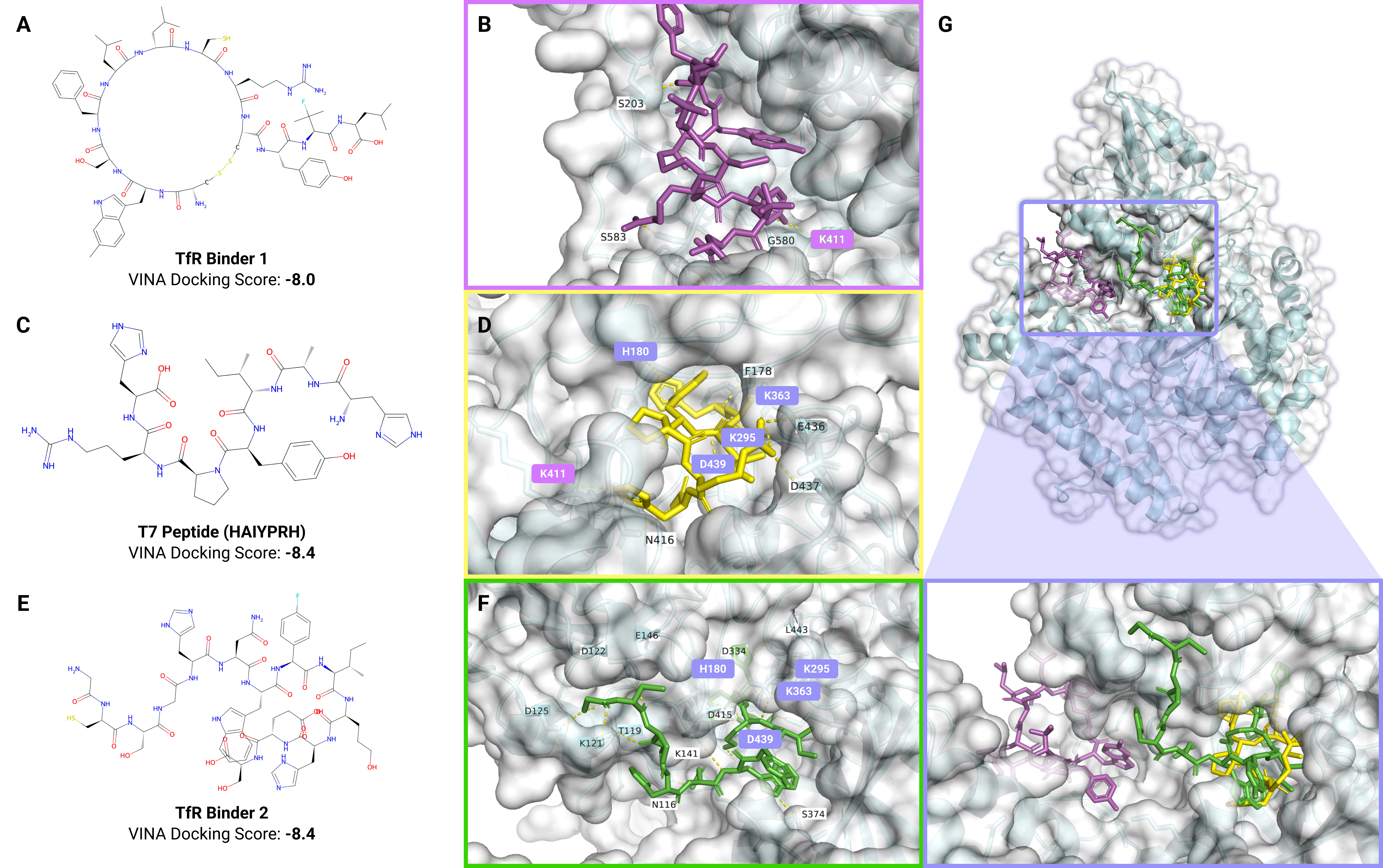}
    \caption{\textbf{Comparison of PepTune-generated peptides and established T7-peptide to TfR.} Two-dimensional chemical structure of (\textbf{A}) PepTune-generated binder 1, (\textbf{C}) established T7 peptide, and (\textbf{E}) PepTune-generated TfR binder 2 and their Vina docking scores to TfR ($\downarrow$). Zoomed-in visualization of the docked binding positions of (\textbf{A}) binder 1, (\textbf{B}) T7, and (\textbf{C}) binder 2 with TfR. Polar contacts within 3.5 \AA\ are annotated, and shared contacts between T7 and binder 1 (purple) and between T7 and binder 2 (blue) are highlighted. (\textbf{C}) Overlay of peptide binders on full TfR protein} 
    \label{fig:t7}
\end{figure}

To further confirm binding affinity to TfR, we compared our peptides to the well-established 7-amino acid peptide T7 (sequence: HAIYPRH) that selectively binds to an alternative site as compared to endogenous Tf on TfR \cite{Lee2001}. T7 has been extensively explored for targeted delivery of nanoparticles to the brain \cite{Kuang2013, Kim2020, Bi2016, Cai2020, Wang2015, Zhao2016, Liang2018}, and has demonstrated 7.89-fold enhanced brain penetration in \textit{in vivo} mice models \cite{Yu2018}. After docking T7 with TfR, we obtained a docking score of -8.4 kcal/mol. Notably, our peptides optimized on all four therapeutic properties, including TfR binding affinity show competitive docking scores to T7 (Figure \ref{fig:t7}A, C, E), suggesting that PepTune is capable of generating promising candidates for \textit{in vivo} targeting and delivery across the BBB. Furthermore, after annotating polar contacts within 3.5 \AA\, we determine that both of the generated peptides with the best binding affinity scores have shared residue contacts when binding with TfR as T7 (Figure \ref{fig:t7}B, D, F), indicating that our generated peptides have similar binding properties to T7, enabling it to bind strongly to a shared binding site. Furthermore, our generated binders have diverse structural features, such as cycles in binder 1 and side-chain modifications in binder 2. Since T7 is known to bind to an alternative site than endogenous Tf \cite{Lee2001}, we show that PepTune can generate viable candidates for non-competitive binding to TfR for BBB-targeting applications. 

\subsection{Targeting Intracellular Proteins} 
\label{appendix:A.2}
\begin{figure}
    \centering
    \includegraphics[width=\linewidth]{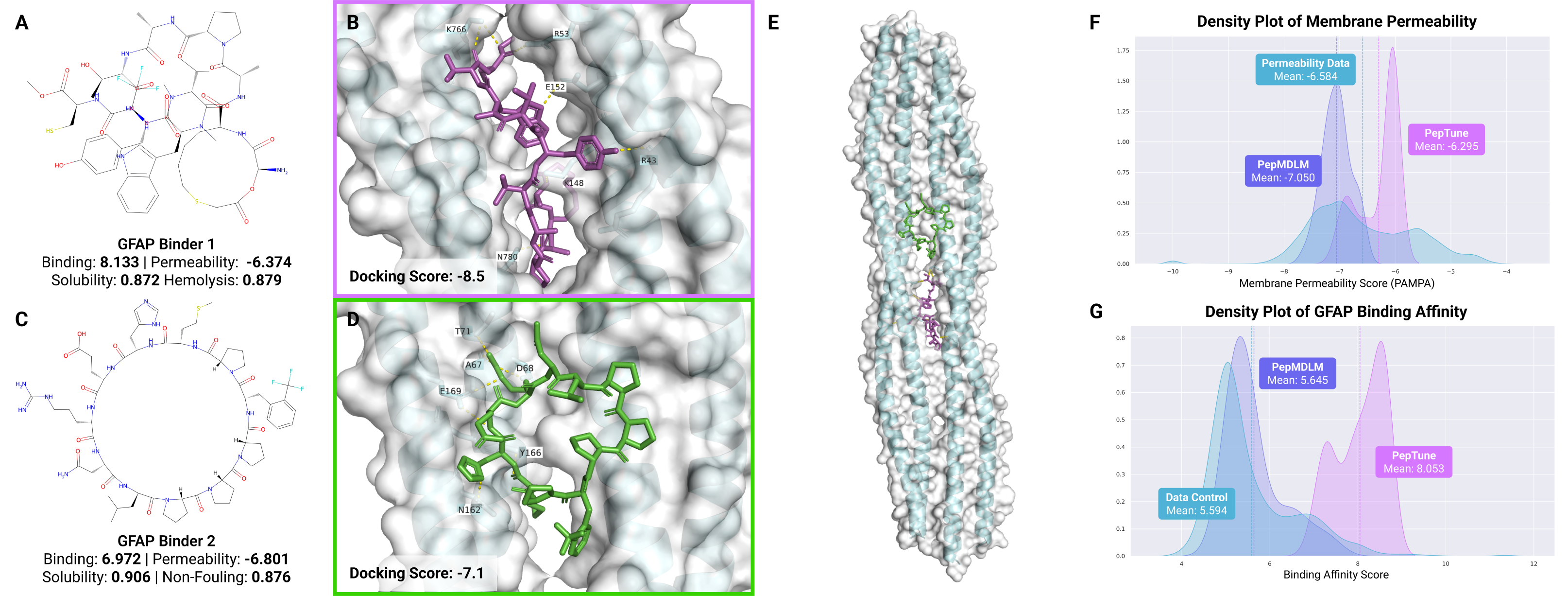}
    \caption{\textbf{PepTune-generated peptide binders to intracellular protein GFAP.} (\textbf{A}, \textbf{C}) Two-dimensional structures of GFAP binder 1 and 2 with predicted property scores, including cell membrane permeability. (\textbf{B}, \textbf{D}) GFAP binders 1 and 2 docked to GFAP with scores of -8.5 kcal/mol and -7.1 kcal/mol, respectively. (\textbf{E}) Full GFAP protein structure with docked binders 1 and 2. (\textbf{F}) The distribution of PAMPA membrane permeability scores from 34,853 experimentally-validated peptides compared to 100 peptides generated using our unconditional PepMDLM model, and 100 peptides generated with PepTune conditioned on both cell membrane permeability and affinity to GFAP. The permeability curve shifted towards higher permeability with a mean of -6.295. \textbf{(G)} Simultaneously, the distribution of predicted binding affinity scores to GFAP for the PepTune-generated peptides is shifted to higher scores with a mean of 8.053 compared to a set of experimentally-tested peptides and unconditional PepMDLM-generated peptides.}
    \label{fig:gfap}
\end{figure}

Glial fibrillary acidic protein (GFAP) is an intracellular protein differentially expressed in astrocytes, a family of glial cells in the brain \cite{Hol2015}. Dysregulation of GFAP expression has been found to cause Rosenthal fibers, astrocytic cytoplasmic inclusions that are responsible for Alexander disease, a fatal neurodegenerative disease affecting infants \cite{Brenner2001, Grossi2024}. Discovering potent binders that inhibit or degrade GFAP proteins can have significant therapeutic implications. However, no established peptide binders exist to GFAP, which motivates their \textit{de novo} design. In addition to achieving high binding affinity with GFAP, we posit that an optimal peptide binder must also cross the astrocyte cell membrane into the cytosol to access GFAP. Therefore, we condition the generation of GFAP binders on five properties: binding affinity to GFAP, solubility, hemolysis, non-fouling, and cell membrane permeability using our permeability regression model, demonstrating optimization across all of these properties (Figure \ref{fig:gfap}). To confirm GFAP engagement, our docking peptides demonstrate strong affinities below -7 kcal/mol, motivating downstream experimental validation in astrocyte cultures (Figure \ref{fig:gfap}B and D).

\subsection{Targeting Proteins Without Existing Binders} 
\label{appendix:A.3}
\begin{figure} [H]
    \centering
    \includegraphics[width=\linewidth]{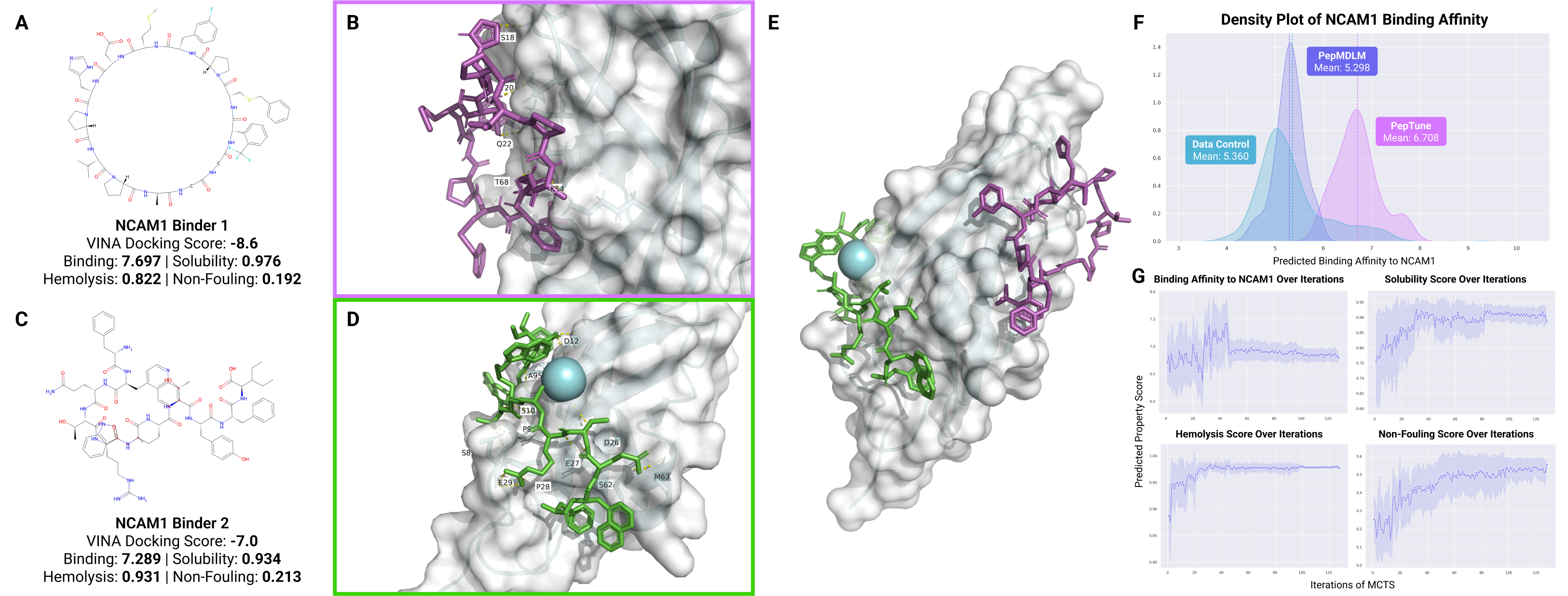}
    \caption{\textbf{PepTune-generated peptide binders to NCAM1.} Two-dimensional structures of (\textbf{A}) binder 1 and (\textbf{C}) binder 2 genered with PepTune. Docking positions of (\textbf{B}) binder 1 and (\textbf{C}) binder 2 on NCAM1 with annotated polar contacts within 3.5 \AA\. (\textbf{G}) Full NCAM1 protein structure with docked peptide binders 1 and 2. (\textbf{H}) (Top) Density plot of NCAM1 binding affinity scores for PepTune (mean: 6.708), PepMDLM (mean: 5.298), and peptides from a control set of experimentally-tested peptide SMILES (mean: 5.360). (Bottom) Plots depicting the average predicted score for NCAM1 binding affinity, solubility, hemolysis, and non-fouling over iterations of MCTS.}
    \label{fig:ncam}
\end{figure}

To test the ability of our model to generate binders to challenging extracellular targets without existing binders, we evaluate PepTune-generated peptides for NCAM1 and AMHR2, two therapeutically relevant receptor proteins. Neural cell adhesion molecule 1 (NCAM1) is a transmembrane protein expressed on the surface of neurons and glial cells \cite{Paratcha2003}. Beyond its roles in neuronal migration and synaptogenesis, NCAM1 is also crucial for memory formation, highlighting its significance in brain development \cite{vukojevic2020ncam}. As NCAM1 is an extracellular protein, we generated a library of peptides with PepTune-optimized NCAM1 binding affinity, solubility, hemolysis, and non-fouling (Figure \ref{fig:ncam}F, G). All properties exhibited an upward trend across optimization iterations. 

We selected two binders with the highest Vina docking scores for visualization (Figure \ref{fig:ncam}A-E). Notably, \textit{in silico} docking analysis revealed that binder 1 exhibits markedly high affinity binding (-8.6 kcal/mol) while binder 2 wraps around the NCAM1 structure via numerous polar contacts, suggesting extensive and specific interactions (Figure \ref{fig:ncam}B and D).

Anti-Müllerian hormone type-2 receptor (AMHR2) is a transmembrane receptor involved in sex differentiation. Mutations in the AMHR2 gene are a leading cause of Persistent Müllerian duct syndrome (PMDS) in males, resulting in the retention of female gonads alongside male reproductive structures \cite{imbeaud199627amhrM1}. In females, polymorphisms of AMHR2 have been associated with infertility \cite{rigon2010amhrF1, lazaros2016amhrF2}. Most interestingly, antagonism of AMHR2 with therapeutic peptides can potentially serve as a specific therapy for polycystic ovarian syndrome (PCOS), which affects an estimated 4\% to 10\% of women globally \cite{Singh2023}, as AMHR2 signaling has been implicated in follicular arrest and dysregulated ovarian function \cite{diClemente2022}.

\begin{figure}
    \centering
    \includegraphics[width=\linewidth]{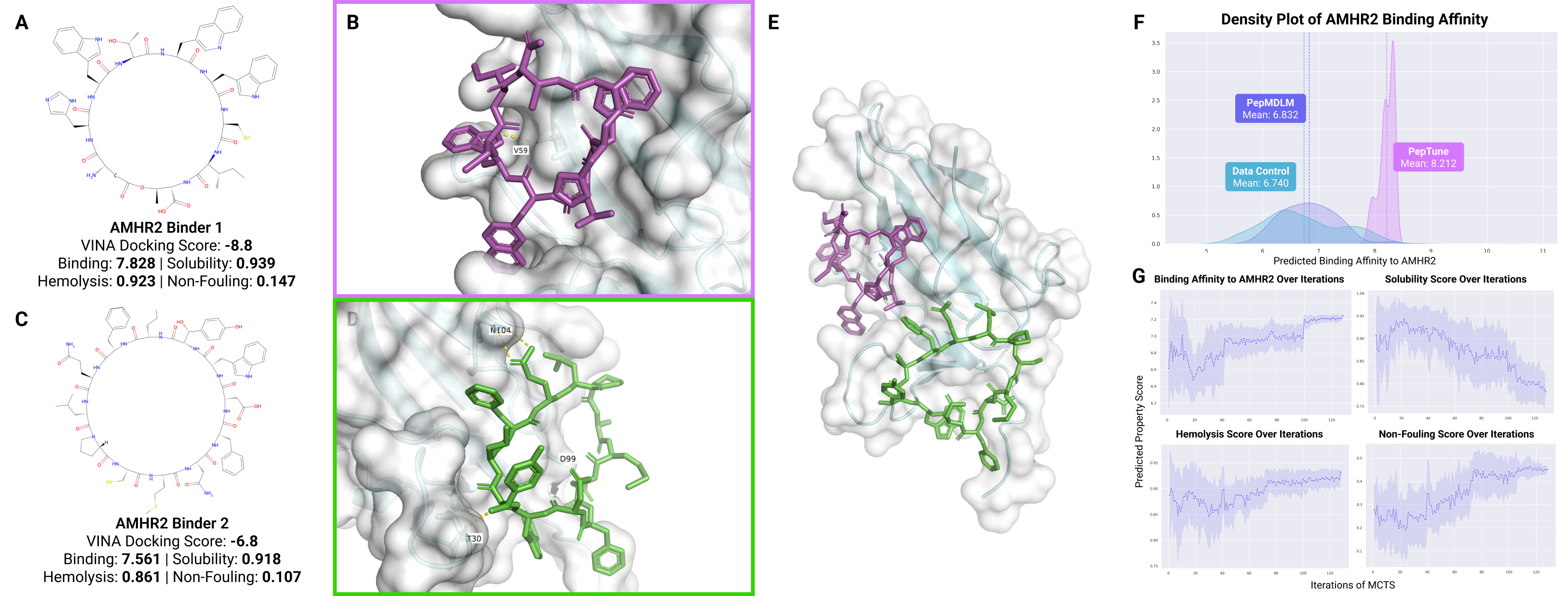}
    \caption{\textbf{PepTune-generated peptides to AMHR2.} Two-dimensional structures of (\textbf{A}) binder 1 and (\textbf{B}) binder 2 generated with PepTune. Docking positions of (\textbf{A}) binder 1 and (\textbf{B}) binder 2 on NCAM1 with annotated polar contacts. (\textbf{G}) Full AMHR2 protein structure with docked peptide binders 1 and 2. (\textbf{H}) (Top) Density plot of AMHR2 binding affinity scores for PepTune (mean: 8.212), PepMDLM (mean: 6.832), and peptides from a control set of experimentally-tested peptide SMILES (mean: 6.740). (Bottom) Plots depicting the average predicted score for AMHR2 binding affinity, solubility, hemolysis, and non-fouling over iterations of MCTS.}
    \label{fig:amhr}
\end{figure}

Following similar computational setups as described previously, we generated \textit{in silico} binders with high Vina predicted binding affinities (<-6 kcal/mol), despite observing a decrease in the predicted solubility along iterations  (Figure \ref{fig:amhr}). However, our observation of reduced solubility upon binder docking can be attributed to the presence of hydrophobic patches within the AMHR2 extracellular domain, particularly near the binding site to its ligand AMH \cite{Hart2021}. This phenomenon highlights the importance of balancing solubility and binding affinity in binder development. With further optimization of their therapeutic properties, we hope to demonstrate the potential of these binders for applications in fertility treatment in the future. 

The examples above demonstrate the versatility of our method, which can be effectively applied to discover peptide binders for single target proteins lacking known ligands, thereby unlocking their therapeutic potential.

\begin{figure}
    \centering
    \includegraphics[width=\linewidth]{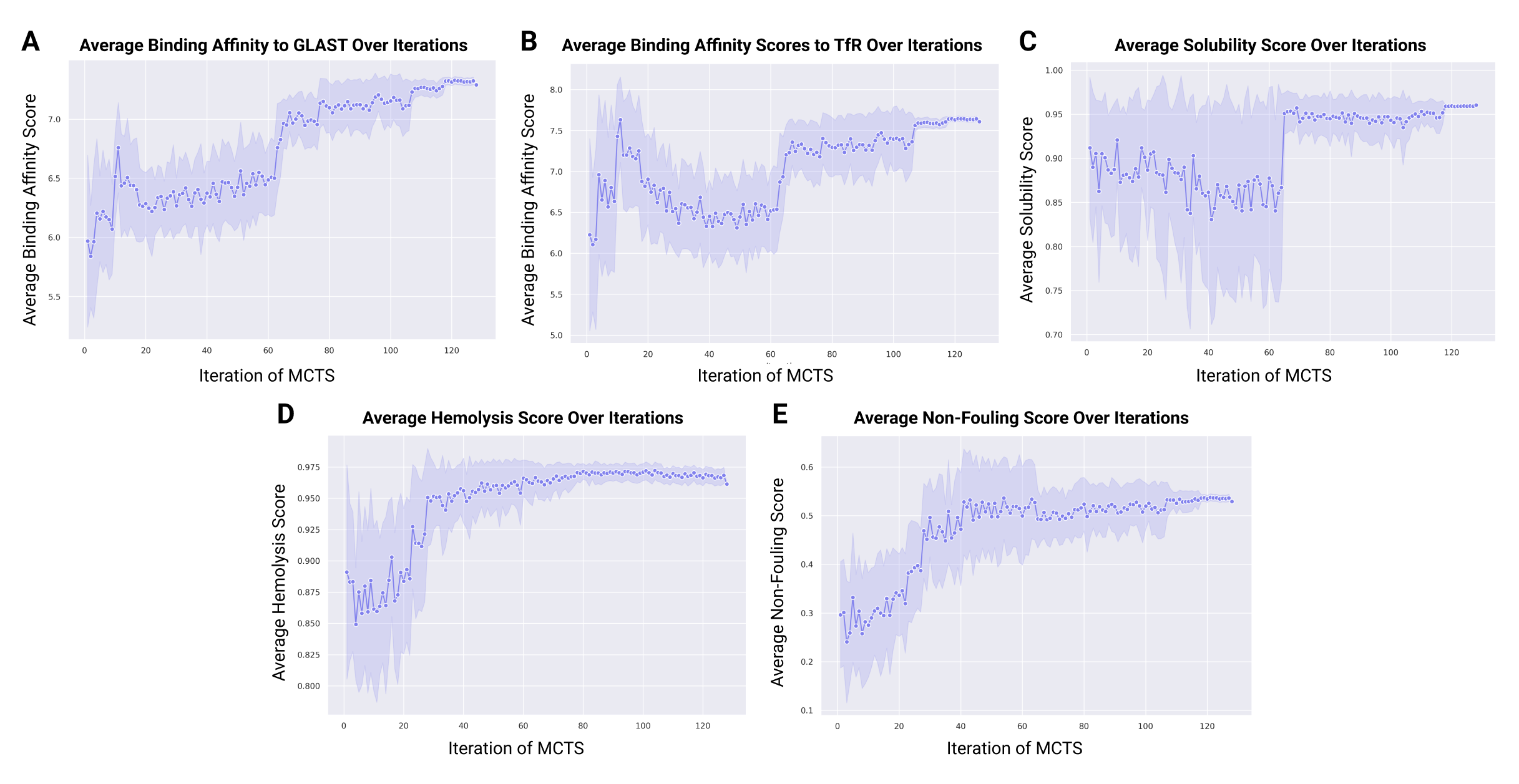}
    \caption{\textbf{Property Scores Over Iteration for Dual-Target Conditioning on TfR and GLAST.} (\textbf{A}) Plot of average predicted binding affinity score to GLAST over iterations. (\textbf{B}) Plot of average predicted binding affinity score to TfR over iterations. (\textbf{C}, \textbf{D}, \textbf{E}) Plot of average predicted solubility, hemolysis, and non-fouling scores over iterations.}
    \label{fig:dual-curves}
\end{figure}

\subsection{Dual-Targeting of GFAP and an E3 Ubiquitin Ligase for Target Protein Degradation} 
\label{appendix:A.4}
\begin{figure}
    \centering
    \includegraphics[width=\linewidth]{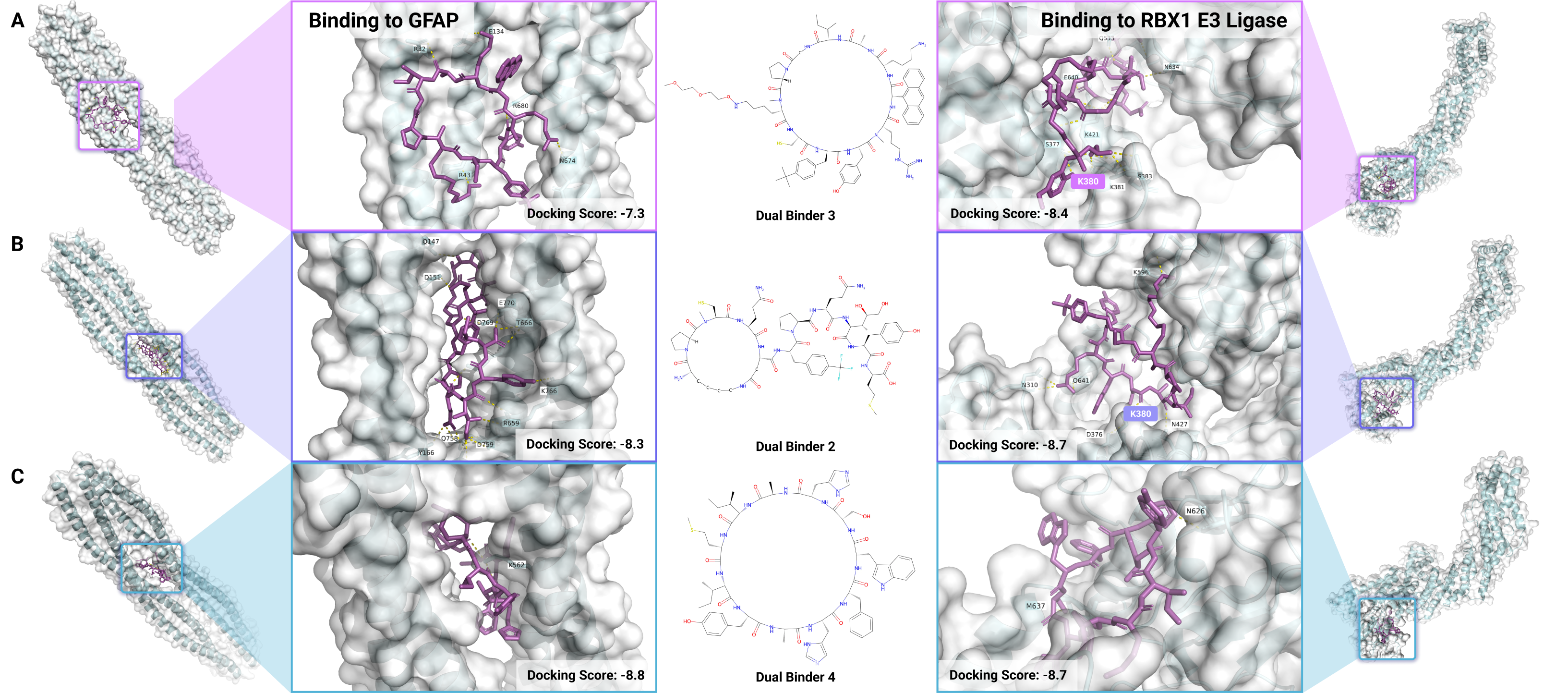}
    \caption{\textbf{PepTune-generated peptides with dual GFAP and RBX1 affinity.} Full protein binding location and close-up binding position for (\textbf{A}) dual binder 3, (\textbf{B}) dual binder 2, and (\textbf{C}) dual binder 4 with GFAP (left) and RBX1 (right). Polar contacts within 3.5 \AA\ are annotated, and shared polar contacts between binders are highlighted.}
    \label{fig:gfap-rbx}
\end{figure}

As another dual-target case study, we used PepTune to generate peptides with high binding affinity to the GFAP protein and an E3 ubiquitin ligase protein RBX1, a protein in the Skp1/Cullin-1/F-box (SCF) E3 ubiquitin ligase complex that recruits E2 to catalyze ubiquitination and subsequent degradation \cite{Yang2021}. A peptide generated for this task would be capable of binding to GFAP proteins overexpressed in Alexander disease and mediate their proteasomal degradation, which could alleviate the production of disease-causing Rosenthal fibers in astrocytes \cite{Sosunov2017}. 
\begin{table}
    \centering
    \caption{\textbf{PepTune-generated dual-target binders to GFAP and RBX1.} The predicted binding affinity scores by our trained classifier are placed in brackets beside the docking score. Larger scores indicate stronger binding for our classifier.}\label{table:gfap-dual}
    \vskip 0.05in
    \resizebox{\textwidth}{!}{%
    \begin{tabular}{@{}lccccc@{}}
    \toprule
     \textbf{Binder ID}& \textbf{GFAP Docking Score} (kcal/mol) ($\downarrow$) & \textbf{E3 Docking Score} (kcal/mol) ($\downarrow$) & \textbf{Solubility} ($\uparrow$)& \textbf{Hemolysis} ($\uparrow$) & \textbf{Non-fouling} ($\uparrow$)\\ \midrule
    Binder 1 & -8.0 (8.384) & -8.4 (7.468) & 0.730 & \textbf{0.894} & 0.111  \\
    Binder 2 & -8.3 (7.395) & \textbf{-9.3} (7.089) & \textbf{0.972} & 0.869 & 0.134\\
    Binder 3 & -7.3 (7.925) & -8.7 (7.158) & 0.935 & 0.812 & 0.143 \\
    Binder 4 & \textbf{-8.8} (7.144) & -8.7 (7.000) & 0.897 & 0.807 & \textbf{0.158} \\
    \bottomrule
    \end{tabular}
    }
\end{table}
After conditioning PepTune generation on binding affinity to GFAP, binding affinity to RBX1, solubility, hemolysis, and non-fouling (Table \ref{table:gfap-dual}), we selected three non-dominated binders with predicted affinities greater than 7.0 for docking experiments. For these Pareto-optimal peptides, we indeed observed strong binding affinities for both GFAP and RBX1 post-docking, indicating their unique potential for multi-target interaction (Figure \ref{fig:gfap-rbx}). GFAP is an intermediate filament protein \cite{Eng1985} and thus forms a unique rod-like structure with a head domain and a tail domain. The docking positions of all three candidates were along the rod domain, binding in the gap between adjacent rods in the filament. Contrarily, docked candidates to RBX1 consistently bound close to its interaction site of Cullin, rather than at the Skp2 F-box adaptor site (Figure \ref{fig:gfap-rbx}), indicating that further motif conditioning, as done with recent peptide design language models \cite{Chen2024}, would benefit PepTune's clinical potential.

\section{Extended Background}
\label{appendix:B}
\subsection{Continuous-Time Discrete Diffusion}
\label{appendix:B.1}
Discrete diffusion models \cite{Austin2021} are a subset of diffusion models where the forward corruption and reverse denoising processes operate in the discrete latent space via categorical probability distributions for discrete variables. 

We denote a token in a sequence from the dataset as a one-hot vector $\mathbf{x}_0^{(\ell)}\in \{0,1\}^{|\mathcal{V}|}$. The discrete-time forward diffusion process involves applying categorical noise to $\mathbf{x}_0$ at varying degrees based on a noise schedule $\sigma(t)$ for a total of $T$ time steps ranging from no noise at $t=0$ to maximum noise at $t=1$. To clearly distinguish each step, we denote the $n$th transition in the forward pass as the transition from $s(n)$ to $t(n)$, where $s(n)=\frac{n-1}{T}$ and $t(n)=\frac{n}{T}$. The marginal noise that transforms the sequence $\mathbf{z}_{s(n)}$ at time $s(n)$ to a progressively noisier sequence $\mathbf{z}_{t(n)}$ at the next time step $t(n)=s(n)+\frac{1}{T}$ is given by a $|\mathcal{V}|\times |\mathcal{V}|$ marginal transition matrix $\mathbf{Q}_{t|s}=\sigma(t)\mathbf{I}_{|\mathcal{V}|}+(1-\sigma(t))\mathbf{1}\mathbf{\pi}^{\top}$, where the $(i,j)$th entry denotes the probability of transitioning from token $i$ to token $j$ at each position in the sequence. 

Therefore, the marginal categorical distribution of $\mathbf{z}_{t(n)}$ in the discrete-time forward-pass diffusion process can be derived as
\begin{align}
    q(\mathbf{z}_{t(n)}|\mathbf{z}_{s(n)})&=\text{Cat}(\mathbf{z}_{t(n)};\mathbf{Q}_{t|s}^{\top} \mathbf{z}_{s(n)})\nonumber\\
    &=\text{Cat}(\mathbf{z}_{t(n)}; \sigma(t(n))\mathbf{z}_{s(n)}+(1-\sigma(t(n)))\mathbf{\pi})
\end{align}
where $\sigma(t(n))$ the marginal probability of a single position in the sequence remaining the same token during the transition $s(n)\to t(n)$ and $\big(1-\sigma(t(n)))$ is the marginal probability of transitioning to a different token based on the token probability distribution $\pi\in \mathbf{\Delta}^{|\mathcal{V}|}$. For simplicity, we denote $\sigma(t(n))=\sigma(n)$.

The cumulative transition from time $0$ to time $t(t)$ is denoted as the product of the marginals  $\mathbf{Q}_t=\prod_{n=0}^t\mathbf{Q}_{t|s}$, which is the product of marginal transitions $s(n)\to t(n)$ for $n$ ranging from 0 to $t$ applied to the clean input sequence $\mathbf{x}_0$.
\begin{align}
    \mathbf{Q}_{t}=\left(\prod_{n=0}^t(1-\sigma(n))\right)\mathbf{I}+\left(1-\prod_{n=0}^t(1-\sigma(n))\right)\mathbf{1}\pi^{\top}
\end{align}
For the continuous-time forward pass diffusion process where $T\to \infty$ and $\frac{1}{T}\to 0$, we can take the limit as $T\to \infty$ to derive an expression for the continuous-time transition probability, $\alpha_t$.
\begin{align}
    \lim_{T\to \infty}\prod_{n=0}^t(1-\sigma(n))&=\lim_{T\to \infty}\exp\left(\log\prod_{n=0}^t(1-\sigma(n))\right)\nonumber\\
    &=\lim_{T\to \infty}\exp\left(\sum_{n=0}^t\log(1-\sigma(n))\right)\nonumber\\
    &\approx \lim_{T\to \infty}\exp\left(\sum_{n=0}^t-\sigma(n)\right)\tag{$\log(1-x)\approx -x$ for small $x$}\nonumber\\
    &=\exp\left(-\int_{n=0}^t\sigma(n)dn\right)\nonumber\\
    &=\exp\left(-\int_{s=0}^{t(t)}\sigma(s)ds\right)\label{eq:masking}
\end{align}
We have shown that the continuous-time forward transition probability from $t=0$ to $t=t(t)$ is $\alpha_{t}=\exp\left(-\int_{s=0}^{t(t)}\sigma(s)ds\right)=\exp (-\bar{\sigma}(t))$ where $\bar{\sigma}(t)=\int_{s=0}^{t(t)}\sigma(s)ds$. Letting $t=t(t)$, we can define the continuous-time cumulative transition matrix $\mathbf{Q}_{t}$ at the limit where $T\to \infty$ and the continuous-time distribution $q(\mathbf{z}_t|\mathbf{x}_0)$ as
\begin{align}
    \mathbf{Q}_{t}&=\alpha_{t}\mathbf{I}+\alpha_t\mathbf{1}\pi^{\top}\\
    q(\mathbf{z}_t|\mathbf{x}_0)&=\text{Cat}(\mathbf{z}_t;\mathbf{Q}_{t} \mathbf{x}_0)\nonumber\\
    &=\text{Cat}(\mathbf{z}_t;\alpha_{t}\mathbf{x}_0+(1-\alpha_{t})\mathbf{\pi})
\end{align}
It follows that the marginal transition $\mathbf{Q}_{s|t}$ is the cumulative transition $\mathbf{Q}_{t}$ divided by all previous transition probabilities, denoted as $\mathbf{Q}_{s}=\prod_{r=0}^s\mathbf{Q}_{s|r}$, so $\alpha_{s|t}=\frac{\alpha_t}{\alpha_s}$. 

Following Austin et al. \cite{Austin2021} and substituting the marginal and cumulative probability distributions, we derive the true reverse transition from $t\to s$ conditioned on a clean sequence $\mathbf{x}_0$ as
\begin{align}
    q(\mathbf{z}_s|\mathbf{z}_t,\mathbf{x}_0)&=\frac{q(\mathbf{z}_t|\mathbf{z}_s, \mathbf{x}_0)q(\mathbf{z}_s|\mathbf{x}_0)}{q(\mathbf{z}_t|\mathbf{x}_0)}\nonumber\\
    &=\text{Cat}\left(\mathbf{z}_s;\frac{\mathbf{Q}_{t|s}\mathbf{z}_t\odot \mathbf{Q}_s^{\top}\mathbf{x}_0}{\mathbf{z}_t^{\top}\mathbf{Q}_t^{\top}\mathbf{x}_0}\right)\nonumber\\
    &=\text{Cat}\left(\mathbf{z}_s;\frac{[\alpha_{t|s}\mathbf{z}_t+(1-\alpha_{t|s})\mathbf{1}\mathbf{\pi}^{\top}\mathbf{z}_t]\odot [\alpha_s\mathbf{x}_0+(1-\alpha_s)\mathbf{\pi}]}{\alpha_t\mathbf{z}_t^{\top}\mathbf{x}_0+(1-\alpha_t)\mathbf{z}_t^{\top}\mathbf{\pi}}\right)
\end{align}
where the numerator is the element-wise product of $|\mathcal{V}|$-dimensional vectors representing the marginal probability distribution of sampling $\mathbf{z}_t$ given $\mathbf{z}_s$ and the cumulative probability distribution for $\mathbf{z}_s$ from the original clean sequence $\mathbf{x}_0$. The denominator is a scalar probability of the specific token $\mathbf{z}_t$ being drawn from the noisy probability distribution at time $t$.

\subsection{Continuous-Time Negative Evidence Lower Bound (NELBO)}
\label{appendix:B.2}
The objective of denoising diffusion probabilistic models (DDPMs) \cite{Hol2015} is to iteratively sample slightly less noisy intermediate sequences $\mathbf{z}_t$ until obtaining a clean sequence $\mathbf{x}$ that has a high probability of being drawn from the data distribution $p(\mathbf{x}_0)$. To train a model that accurately samples from $p(\mathbf{x}_0)$, we maximize the Evidence Lower Bound (ELBO) of $\log p_{\theta}(\mathbf{x}_0)$ which measures how accurately the model parameterized by $\theta$ generates true samples $\mathbf{x}_0$ given a corrupted sequence $\mathbf{z}_T$ by iterative sampling from the reverse posterior $p_{\theta}(\mathbf{z}_s|\mathbf{z}_t)$. The ELBO is maximized when $p_{\theta}(\mathbf{x}_0)=1$ and $\log(p_{\theta}(\mathbf{x}_0))=0$ for all sequences $\mathbf{x}_0$ in the dataset, which supports the objective of accurately generating sequences from the data distribution. To convert this into a loss minimization objective, we define the loss function as the negative ELBO (NELBO). First, we compute $\log p_{\theta}(\mathbf{x}_0)$ by integrating over the joint probability of all possible paths of intermediate states from the noisy state $\mathbf{z}_T$ at $t = T$ to the clean sample $x_0$ at $t=0$, denoted by $p_{\theta}(\mathbf{x}_{0:T})$. Since our goal is to reverse the forward masking of the clean sample $x_0$ from all time steps, we introduce an encoder term $q(\mathbf{z}_{1:T}|\mathbf{x}_0)$ denoting the combined distribution of obtaining any noisy sequence between times $t=1$ to $t=T$ from the clean sequence $\mathbf{x}_0$.
\begin{align}
    \log p_{\theta}(\mathbf{x}_0) &= \log \int p_{\theta}(\mathbf{z}_{0:T})d\mathbf{z}_{1:T}\nonumber\\
    &= \log \int q(\mathbf{z}_{1:T}|\mathbf{x}_0)\left[\frac{p_{\theta}(\mathbf{z}_{0:T})}{q(\mathbf{z}_{1:T}|\mathbf{x}_0)}\right]d\mathbf{z}_{1:T}\nonumber\\
    &= \log\left(\mathbb{E}_{q(\mathbf{z}_{1:T}|\mathbf{x}_0)}\left[\frac{p_{\theta}(\mathbf{z}_{0:T})}{q(\mathbf{z}_{1:T}|\mathbf{x}_{0})}\right]\right)
\end{align}
where $\mathbf{z}_{0:T}$ includes $\mathbf{x}_0$.

Using Jenson's inequality, we move the logarithm inside the expectation and reverse the sign to get the NELBO for $\log p_{\theta}(\mathbf{x}_0)$. We split the terms associated with the forward noising process $q(\mathbf{z}_{1:T}|\mathbf{x}_0)$ and the reverse denoising model $p_{\theta}(\mathbf{z}_{0:T})$ into reconstruction term, the prior term, and the intermediate diffusion term. 
\begin{align}
    \mathcal{L}_{\text{NELBO}}
    &= \mathbb{E}_{q(\mathbf{z}_{1:T}|\mathbf{x}_0)}\left[-\log \frac{p_{\theta}(\mathbf{z}_{0:T})}{q(\mathbf{z}_{1:T}|\mathbf{x}_0)}\right] \nonumber\\
    &= \mathbb{E}_{q(\mathbf{z}_{1:T}|\mathbf{x}_0)}\left[
    -\log\frac{p_{\theta}(\mathbf{x}_{0}|\mathbf{z}_{t(1)})p_{\theta}(\mathbf{z}_{t(T)})\prod_{n=1}^{T-1}p_{\theta}(\mathbf{z}_{s(n)}|\mathbf{z}_{t(n)})}{q(\mathbf{z}_{t(T)}|\mathbf{z}_{t(T-1)})\prod_{n=1}^{T-1}q(\mathbf{z}_{t(n)}|\mathbf{z}_{s(n)})}\right] \nonumber\\
    &= \mathbb{E}_{q(\mathbf{z}_{1:T}|\mathbf{x}_0)}\left[
    -\log p_{\theta}(\mathbf{x}_{0}|\mathbf{z}_{t(1)}) - \log\frac{p_{\theta}(\mathbf{z}_{t(T)})}{q(\mathbf{z}_{t(T)}|\mathbf{z}_{s(T)})} - \log\frac{\prod_{t=1}^{T-1}p_{\theta}(\mathbf{z}_{s(n)}|\mathbf{z}_{t(n)})}{\prod_{n=1}^{T-1}q(\mathbf{z}_{t(n)}|\mathbf{z}_{s(n)})}\right] \nonumber\\
    &= \mathbb{E}_{q(\mathbf{z}_{1:T}|\mathbf{x}_0)}\left[
    -\log p_{\theta}(\mathbf{x}_{0}|\mathbf{z}_{t(1)}) - \log\frac{p_{\theta}(\mathbf{z}_{t(T)})}{q(\mathbf{z}_{t(T)}|\mathbf{z}_{s(T)})} - \sum_{n=1}^{T-1}\log\frac{p_{\theta}(\mathbf{z}_{s(n)}|\mathbf{z}_{s(n)})}{q(\mathbf{z}_{t(n)}|\mathbf{z}_{s(n)})}\right] \nonumber 
\end{align}
\begin{align}
    &= \mathbb{E}_{q(\mathbf{z}_{1:T}|\mathbf{x}_0)}\bigg[-\log p_{\theta}(\mathbf{x}_{0}|\mathbf{z}_{t(1)})\bigg] 
    + \mathbb{E}_{q(\mathbf{z}_{1:T}|\mathbf{x}_0)}\left[-\log\frac{p_{\theta}(\mathbf{z}_{t(T)})}{q(\mathbf{z}_{t(T)}|\mathbf{z}_{s(T)})}\right] \nonumber \\
    &\quad + \sum_{n=1}^{T-1}\mathbb{E}_{q(\mathbf{z}_{1:T}|\mathbf{x}_0)}\bigg[-\log\frac{p_{\theta}(\mathbf{z}_{s(n)}|\mathbf{z}_{t(n)})}{q(\mathbf{z}_{t(n)}|\mathbf{z}_{s(n)})}\bigg] \nonumber\\
    &= \underbrace{\mathbb{E}_{q(\mathbf{z}_{t(1)}|\mathbf{x}_0)}\bigg[-\log p_{\theta}(\mathbf{x}_{0}|\mathbf{z}_{t(1)})\bigg]}_{\text{reconstruction loss}} 
    + \underbrace{\mathbb{E}_{q(\mathbf{z}_{t(T)}, \mathbf{z}_{s(T)}|\mathbf{x}_0)}\left[-\log\frac{p_{\theta}(\mathbf{z}_{t(T)})}{q(\mathbf{z}_{t(T)}|\mathbf{z}_{s(T)})}\right]}_{\text{prior loss}} \nonumber \\
    &\quad + \underbrace{\sum_{n=1}^{T-1}\mathbb{E}_{q(\mathbf{z}_{s(n)},\mathbf{z}_{t(n)},\mathbf{z}_{t(n+1)}|\mathbf{x}_0)}\bigg[-\log\frac{p_{\theta}(\mathbf{z}_{s(n)}|\mathbf{z}_{t(n)})}{q(\mathbf{z}_{t(n)}|\mathbf{z}_{s(n)})}\bigg]}_{\text{diffusion loss}}
\end{align}
Now, we can take the limit for each of the loss terms as $T\to \infty$ to derive the continuous-time MDLM objective. 

\paragraph{Reconstruction Loss $\mathcal{L}_{\text{reconst}}$} The reconstruction loss evaluates the final step of the reverse diffusion process that denoises the sequence at time $t(1)$ to the clean sequence at time $t=0$. Since $t(0)=\frac{1}{T}$, the distribution that the sequence $\mathbf{z}_{t(1)}$ is drawn from in the forward pass diffusion is given by
\begin{align}
    p(\mathbf{z}_{t(1)}|\mathbf{x}_0)=\text{Cat}(\mathbf{z}_{t(1)};\alpha_{t(1)}(\mathbf{x}_0)\mathbf{x}_0+(1-\alpha_{t(1)}(\mathbf{x}_0))\mathbf{m})
\end{align}
Since we have $\alpha_t(\mathbf{x}_0)=1-t^w$ for $\mathbf{x}_0=\mathbf{b}$ and $\alpha_t(\mathbf{x}_0)=1-t$ for $\mathbf{x}_0\neq\mathbf{b}$, we can write
\begin{align}
    \alpha_{t(1)}(\mathbf{x}_0)\mathbf{x}_0+(1-\alpha_{t(1)}(\mathbf{x}_0))\mathbf{m}&=\begin{cases}\left(1-\frac{1}{T^w}\right)\mathbf{x}_0+\frac{1}{T^w}\mathbf{m}&\mathbf{x}_0=\mathbf{b}\\\left(1-\frac{1}{T}\right)\mathbf{x}_0+\frac{1}{T}\mathbf{m}&\mathbf{x}_0\neq \mathbf{b}\end{cases}
\end{align}
In the limit as $T\to \infty$, both cases converge to $\mathbf{x}_0$, so we have $\mathbf{z}_{t(1)}\sim \text{Cat}(\mathbf{z}_{t(1)}; \mathbf{x}_0)$ and $\mathbf{z}_{t(1)}=\mathbf{x}_0$. Since $q(\mathbf{z}_{t(1)}|\mathbf{x}_0)=\mathbf{x}_0$ in the forward pass, by parameterizing the reverse posterior to copy-over unmasked tokens, we get $p_{\theta}(\mathbf{x}_0|\mathbf{z}_{t(1)})=\mathbf{x}_0$. Therefore, the reconstruction loss reduces to 0.
\begin{align}
    \mathbb{E}_{q(\mathbf{z}_{t(1)}|\mathbf{x}_0)}\bigg[-\log p_{\theta}(\mathbf{x}_{0}|\mathbf{z}_{t(1)})\bigg]&=\mathbb{E}_{q(\mathbf{z}_{t(1)}|\mathbf{x}_0)}\bigg[-\log p_{\theta}(\mathbf{x}_{0}|\mathbf{x}_0)\bigg]\nonumber\\
    &=0\nonumber
\end{align}

\paragraph{Prior Loss $\mathcal{L}_{\text{prior}}$} 
The prior loss measures the first reverse transition from the fully masked sequence $\mathbf{z}_{t(T)}$ to a slightly unmasked sequence $\mathbf{z}_{s(T)}$. 
\begin{align}
    \mathbb{E}_{q(\mathbf{z}_{t(T)}, \mathbf{z}_{s(T)}|\mathbf{x}_0)}\left[-\log\frac{p(\mathbf{z}_{t(T)})}{q(\mathbf{z}_{t(T)}|\mathbf{z}_{s(T)})}\right]&=-\mathbb{E}_{q( \mathbf{z}_{s(T)}|\mathbf{x}_0)}\underbrace{\mathbb{E}_{q( \mathbf{z}_{t(T)}|\mathbf{z}_{s(T)})}\left[\log\frac{p(\mathbf{z}_{t(T)})}{q(\mathbf{z}_{t(T)}|\mathbf{z}_{s(T)})}\right]}_{\text{KL Divergence}}\nonumber\\
    &= -\mathbb{E}_{q( \mathbf{z}_{s(T)}|\mathbf{x}_0)}\bigg[\text{KL}\bigg(q( \mathbf{z}_{t(T)}|\mathbf{z}_{s(T)})||p_{\theta}(\mathbf{z}_{t(T)})\bigg)\bigg]
\end{align}
Since $t(T)=1$, we have $\alpha_{t(T)}(\mathbf{x}_0)=1-1=0$. Therefore, we derive
\begin{align}
    q(\mathbf{z}_{t(T)}|\mathbf{x}_0)&=\text{Cat}\big(\mathbf{z}_{t(T)}\;;\; \alpha_{t(T)}(\mathbf{x}_0)\mathbf{x}_0+(1-\alpha_{t(T)}(\mathbf{x}_0))\mathbf{m}\big)\nonumber\\
    &=\text{Cat}\big(\mathbf{z}_{t(T)}\;;\; 0\mathbf{x}_0+(1-0)\mathbf{m}\big)\nonumber\\
    &=\text{Cat}\big(\mathbf{z}_{t(T)};\mathbf{m}\big)
\end{align}
Since all sequences are completely masked at time $T$, it follows that the marginal distribution $q(\mathbf{z}_{t(T)}|\mathbf{z}_{s(T)})=\text{Cat}\big(\mathbf{z}_{t(T)};\mathbf{m}\big)$ and the prior distribution $p_{\theta}(\mathbf{z}_{t(T)})=\text{Cat}\big(\mathbf{z}_{t(T)};\mathbf{m}\big)$, so the KL divergence reduces to 0.

\paragraph{Diffusion Loss $\mathcal{L}_T$} 
The diffusion loss measures the consistency of each predicted reverse transition with the forward marginal transition for all $T$ time steps. 
\begin{align}
    \sum_{n=1}^{T-1}\mathbb{E}_{q(\mathbf{z}_{s(n)},\mathbf{z}_{t(n)},\mathbf{z}_{t(n+1)}|\mathbf{x}_0)}&\bigg[-\log\frac{p_{\theta}(\mathbf{z}_{s(n)}|\mathbf{z}_{t(n)})}{q(\mathbf{z}_{t(n)}|\mathbf{z}_{s(n)})}\bigg]\\&=-\sum_{n=1}^{T-1}\mathbb{E}_{q(\mathbf{z}_{s(n)}, \mathbf{z}_{t(n+1)}|\mathbf{x}_0)}\underbrace{\mathbb{E}_{q(\mathbf{z}_{t(n)}|\mathbf{z}_{s(n)})}\bigg[\log\frac{p_{\theta}(\mathbf{z}_{s(n)}|\mathbf{z}_{t(n)})}{q(\mathbf{z}_{t(n)}|\mathbf{z}_{s(n)})}\bigg]}_{\text{KL divergence}}\nonumber\\
    &= -\sum_{n=1}^{T-1}\mathbb{E}_{q(\mathbf{z}_{s(n)}, \mathbf{z}_{t(n+1)}|\mathbf{x}_0)}\bigg[\text{KL}\bigg(q(\mathbf{z}_{t(n)}|\mathbf{z}_{s(n)})||p_{\theta}(\mathbf{z}_{s(n)}|\mathbf{z}_{t(n)})\bigg)\bigg]\nonumber
\end{align}
Since the objective is to accurate predict $\mathbf{z}_{s(n)}$ given $\mathbf{z}_{t(n)}$, we cannot condition on the term $\mathbf{z}_{s(n)}$. Instead, we can condition $q(\mathbf{z}_{t(n)}|\mathbf{z}_{s(n)})$ on $\mathbf{x}_0$ and use Bayes' theorem to derive
\begin{align}
    q(\mathbf{z}_{t(n)}|\mathbf{z}_{s(n)}, \mathbf{x}_0)&=\frac{q(\mathbf{z}_{s(n)}|\mathbf{z}_{t(n)}, \mathbf{x}_0)q(\mathbf{z}_{t(n)}|\mathbf{x}_0)}{q(\mathbf{z}_{s(n)}|\mathbf{x}_0)}\nonumber
\end{align}
Rearranging the terms we get an expression for the true reverse transition $q(\mathbf{z}_{s(n)}|\mathbf{z}_{t(n)},\mathbf{x}_0)$ conditioned on the clean data $\mathbf{x}_0$. By minimizing the KL divergence between the learned reverse transition $p_{\theta}(\mathbf{z}_{s(n)}|\mathbf{z}_{t(n)})$ and $q(\mathbf{z}_{s(n)}|\mathbf{z}_{t(n)},\mathbf{x}_0)$, we can rewrite the diffusion loss as
\begin{align}
    \mathcal{L}_T=\sum_{n=1}^{T-1}\mathbb{E}_{q(\mathbf{z}_{t(n)}|\mathbf{x}_0)}\bigg[\text{KL}\bigg(q(\mathbf{z}_{s(n)}|\mathbf{z}_{t(n)}, \mathbf{x}_0)||p_{\theta}(\mathbf{z}_{s(n)}|\mathbf{z}_{t(n)})\bigg)\bigg]
\end{align}
In Appendix \ref{appendix:G.3}, we derive the bond-dependent continuous-time NELBO loss from its general form above.

\subsection{Guided Diffusion Models}
\label{appendix:B.3}
Guided diffusion aims to sample from the data distribution conditioned on some property $y$, $\mathbf{x}\sim q(\mathbf{x}_0, y)$, such that $q(y|\mathbf{x})$ is maximized. Therefore, the marginal reverse transition aims to sample $\mathbf{z}_s$ from a distribution $q(\mathbf{z}_s|\mathbf{z}_t, y)$ conditioned on the current sequence $\mathbf{z}_t$ and a property $y$. Using Bayes' theorem, we can decompose the guided conditional distribution as
\begin{align}
    q(\mathbf{z}_s|\mathbf{z}_t,y)&= \frac{q(y|\mathbf{z}_s, \mathbf{z}_t)q(\mathbf{z}_s|\mathbf{z}_t)}{q(y|\mathbf{z}_t)}\label{eq:reverse-posterior}
\end{align}
There are two strategies to generate samples from this conditional distribution: classifier-free and classifier-based guidance.

\paragraph{Classifier-Free Guidance}
Classifier-free guidance strategies aim to model the conditional distribution $q(\mathbf{z}_s|\mathbf{z}_t,y)$ by directly training the diffusion model on a subset of the unconditional data with property $y$, such that after training, the model samples from a learned distribution $p_{\theta}(\mathbf{z}_s|\mathbf{z}_t,y)$. However, classifier-free guidance fails at tasks where quality-annotated data is scarce, including peptide sequences. Furthermore, this strategy cannot scale to multiple objectives, which would require data conditioned on more than one property. 

\paragraph{Classifier-Based Guidance} 
Classifier-based guidance trains an unconditional diffusion model $p_{\theta}(\mathbf{z}_s|\mathbf{z}_t)$ and a classifier model $p_{\phi}(y|\mathbf{z}_s)$ with learned parameters $\phi$ that generates a score measuring the probability that the intermediate sequence $\mathbf{z}_s$ has property $y$. By Bayes' theorem, we can model the conditional distribution as
\begin{align}
    p_{\theta, \phi}(\mathbf{z}_s|\mathbf{z}_t,y)= \frac{p_{\phi}(y|\mathbf{z}_s)p_{\theta}(\mathbf{z}_s|\mathbf{z}_t)}{p_{\phi}(y|\mathbf{z}_t)}
\end{align}
Since the model parameters implicitly learn the normalized distribution, we can drop the $p_{\phi}(y|\mathbf{z}_t)$ term. Then, at each iteration, we update the parameters $\theta, \phi$ in the direction of the gradient of $\log p_{\theta, \phi}(\mathbf{z}_s|\mathbf{z}_t,y)$ obtained as the sum of the gradients of the unconditional distribution $\log p_{\theta}(\mathbf{z}_s|\mathbf{z}_t)$ and classifier probability $p_{\phi}(y|\mathbf{z}_s)$ with respect to the sampled sequence $\mathbf{z}_s$. 
\begin{align}
    \nabla_{\mathbf{z}_s}\log p_{\theta, \phi}(\mathbf{z}_s|\mathbf{z}_t,y)= \nabla_{\mathbf{z}_s}\log p_{\phi}(y|\mathbf{z}_s)+\nabla_{\mathbf{z}_s} \log p_{\theta}(\mathbf{z}_s|\mathbf{z}_t)
\end{align}
After joint training with the classifier and unconditional data distribution, we can sample from the learned conditional distribution $p_{\theta, \phi}(\mathbf{z}_s|\mathbf{z}_t,y)$.

Unlike classifier-free guidance, classifier-based guidance does not require training a generative model from a conditioned dataset. However, the gradient-based strategy for classifier-based guidance is not directly applicable to discrete state spaces due to the lack of a defined gradient. To mimic gradient-based updates to each sampling step, Gruver et al. \cite{Gruver2023} leveraged iterative guidance steps on continuous latent embeddings for each token before decoding back to a discrete sequence at each time step.  However, projecting to and from the continuous and discrete spaces results in inconsistencies in the guided diffusion process, where optimized hidden embeddings do not always map to optimal tokens. 

\paragraph{Guidance in the Discrete State Space} 
To directly guide the diffusion objective in the discrete space, we must compute the optimality of a single discrete reverse transition $\mathbf{z}_s$ against all other possible transitions to maximize the conditional probability $p(y| \mathbf{z}_s, \mathbf{z}_t)$. That is, we need to compute Equation (\ref{eq:reverse-posterior}) with the denominator expanded to represent all possible transitions from $\mathbf{z}_t$.
\begin{align}
    p_{\theta, \phi}(\mathbf{z}_s|\mathbf{z}_t,y)= \frac{p_{\phi}(y|\mathbf{z}_s)p_{\theta}(\mathbf{z}_s|\mathbf{z}_t)}{\sum_{\mathbf{z}'_s}p_{\phi}(y|\mathbf{z}'_s)p_{\theta}(\mathbf{z}'_s|\mathbf{z}_t)}
\end{align}
However, computing $p_{\phi}(y|\mathbf{z}'_s)$ for all the possible transitions from state $\mathbf{z}_t$ is computationally intractable. Previous work has bypassed this limitation by approximation. For continuous and differentiable classifier functions $p(y|\mathbf{x}): \mathbb{R}^{L\times |\mathcal{V}|}\to \mathbb{R}$, we can approximate the denominator using the first-order Taylor expansion given by
\begin{align}
    \log p_{\phi}(y|\mathbf{z}_s, \mathbf{z}_t)&\approx \log p_{\phi}(y|\mathbf{z}_t)+(\mathbf{z}_s-\mathbf{z}_t)^{\top}\nabla_{\mathbf{z}}\log p(y|\mathbf{z})|_{\mathbf{z}=\mathbf{z}_t}
\end{align}
which approximates the likelihood of observing property $y$ at the slightly denoised state $\mathbf{z}_s=\mathbf{z}_{t-\frac{1}{T}}$ given the known log-probability of observing $y$ for state $\mathbf{z}_t$. This eliminates the need to explicitly sample $\mathbf{z}_s$ for all possible state transitions and compute $\log p_{\phi}(y|\mathbf{z}_s, \mathbf{z}_t)$. 

Digress \cite{Vignac2022} has used this approximation strategy for guided discrete diffusion on categorical graph generation. Furthermore, Nisonoff et al. \cite{Nisinoff2024} uses the Taylor-approximated conditional distribution $\log p_{\phi}(y|\mathbf{z}_s)$ to adjust the unconditional transition rates $R_t(\mathbf{z}_t, \mathbf{z}_s|y)$ given the unconditional rates $R_t(\mathbf{z}_t, \mathbf{z}_s)$ for predictor-guidance of Continuous-Time Markov Chains (CTMCs) in the discrete state space.
\begin{align}
    R_t(\mathbf{z}_t, \mathbf{z}_s|y)&=R_t(\mathbf{z}_t, \mathbf{z}_s)\frac{\log p_{\phi}(y|\mathbf{z}_s, \mathbf{z}_t)}{\log p_{\phi}(y|\mathbf{z}_t)}
\end{align}
where $R_t(\mathbf{z}_t, \mathbf{z}_s|y)$ is the predictor-adjusted rate of transitioning from state $\mathbf{z}_t$ to state $\mathbf{z}_s$

However, this strategy fails to scale to multiple objective guidance since it would require computing the joint probability over $K$ objectives $p_{\phi}(y_1, y_2, \dots, y_K|\mathbf{z}_s, \mathbf{z}_t)$ for some $K> 1$. If all properties are mutually independent, we can factorize the distribution and compute the estimated probability of each objective and take their product $p_{\phi}(y_1, y_2, \dots, y_K|\mathbf{z}_s, \mathbf{z}_t)=\prod_{k=1}^Kp_{\phi}(y_k|\mathbf{z}_s, \mathbf{z}_t)$. For the majority of multi-objective tasks, including therapeutic peptide generation, independence across properties is not a reasonable assumption, and computing the joint distribution is required. Moreover, for objectives that guide toward contradictory optimal rates or transitions, training a model conditioned on these objectives could prevent the model from generating optimal sequences for either objective. Given these limitations, there remains a gap for efficient classifier-based conditioning for discrete diffusion that is robust to multi-objective tasks, which we address in this work. 

\section{Data Curation and Tokenization}
\label{appendix:C}
\subsection{PepMDLM Training Data}  
\label{appendix:C.1}
To train the unconditional masked diffusion language model generator, we collected 11 million peptide SMILES consisting of 7451 sequences from the CycPeptMPDB database \cite{Li2023}, 825,632 unique peptides from SmProt \cite{Li2021}, and approximately 10 million modified peptides generated from CycloPs \cite{Duffy2011, Feller2024}, which consists of 90\% canonical amino acids, 10\% unnatural amino acids from SwissSidechain \cite{Gfeller2012}, 10\% dextro-chiral alpha carbons, 20\% N-methylated amine backbone atoms, and  10\% PEGylated peptides. All possible cyclization conformations were attempted on the peptides generated with CycloPs. We used SELFIES \cite{krenn2020self} to check the integrity of the SMILES sequences.

We split our data by $k$-means clustering into 1000 groups of sequences with similar chemical properties based on their Morgan fingerprint \cite{Rogers2010}, which is a bit-vector representation of the full peptide sequence where each bit encodes a feature relating to the SMILES atom types, connectivity, and bonding environment. The final dataset was a 0.8 to 0.2 split based on the clusters, maintaining similar diversities of the SMILES strings. Since the degree of masking is evenly spread between $t=0$ to $t=1$ within each training batch, grouping similar SMILES in the same batch ensures the model learns to reconstruct a diverse set of peptide SMILES from various degrees of masking. 

\subsection{Dynamic Batching}\label{appendix:C.2}  
We applied dynamic batching to handle variable-length token sequences and increase computational efficiency. Inspired by ESM-2's dynamic batching technique \cite{Lin2023-gh}, input SMILES are sorted by length to maximize the utility of GPU memory. The maximum tokens per GPU was set to 16,000.

\subsection{SMILES Tokenization} 
\label{appendix:C.3}
To enable the novel generation of non-natural amino acids containing cyclizations and diverse backbone and side-chain modifications, we trained our generative diffusion model on Simplified Molecular-Input Line-Entry System (SMILES) \cite{Weininger1988} representations of peptides. We experimented with several tokenization schemes that capture common motifs in the training data to enhance the generation of valid peptide SMILES. We find that the SMILES Pair Encoding (SPE) tokenization scheme \cite{Li2021} with the PeptideCLM \cite{Feller2024} vocabulary of 581 SMILES tokens and 5 special tokens with an average length of four characters per token, demonstrated superior performance, generating precise but valid peptides (Appendix \ref{appendix:Tokenization Scheme.}).

\section{PepMDLM Implementation Details}
\label{appendix:D}

\subsection{Notation} 
\label{appendix:D.1}
Let $\mathbf{x}_0\in \{0, 1\}^{|\mathcal{V}|}$ represent the one-hot vector of a token in a sequence in the training data and $\mathbf{x}_{\theta}(\mathbf{z}_t,t)\in \Delta^{|\mathcal{V}|}$ be the vector of predicted token probabilities across the vocabulary $\mathcal{V}$ given the current state $\mathbf{z}_t$ at time $t$. In most contexts, $\mathbf{x}_0$ will be used to denote a single token, but when discussing the full sequence, $\mathbf{x}_0^{(\ell)}$ is used to denote the token at position $\ell$ in the sequence. Let $T$ denote the total number of time steps in the discrete forward and reverse diffusion processes. In the context of all time steps, we expand $t$ to $t(n)\in (0, 1]$ when denoting a single time step in the forward and backward diffusion process with $n\in [1\dots T-1]$. Let $s(n)=t(n)-\frac{1}{T}$ denote the previous time step in the forward process. Then, let $\mathbf{z}_{t(n)}$ and $\mathbf{z}_{s(n)}$ denote the state of a specific token at time $t(n)$ and $s(n)$ in the diffusion process, respectively. Let $\alpha_t(\mathbf{x}_0):\mathbb{R}^{|\mathcal{V}|}\to \mathbb{R}$ denote a function that takes the unmasked token $\mathbf{x}_0$ and outputs a value in [0,1] representing the probability of remaining unmasked at time $t$ in the forward diffusion process. Let $\mathbf{b}\in \mathbb{R}^{|\mathcal{V}|}$ denote a vector with ones at indices of peptide bond tokens and zeroes at all remaining indices, and let $\mathbf{x}_0=\mathbf{b}$ indicate that $\mathbf{x}_0$ is a peptide-bond token. 

\subsection{Model Architecture}
\label{appendix:D.2}
To predict the token probabilities at each reverse step $\mathbf{x}_{\theta}(\mathbf{z}_t, t)$, we trained a RoFormer model \cite{Roformer} that leverages rotary positional embeddings (RoPE) robust to varying input lengths and long-range dependencies between tokens. The specific hyperparameters of our model are given below.
\begin{table}[H]
  \caption{\textbf{Roformer Architecture Hyperparameters}}
  \label{roformer-table}
  \vskip 0.05in
  \centering
  \begin{tabular}{ll}
    \toprule
     \textbf{Hyperparameter}    & \textbf{PepTune}\\
    \midrule
    Input Dimension  & 581 (vocab size)  \\
    Hidden Dimension & 768 \\
    Intermediate Dimension & 3072 \\
    Number of Layers & 8 \\
    Attention Heads & 8 \\
    Max Positional Embeddings & 1035 \\
    Hidden and Attention Dropout Probability & 0.1\\
    \bottomrule
  \end{tabular}
\end{table}

\subsection{Unconditional Generation Results} 
\label{appendix:D.3}
Here, we provide additional tables and figures demonstrating the performance of our unconditional PepMDLM model. Table \ref{table:loss} shows the training and validation metrics after 8 epochs of training on 11 million peptide SMILES. Table \ref{table:benchmark} presents our benchmark results in comparison to the HELM-GPT model \cite{Xu2024}. We demonstrate that PepMDLM generates peptides with higher uniqueness, diversity, and lower similarity to their nearest neighbor (SNN) (details on metrics are provided in Appendix \ref{appendix:F.2}). While the percentage of valid unconditionally-generated peptides is lower than HELM-GPT, peptides generated by HELM-GPT are represented in HELM notation \cite{Zhang2012}, where each token corresponds to an amino acid. In contrast, PepMDLM is trained on SMILES tokens that decompose amino acids into smaller tokens that can be pieced together into valid amino acids during generation. This enables us to represent a greater diversity of non-natural amino acids, but even a single token generated in the wrong position can result in an invalid peptide sequence, resulting in a lower validity rate. However, we note that our MCTS guidance strategy increases the validity rate to 100\% in only ~20 iterations due to its iterative unmasking process that is rewarded on high-scoring and valid peptides.

\begin{figure}[H]
    \centering
    \begin{minipage}{0.6\textwidth}
        \centering
        \includegraphics[width=\linewidth]{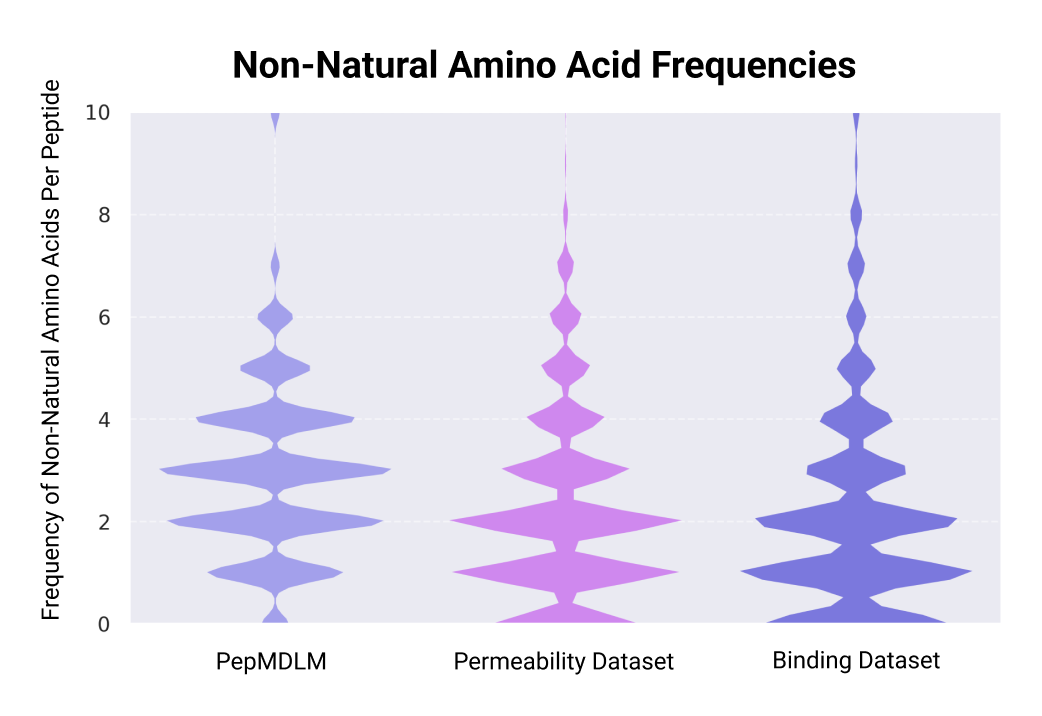} 
    \end{minipage}
    \hfill
    \begin{minipage}{0.6\textwidth}
        \centering
        
        \begin{table}[H]
          \vskip 0.05in
          \centering
          \resizebox{\textwidth}{!}{%
            \begin{tabular}{lccc}
                \toprule
                     & \textbf{Permeability Data} & \textbf{Binding Data} & \textbf{PepMDLM} \\
                \midrule
                Mean nAAs Per Peptide           & 2.215 & 2.150 & 2.940 \\
                Cyclic Peptides (\%)  & 0.467 & 0.027 & 0.100 \\
                \bottomrule
            \end{tabular}%
        }
        \end{table}
    \end{minipage}
    \caption{\textbf{PepMDLM generates cyclic and modified peptides. }(Above) Distribution comparison of non-natural amino acid frequency for 100 unconditionally-generated peptide SMILES with the peptide SMILES dataset of experimentally-validated peptides for membrane permeability (PAMPA) and binding affinity (Appendix \ref{appendix:B.1}). (Bottom) Per peptide frequency of non-natural amino acids (nAAs) and percentage of cyclic peptides in PepMDLM-generated sequences and experimentally validated membrane-permeable peptides.}
    \label{fig:nAA}
\end{figure}

In Figure \ref{fig:nAA}, we show that the frequency of non-natural amino acids from Swiss-Sidechain \cite{Gfeller2012} present in our PepMDLM-generated peptides is similar to the membrane permeability and binding datasets containing experimentally-validtated modified peptides that we used to train our property classifiers (Appendix \ref{appendix:B.1}). In addition, we show that 10\% of unconditionally generated peptides from PepMDLM contain modifications that result in cyclicization (Figure \ref{fig:nAA}). In total, we demonstrate PepMDLM's unique capability of generating diverse non-natural and cyclic peptides.

\begin{table}
  \centering
  \caption{\textbf{Training and Validation Loss of PepMDLM.} Loss values are taken after convergence at 8 epochs when training PepMDLM on 11 million peptide SMILES with bond-dependent masking and invalid loss.}
  \label{table:loss}
  \vskip 0.05in
  \adjustbox{max width=\textwidth}{%
    \begin{tabular}{lcccc}
      \toprule
      \textbf{Model} & \textbf{Train Loss ($\downarrow$)} & \textbf{Train PPL ($\downarrow$)} & \textbf{Val Loss ($\downarrow$)} & \textbf{Val PPL ($\downarrow$)} \\
      \midrule
      PepMDLM & 0.832 & 2.460 & 0.880 & 2.277 \\
      \bottomrule
    \end{tabular}%
  }
\end{table}

\begin{table}[h!]
\centering
\caption{\textbf{Benchmark of PepMDLM unconditional model against HELM-GPT. }The best scores are bolded. HELM-GPT was trained on HELM notation, where each token is a monomer encoding natural and modified residues. Since there are no existing peptide SMILES generative models, we chose HELM-GPT as the closest comparison. The validity is assessed differently, as all HELM sequences correspond to a valid peptide, while not all SMILES sequences can be decoded into a peptide. }
\label{table:benchmark}
\vskip 0.05in
\begin{tabular}{@{}lcccc@{}}
\toprule
\textbf{Model} & \textbf{Validity ($\uparrow$)} & \textbf{Uniqueness} ($\uparrow$)& \textbf{Diversity} ($\uparrow$)& \textbf{SNN} ($\downarrow$)\\ \midrule
HELM-GPT & \textbf{0.839} & 0.913 & 0.595 & 0.975  \\
PepMDLM & 0.450 & \textbf{1.000} & \textbf{0.705} & \textbf{0.513}\\
\bottomrule
\end{tabular}
\end{table}

\section{Therapeutic Property Prediction for Peptide SMILES}
\label{appendix:E}
While several classifiers exist for predicting properties of small-molecule SMILES sequences and amino-acid representations of peptides, there exists a gap in high-quality property models trained specifically on peptide SMILES data. To fill this gap, we train regression models for target-binding affinity and cell membrane permeability and binary classification models for solubility, hemolysis, and non-fouling specifically on peptide SMILES data (Table \ref{table:solubility_hemolysis_non_fouling}). 

\subsection{Protein Target-Binding Prediction}
\label{appendix:E.1}
\begin{figure}[h!]
    \centering
    \includegraphics[width=\linewidth]{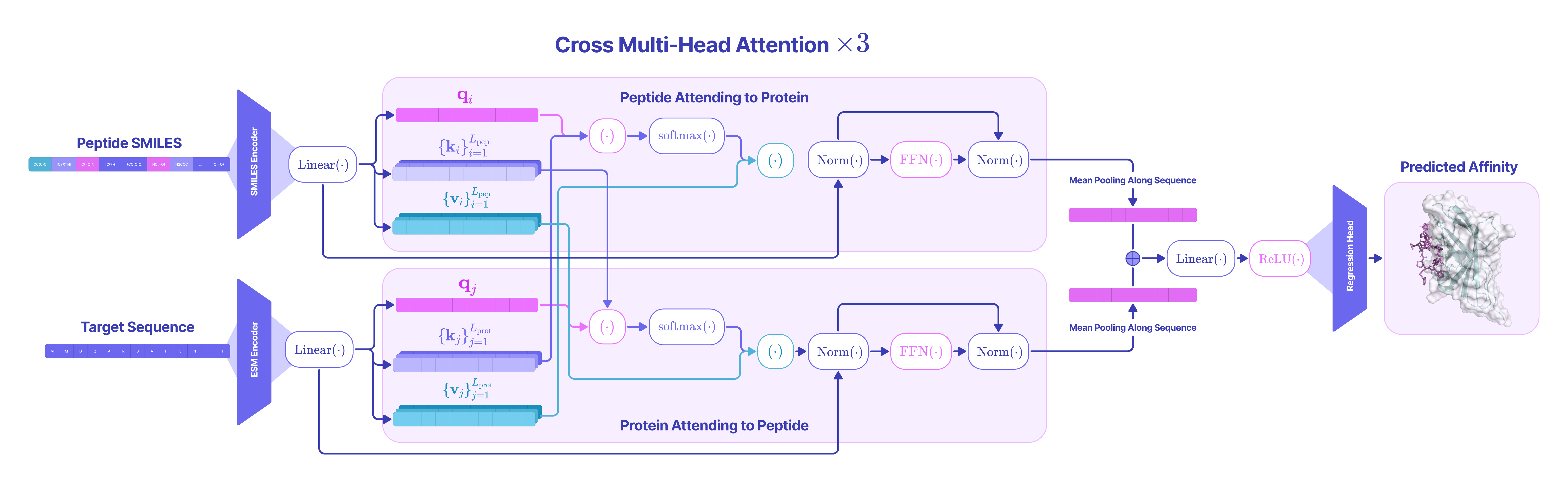}
    \caption{\textbf{Architecture of binding affinity regression model.} Embeddings for the target protein sequence are generated with ESM-2, and embeddings for the peptide SMILES are generated using PeptideCLM. Cross multi-head attention layers combine the embeddings and predict a binding affinity score.}
    \label{fig:affinity}
\end{figure}

To guide the generation of peptides with high binding affinity to a given protein target, we trained a Transformer-based model with cross multi-head attention layers that learn the joint latent space of ESM-2-650M \cite{Lin2023-gh} embeddings of the protein amino acid sequence and PeptideCLM \cite{Feller2024} embeddings of the peptide SMILES sequence (Figure \ref{fig:affinity}; Table \ref{table:affinity-predictor}). 

We trained on 1806 sequences from the PepLand \cite{Zhang2023} canonical and non-canonical binding datasets containing the protein-target sequence, peptide SMILES sequence, and the experimentally-validated $K_d/K_i/IC50$ binding affinity score. Given a peptide SMILES sequence and a protein amino acid sequence, the model predicts a score that indicates weak binding ($< 6.0$), medium binding ($6.0-7.5$), and high binding ($\geq 7.5$). After training for 50 epochs, our regression model achieved a strong Spearman correlation coefficient of $0.869$ on the training data and $0.633$ on the held-out validation data. 

\begin{table}[h!]
  \caption{\textbf{Cross-Attention Model Architecture for Target-Binding Affinity Prediction}}
  \label{table:affinity-predictor}
  \vskip 0.05in
  \centering
  \begin{tabular}{lcc}
    \toprule
     \textbf{Layers} & \textbf{Protein Dimension} & \textbf{Peptide Dimension} \\
    \midrule
    Embedding Module  & 1280 &  768\\
    Linear Layer & 512 & 512 \\
    Layer Norm & 512 & 512 \\
    \textbf{Cross-Attention} $\times 3$ & &\\
    \hspace{5mm} Multi-Head Attention ($h=8$) & 512  & 512\\
     \hspace{5mm} Linear Layer &  2048& 2048\\
    \hspace{5mm} ReLU & 2048  & 2048\\
    \hspace{5mm} Dropout & 2048  & 2048\\
    \hspace{5mm} Linear Layer &  512& 512\\
    \textbf{Shared Prediction Head} & &\\
    \hspace{5mm} Linear Layer & \multicolumn{2}{c}{1024} \\
    \hspace{5mm} ReLU & \multicolumn{2}{c}{1024} \\
    \hspace{5mm} Dropout & \multicolumn{2}{c}{1024} \\
    \textbf{Regression Head} &  \multicolumn{2}{c}{1} \\
    \textbf{Classification Head } &  \multicolumn{2}{c}{3} \\
    \bottomrule
  \end{tabular}
\end{table}

\subsection{Cell Membrane Permeability Prediction}
\label{appendix:E.2}
For cell membrane permeability, we trained an XGBoost \cite{chen2016xgboost} boosted tree regression model on PeptideCLM \cite{Feller2024} embeddings which returns the predicted PAMPA lipophilicity score ($\log P$) given a peptide SMILES sequence, where values $\geq -6.0$ indicate strong permeability and values $< 6.0$ indicate weak permeability. The XGBoost regression parameters were optimized with 50 trials of OPTUNA search and are provided in Table \ref{tab:xgboost_params_combined}.

\begin{figure}[h!]
    \centering
    \includegraphics[width=0.8\linewidth]{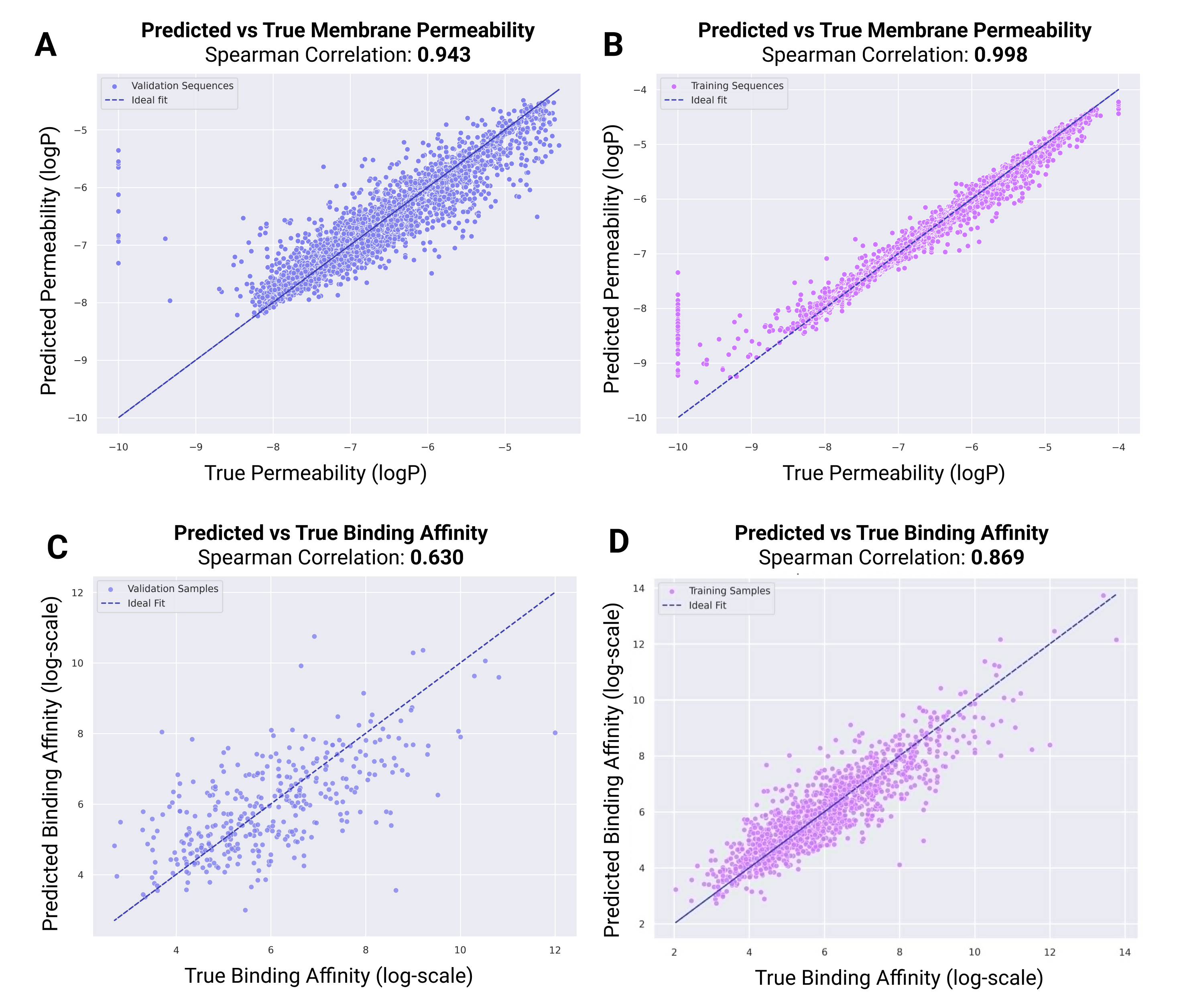}
    \caption{\textbf{Correlation plots for binding affinity and membrane permeability classifiers.} Plot of true permeability (logP) on the $x$-axis and predicted permeability on the $y$-axis for the (\textbf{A}) validation set and (\textbf{B}) training set. Plot of true binding affinity (log-scale) on the $x$-axis and predicted permeability on the $y$-axis for the (\textbf{C}) validation set and (\textbf{D}) training set. }
    \label{fig:binding_affinity and permeability}
\end{figure}

The dataset contains 34,853 experimentally validated peptide SMILES, consisting of 22,040 SMILES sequences obtained from the ChEMBL database \cite{Mendez2018} and 7451 sequences from the CycPeptMPDB database \cite{Li2023}. Data was randomly shuffled and split into a 0.8/0.1/0.1 ratio for train, validation, and test. Our model achieved a strong Spearman correlation coefficient of 0.998 on the training dataset and 0.943 on the test dataset (Figure \ref{fig:binding_affinity and permeability}, Table \ref{table:binding affinity and permeability}). 

\begin{table}[h!]
\centering
\caption{\textbf{Held-out validation performance of binding affinity and membrane permeability regression models trained on peptide SMILES.} Spearman rank correlation and MSE were calculated on the 20 percent held-out validation set.}
\vskip 0.05in
\label{table:binding affinity and permeability}
\resizebox{0.7\textwidth}{!}{
\begin{tabular}{@{}lccc@{}}
\toprule
\textbf{Metric} & \textbf{Binding Affinity} & \textbf{Membrane Permeability} \\ \midrule
Spearman Rank Correlation & 0.633 & 0.943 \\
MSE & 0.566 & 0.088 \\
\bottomrule
\end{tabular}
}
\end{table}

\subsection{Solubility and Toxicity Prediction} 
\label{appendix:E.3} 
 For solubility, hemolysis, and non-fouling, we collected binary data from the PepLand and PeptideBERT datasets \cite{Zhang2023, Guntuboina2023}, where 1 indicates the positive class and 0 indicates the negative class. Since increased solubility improves drug-loading, we seek to maximize the probability of the positive class. Since positive hemolysis indicates that the peptide sequence induces destruction of red blood cells, we seek to minimize the probability of the positive class, or maximize the inverse. Positive non-fouling indicates lower off-target binding, so we seek to maximize positive non-fouling.
 
 We leveraged PeptideCLM embedding representations of the SMILES tokens and trained XGBoost models for binary classification. The optimal thresholds for the positive class were determined to be 0.500 for solubility, 0.800 for hemolysis (non-hemolysis), and 0.450 for non-fouling. The XGBoost binary classification parameters were optimized with 50 trials of OPTUNA search and are provided in Table \ref{tab:xgboost_params_combined}. We achieved strong Spearman correlations across all three tasks, showing improvements against the state-of-the-art PeptideBERT \cite{Guntuboina2023} baseline model (Table \ref{table:solubility_hemolysis_non_fouling}).

\begin{table}[h!]
\centering
\caption{\textbf{XGBoost Hyperparameters for Classification and Regression.} Classification hyperparameters were used for the solubility, hemolysis, and non-fouling binary classification models (left). Regression hyperparameters were used for the membrane permeability regression model (right). }
\label{tab:xgboost_params_combined}
\vskip 0.05in
\begin{tabular}{@{}ll|ll@{}}
\toprule
\multicolumn{2}{c}{\textbf{Classification Hyperparameters}} & \multicolumn{2}{c}{\textbf{Regression Hyperparameters}} \\ \midrule
\textbf{Hyperparameter}       & \textbf{Value/Range}          & \textbf{Hyperparameter}       & \textbf{Value/Range}          \\ \midrule
Objective                     & \texttt{binary:logistic}      & Objective                     & \texttt{reg:squarederror}     \\
Lambda                        & \([1\text{e}{-8}, 10.0]\)     & Lambda                        & \([0.1, 10.0]\) (log scale)   \\
Alpha                         & \([1\text{e}{-8}, 10.0]\)     & Alpha                         & \([0.1, 10.0]\) (log scale)   \\
Colsample by Tree             & \([0.1, 1.0]\)               & Gamma                         & \([0, 5]\)                   \\
Subsample                     & \([0.1, 1.0]\)               & Colsample by Tree             & \([0.5, 1.0]\)               \\
Learning Rate                 & \([0.01, 0.3]\)              & Subsample                     & \([0.6, 0.9]\)               \\
Max Depth                     & \([2, 30]\)                  & Learning Rate                 & \([1\text{e}{-5}, 0.1]\)     \\
Min Child Weight              & \([1, 20]\)                  & Max Depth                     & \([2, 30]\)                  \\
Tree Method                   & \texttt{hist}                & Min Child Weight              & \([1, 20]\)                  \\
                   &              & Tree Method                   & \texttt{hist}                \\
                              &                               & Scale Pos Weight              & \([0.5, 10.0]\) (log scale)  \\ \bottomrule
\end{tabular}
\end{table}

\section{Evaluation}
\label{appendix:F}

\subsection{Peptide Validity Filter}  
\label{appendix:F.1} 
Among the sequential representations of peptides, including amino acid sequences, HELM \cite{Zhang2012}, and SMILES \cite{Weininger1988}, SMILES is the most intricate representation of peptide sequences. Although this enables the representation of non-natural amino acids, diverse side chains, and backbone modifications, and cyclic peptides, it also means that the vast majority of SMILES strings are not synthesizable peptides. Therefore, we devised an algorithm that determines whether a SMILES string is a valid peptide, characterized by peptide bonds and central carbon atoms. The filter first checks if the SMILES sequence is a valid molecule using RDKit \cite{rdkit}. 

Then, we use regular expressions to detect bond patterns for peptide bonds, N-methylated peptide bonds, reversed peptide bonds, and ester bonds, along the sequence to split the sequence into several segments with a bond before and after each segment. The filter checks each segment for chemical modifications based on their bond type, including N-methylation (N-Me) and O-linked glycosylation. The remaining segment is matched to the corresponding natural or non-natural amino acid side chains (Algorithm \ref{alg:7}). Our filter is capable of detecting a library of over 200 nAAs from SwissSidechain \cite{Gfeller2012} and can classify a peptide SMILES as cyclic or non-cyclic. The tool is freely available on HuggingFace: \url{https://huggingface.co/spaces/ChatterjeeLab/SMILES2PEPTIDE}. 

\subsection{Metrics}
\label{appendix:F.2}
To evaluate the generation quality of our unconditional MDLM, PepMDLM, and our MCTS-guided MDLM, PepTune, we leverage the Moses metrics, including validity, uniqueness, diversity, similarity to nearest neighbor (SNN), and novelty \cite{Polykovskiy2020}.

\paragraph{Validity} is defined as the fraction of peptide SMILES that pass our SMILES2PEPTIDE filter (Algorithm \ref{alg:7}), indicating that it translates to a synthesizable peptide.

\paragraph{Uniqueness} is defined as the fraction of mutually distinct peptide SMILES.

\paragraph{Diversity} is defined as one minus the average Tanimoto similarity between the Morgan fingerprints of every pair of generated sequences, which measures the similarity in structure across generated peptides. 
\begin{align}
    \text{Diversity}=1-\frac{1}{\binom{N_{\text{generated}}}{2}}\sum_{i,j}\frac{\mathbf{f}(\mathbf{x}_i)\cdot \mathbf{f}(\mathbf{x}_j)}{|\mathbf{f}(\mathbf{x}_i)|+|\mathbf{f}(\mathbf{x}_j)|-\mathbf{f}(\mathbf{x}_i)\cdot \mathbf{f}(\mathbf{x}_j)}
\end{align}
where $\mathbf{f}(\mathbf{x}_i)$ and $\mathbf{f}(\mathbf{x}_j)$ are the 2048-dimensional Morgan fingerprint with radius 3 for a pair of generated sequences $\mathbf{x}_i$ and $\mathbf{x}_j$. 

\paragraph{Similarity to Nearest Neighbor (SNN)} is defined as the maximum Tanimoto similarity between a generated sequence $\mathbf{x}_i$ with a sequence in the dataset $\tilde{\mathbf{x}}_j$.
\begin{align}
    \text{SNN}=\max_{j\in |\mathcal{D}|}\left(\frac{\mathbf{f}(\mathbf{x}_i)\cdot \mathbf{f}(\tilde{\mathbf{x}}_j)}{|\mathbf{f}(\mathbf{x}_i)|+|\mathbf{f}(\tilde{\mathbf{x}}_j)|-\mathbf{f}(\mathbf{x}_i)\cdot \mathbf{f}(\tilde{\mathbf{x}}_j)}\right)
\end{align}
 
\paragraph{Randomness} is defined as the Shannon Entropy \cite{lin1991divergence} on tokenized sequences given as:
\begin{align}
    E = -\sum_{i}^{L}p_{i}\log_{2}(p_{i})
\end{align}
where $p_{i}$ is the probability of $i$-th unique token divided by the total number of tokens $L$ in the sequence.

\paragraph{KL-Divergence} is defined as the divergence between the token distribution in the generated peptide SMILES $p_i$ and the token distribution in the training data.
\begin{align}
    \text{KL}(P\big|\big|Q) &=\sum_{i\in \mathcal{V}}\begin{cases}
        p_{i}\log_{2}(\frac{p_{i}}{q_{i}})& \text{if} \ q_{i} > 0\\
        p_{i}\log_{2}(\frac{p_{i}}{10_{-9}})& \text{if} \ q_{i}= 0
    \end{cases}
\end{align}
where $p_{i}$ is the probability of token $i$ in the training data, and $q_{i}$ is the probability of token $i$ in the generated data.

Due to the limit of memory and CPU time required to load all the training dataset of 11 million peptide SMILES, we chose to sample a subset of 1000 batches randomly ($\sim$100k sequences) for novelty and SNN calculation. To assess the novelty of generated sequences, we employed Shannon entropy \cite{lin1991divergence} to quantify the SMILES token randomness between 100 PepTune-generated and 100 PepMDLM-generated sequences and the same randomly sampled 1000 subsets from the training set. Then, Kullback-Leibler (KL) divergence was used to evaluate divergence across token distributions from the generated peptides compared to the training data. 

\subsection{Docking}
\label{appendix:F.3} 
For valid generated peptide SMILES with non-dominated scores across objectives, we used Autodock Vina \cite{eberhardt2021autodock} (v 1.1.2) for \textit{in silico} docking of the peptide binders to their target proteins (Appendix \ref{pdb-table}) to confirm binding affinity. Targets were preprocessed with MGITools \cite{morris2009autodock4} (v 1.5.7) and the conformations of the SMILES were optimized by ETKDG from RDKit \cite{eberhardt2021autodock, wang2020etkdg}. The final results were visualized in PyMol \cite{PyMOL} (v 3.1), where the residues in the protein targets with polar contacts to the peptide binder with distances closer than 3.5 \AA\ are annotated. 

\begin{table}[H]
  \caption{\textbf{PDB structures of target proteins used for docking.}}
  \label{pdb-table}
  \vskip 0.05in
  \centering
  \begin{tabular}{ll}
    \toprule
     \textbf{Protein}    & \textbf{PDB}\\
    \midrule
    GFAP  & 6A9P  \\
    TfR & 3KAS \\
    GLP-1R & 3C5T \\
    AMHR2 & 7L0J \\
    GLAST & 5LM4 \\
    NCAM1 & 2HAZ \\
    RBX1 & 1LDJ \\
    \bottomrule
  \end{tabular}
\end{table}

\section{Theoretical Details}
\label{appendix:G}
\subsection{Bond-Dependent Masking Schedule}\label{appendix:Bond-Dependent Masking Schedule}
From Equation (\ref{eq:masking}), we define the continuous-time forward masking probability $1-\alpha_t$ at time $t$ with $\alpha_t=\exp(-\bar{\sigma}(t))$ , where $\bar{\sigma}:[0,1]\to \mathbb{R}^+$ is the cumulative discrete-time masking schedule. Following Lou et al. \cite{Lou2024}, we apply a log-linear masking schedule $\bar{\sigma}(t)=-\log(1-t)$ for the forward diffusion process, which has been shown to result in the lowest variance in the NELBO loss \cite{Sahoo2024}. Therefore, the continuous-time probability of remaining unmasked at time $t$ is equal to $\alpha_t=\exp\big(-(-\log(1-t))\big)=1-t$ and the weight that scales the negative log loss (NLL) is given by $\frac{1}{t}$ by our derivation in Appendix \ref{appendix:G.3}. 

For peptide-bond tokens, we alter the masking schedule such that peptide-bonds are masked at a slower rate at earlier time steps by defining a log-polynomial masking schedule $\bar{\sigma}(t)=-\log(1-t^w)$, for some constant exponent $w> 1$. Note that when $w=1$, the log-polynomial schedule reduces to the log-linear schedule. Therefore, the probability of remaining unmasked becomes $\alpha_t=\big(-(-\log(1-t^w))\big)=1-t^w$ and the weight that scales the negative log loss (NLL) is given by $\frac{w}{t}$ by our derivation in Appendix \ref{appendix:G.3}. 

Since $t\in(0, 1]$, the probability that a peptide-bond token remains unmasked at time $t$ is equal to $\alpha_t=1-t^w$, which is larger than the log-linear schedule for $w>1$. Conversely, the probability that a peptide-bond token is masked before $t$ is $1-\alpha_t=t^w$, which is smaller than the log-linear schedule for $w> 1$. As $t\to 1$, $\alpha_t\to 0$ for both the log-linear and log-polynomial time schedules, which means that both peptide-bond and non-peptide bond tokens will have a high probability of being masked in later times in the forward pass diffusion process. 

The NLL of the peptide-bond tokens is weighted more heavily than non-peptide bond tokens for $t$ close to 1. As $t\to 0$, the NLL weight approaches $\infty$ for all tokens. This biases the reverse diffusion process to unmask peptide bond tokens earlier since it was trained to minimize the loss associated with each unmasking step. As $t\to 0$, the large NLL weight ensures that the final unmasking steps during the reverse diffusion process result in an unmasked sequence that lies within the space of valid peptide SMILES. 

\begin{figure}
    \centering
    \includegraphics[width=\linewidth]{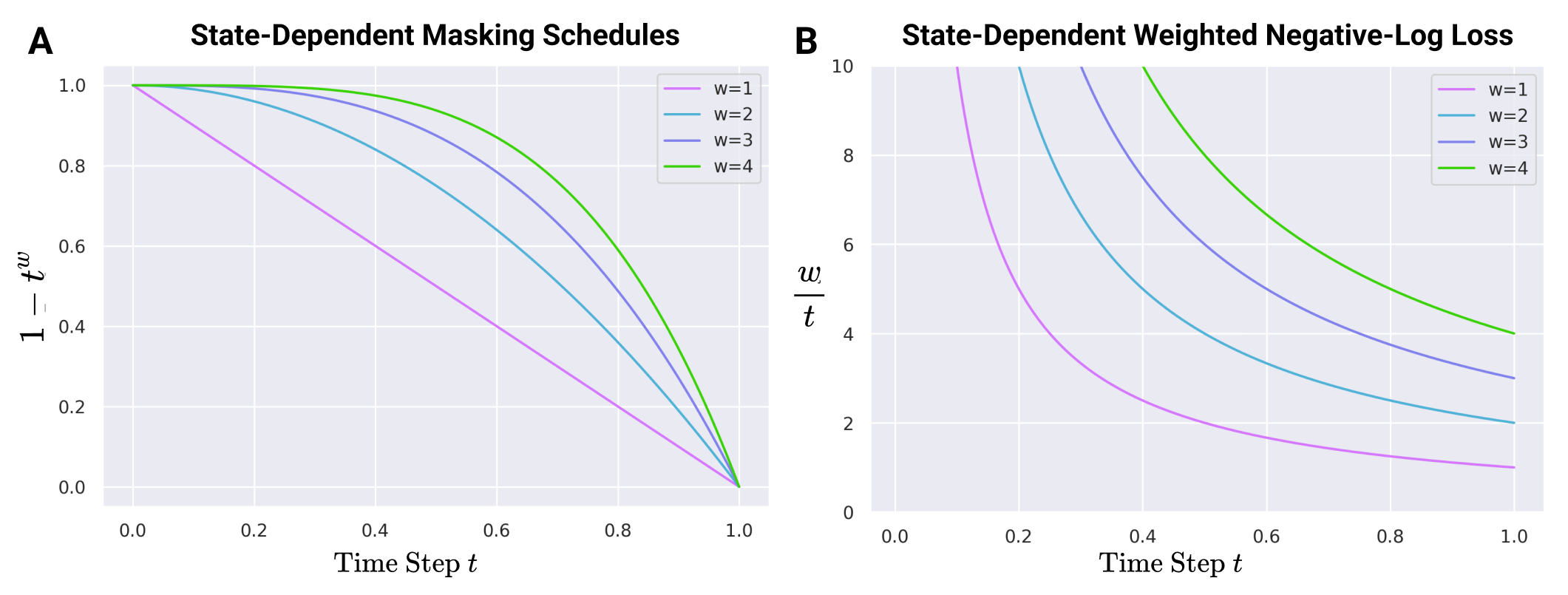}
    \caption{\textbf{Plots of bond-dependent masking schedules.} (\textbf{A}) The probability of remaining unmasked during the continuous-time forward diffusion process over time $t$ given different values of $w$ as the exponent of the masking schedule $\alpha_t=1-t^w$. We use $w=1$ for non-peptide bond tokens and $w=3$ for peptide bond tokens, resulting in slower masking of peptide-bond tokens. (\textbf{B}) The weight of the negative log-loss for different exponents $w$ in the log-polynomial masking schedule. The weight of the loss is higher for larger $w$ in earlier time steps, which results in a higher penalty for inaccurate predictions of peptide bond tokens compared to other tokens. }
    \label{fig:masking}
\end{figure}

\subsection{Derivation of Bond-Dependent Reverse Posterior}
\label{appendix:G.2}
\paragraph{Proposition 2.1} The reverse posterior defining the probability distribution of the token $\mathbf{z}_s$ at time $s=t-\Delta t$ given the token $\mathbf{z}_t$ at time $t$ with our bond-dependent forward masking schedule is defined as 
\begin{align}
    q(\mathbf{z}_s|\mathbf{z}_t, \mathbf{x}_0)=\begin{cases}
        \left\langle\left(\frac{s}{t}-\frac{s^w}{t^w}\right)\mathbf{b}+\frac{t-s}{t}\mathbf{1}, \mathbf{x}_0\right\rangle \mathbf{x}_0+\left\langle \left(\frac{s^w}{t^w}-\frac{s}{t}\right)\mathbf{b}+\frac{s}{t}\mathbf{1}, \mathbf{x}_0\right\rangle \mathbf{m}&\mathbf{z}_t=\mathbf{m}\\
        \mathbf{z}_t&\mathbf{z}_t\neq \mathbf{m}
    \end{cases}
\end{align}
When the clean token is a peptide bond token (i.e. $\mathbf{x}_0=\mathbf{b}$), the transition distribution for a masked token $\mathbf{z}_s=\mathbf{m}$ reduces to $q(\mathbf{z}_s|\mathbf{z}_t=\mathbf{m}, \mathbf{x}_0=\mathbf{b})=\left(1-\frac{s^w}{t^w}\right)\mathbf{x}_0+\left(\frac{s^w}{t^w}\right)\mathbf{m}$. When the clean token is not a peptide bond token (i.e. $\mathbf{x}_0\neq \mathbf{b}$), the transition distribution for a masked token $\mathbf{z}_s=\mathbf{m}$ reduces to $q(\mathbf{z}_s|\mathbf{z}_t=\mathbf{m}, \mathbf{x}_0\neq \mathbf{b})=\left(1-\frac{s}{t}\right)\mathbf{x}_0+\left(\frac{s}{t}\right)\mathbf{m}$. If the token is already unmasked, it remains unmasked at the same token with probability 1.

\textit{Proof.} For a single token, the bond-dependent forward diffusion process is defined by the probability distribution $q(\mathbf{z}_t|\mathbf{x}_0)$ which transforms the clean inputs to sequences with varying degrees of masking based on a probability distribution $\alpha_t(\mathbf{x}_0)$. We define $\alpha_t(\mathbf{x}_0): \mathbb{R}^{|\mathcal{V}|}\to \mathbb{R}$ as a function that takes the clean token encoding $\mathbf{x}_0$ and outputs the probability of remaining unmasked at time $t$ depending on whether $\mathbf{x}_0$ encodes a peptide bond token. 
\begin{align}
    q(\mathbf{z}_t|\mathbf{x}_0)=\text{Cat}(\mathbf{z}_t;(\alpha_t(\mathbf{x}_0))\mathbf{x}_0+(1-\alpha_t(\mathbf{x}_0))\mathbf{m})
\end{align}
Then, the marginal forward transition from time $s(n)\to t(n)$ is defined as
\begin{align}
    q(\mathbf{z}_{t(n)}|\mathbf{z}_{s(n)})=\text{Cat}\left(\mathbf{z}_{t(n)};\left(\frac{\alpha_t(\mathbf{x}_0)}{\alpha_s(\mathbf{x}_0)}\right)\mathbf{x}_0+\left(\mathbf{1}-\frac{\alpha_t(\mathbf{x}_0)}{\alpha_s(\mathbf{x}_0)}\right)\mathbf{m}\right)
\end{align}
In this work, we classify each token into one of two states: peptide-bond tokens and non-peptide-bond tokens, which represent amino acid side chains and modifications. We define a function that generates a mask with values of 1 indicating tokens containing or contained within a peptide bond, and 0 otherwise (Algorithm \ref{alg:6}). Let $\mathbf{b}\in \mathbb{R}^{|\mathcal{V}|}$ denote a vector with ones at indices of peptide-bond tokens. For derivation purposes, we let $\mathbf{b}^{\top}\mathbf{x}_0^{(\ell)}=1$ and $\mathbf{x}_0^{(\ell)}=\mathbf{b}$ when a token at position $\ell$ is a peptide bond token. Note that $\mathbf{b}$ is defined differently depending on the context of the token in the full sequence, which is handled by the \textsc{BondMask} function. Then, we have $\alpha_t(\mathbf{x}_0)= \big(\mathbf{1}-\mathbf{b}^{\top}\mathbf{x}_0\big)(1-t)+\mathbf{b}^{\top}\mathbf{x}_0(1-t^w)$ or equivalently we can write
\begin{align}
    \alpha_t(\mathbf{x}_0)=\begin{cases}
        1-t^w&\mathbf{x}_0=\mathbf{b}\\
        1-t&\mathbf{x}_0\neq \mathbf{b}
    \end{cases}
\end{align}
By Bayes' rule, the general state-independent form of the true reverse posterior is given by
\begin{align}
    q(\mathbf{z}_s|\mathbf{z}_t, \mathbf{x}_0)&=\frac{q(\mathbf{z}_t|\mathbf{z}_s)q(\mathbf{z}_s|\mathbf{x}_0)}{q(\mathbf{z}_t|\mathbf{x}_0)}\nonumber\\
    &=\frac{\left[\left(\frac{\alpha_t}{\alpha_s}\right)\mathbf{z}_s^{\top}\mathbf{z}_t+\left(1-\frac{\alpha_t}{\alpha_s}\right)\mathbf{m}^{\top}\mathbf{z}_t\right]\left[\alpha_s\mathbf{x}_0^{\top}\mathbf{z}_t+\left(1-\alpha_s\right)\mathbf{m}^{\top}\mathbf{z}_t\right]}{\left[\alpha_t\mathbf{x}_0^{\top}\mathbf{z}_t+\left(1-\alpha_t\right)\mathbf{m}^{\top}\mathbf{z}_t\right]}
\end{align}
With bond-dependent masking, the value of $\alpha_t(\mathbf{x}_0)$ and $\alpha_s(\mathbf{x}_0)$ depend on the state of $\mathbf{x}_0$, so the bond-dependent reverse posterior becomes
\begin{align}
    q(\mathbf{z}_s|\mathbf{z}_t, \mathbf{x}_0)&=\frac{\left[\left(\frac{\alpha_t(\mathbf{z}_s)}{\alpha_s(\mathbf{z}_s)}\right)\mathbf{z}_s^{\top}\mathbf{z}_t+\left(1-\frac{\alpha_t(\mathbf{z}_s)}{\alpha_s(\mathbf{z}_s)}\right)\mathbf{m}^{\top}\mathbf{z}_t\right]\left[\alpha_s(\mathbf{x}_0)\mathbf{x}_0^{\top}\mathbf{z}_t+\left(1-\alpha_s(\mathbf{x}_0)\right)\mathbf{m}^{\top}\mathbf{z}_t\right]}{\left[\alpha_t(\mathbf{x}_0)\mathbf{x}_0^{\top}\mathbf{z}_t+\left(1-\alpha_t(\mathbf{x}_0)\right)\mathbf{m}^{\top}\mathbf{z}_t\right]}
\end{align}
When $\mathbf{z}_t=\mathbf{x}_0$, the true reverse posterior simplifies to
\begin{align}
    q(\mathbf{z}_s|\mathbf{z}_t=\mathbf{x}_0, \mathbf{x}_0)&=\frac{\left[\left(\frac{\alpha_t(\mathbf{z}_s)}{\alpha_s(\mathbf{z}_s)}\right)\mathbf{z}_s^{\top}\mathbf{x}_0+\left(1-\frac{\alpha_t(\mathbf{z}_s)}{\alpha_s(\mathbf{z}_s)}\right)\mathbf{m}^{\top}\mathbf{x}_0\right]\left[\alpha_s(\mathbf{x}_0)\mathbf{x}_0^{\top}\mathbf{x}_0+\left(1-\alpha_s(\mathbf{x}_0)\right)\mathbf{m}^{\top}\mathbf{x}_0\right]}{\left[\alpha_t(\mathbf{x}_0)\mathbf{x}_0^{\top}\mathbf{x}_0+\left(1-\alpha_t(\mathbf{x}_0)\right)\mathbf{m}^{\top}\mathbf{x}_0\right]}\nonumber\\
    &=\frac{\left[\left(\frac{\alpha_t(\mathbf{z}_s)}{\alpha_s(\mathbf{z}_s)}\right)\mathbf{z}_s^{\top}\mathbf{x}_0\right]\left[\alpha_s(\mathbf{x}_0)\right]}{\alpha_t(\mathbf{x}_0)}
\end{align}
When $\mathbf{z}_s \neq \mathbf{x}_0$, $\mathbf{z}_s^{\top} \mathbf{x}_0=0$ so $q(\mathbf{z}_s\neq\mathbf{x}_0|\mathbf{z}_t=\mathbf{x}_0, \mathbf{x}_0)=0$. When $\mathbf{z}_s=\mathbf{x}_0$, we have
\begin{align}
    q(\mathbf{z}_s=\mathbf{x}_0|\mathbf{z}_t=\mathbf{x}_0, \mathbf{x}_0)&=\frac{\left[\left(\frac{\alpha_t(\mathbf{x}_0)}{\alpha_s(\mathbf{x}_0)}\right)\mathbf{x}_0^{\top}\mathbf{x}_0\right]\left[\alpha_s(\mathbf{x}_0)\right]}{\alpha_t(\mathbf{x}_0)}\nonumber\\
    &=\left(\frac{\alpha_t(\mathbf{x}_0)}{\alpha_s(\mathbf{x}_0)}\right)\left(\frac{\alpha_s(\mathbf{x}_0)}{\alpha_t(\mathbf{x}_0)}\right)\nonumber\\
    &=1\nonumber
\end{align}
which means that $\mathbf{z}_t$ remains unchanged after unmasking. This supports the carry-over unmasking scheme, which explicitly sets the probability of changing an unmasked token equal to $-\infty$. 

In the forward diffusion process, a token either remains unchanged or is masked, so the only other case we need to consider is $\mathbf{z}_t=\mathbf{m}$. Since the masking schedule differs only when the ground truth token is a peptide bond token, or $\mathbf{x}_0=\mathbf{b}$, we can consider two cases: first, when $\mathbf{x}_0=\mathbf{b}$ and second, when $\mathbf{x}_0\neq \mathbf{b}$.

\textbf{Case 1.} Consider the case when $\mathbf{x}_0=\mathbf{b}$ or the ground truth token $\mathbf{x}_0$ is a peptide-bond token. From our modified masking schedule, we have $\alpha_t(\mathbf{b})=1-t^w$. Therefore, we can write the probability distribution for unmasking a peptide-bond token as
\begin{align}
    q(\mathbf{z}_s|\mathbf{z}_t=\mathbf{m}, \mathbf{x}_0=\mathbf{b})&=\frac{\left[\left(\frac{\alpha_t(\mathbf{z}_s)}{\alpha_s(\mathbf{z}_s)}\right)\mathbf{z}_s^{\top}\mathbf{m}+\left(1-\frac{\alpha_t(\mathbf{z}_s)}{\alpha_s(\mathbf{z}_s)}\right)\mathbf{m}^{\top}\mathbf{m}\right]\left[\alpha_s(\mathbf{b})\mathbf{b}^{\top}\mathbf{z}_s+\left(1-\alpha_s(\mathbf{b})\right)\mathbf{m}^{\top}\mathbf{z}_s\right]}{\left[\alpha_t(\mathbf{b})\mathbf{b}^{\top}\mathbf{m}+\left(1-\alpha_t(\mathbf{b})\right)\mathbf{m}^{\top}\mathbf{m}\right]}\nonumber\\
    &=\frac{\left[\left(\frac{\alpha_t(\mathbf{z}_s)}{\alpha_s(\mathbf{z}_s)}\right)\mathbf{z}_s^{\top}\mathbf{m}+\left(1-\frac{\alpha_t(\mathbf{z}_s)}{\alpha_s(\mathbf{z}_s)}\right)\right]\left[\alpha_s(\mathbf{b})\mathbf{b}^{\top}\mathbf{z}_s+\left(1-\alpha_s(\mathbf{b})\right)\mathbf{m}^{\top}\mathbf{z}_s\right]}{\left(1-\alpha_t(\mathbf{b})\right)}
\end{align}
The probability of transitioning from a masked state to a peptide-bond token is simplified to
\begin{align}
    q(\mathbf{z}_s=\mathbf{b}|\mathbf{z}_t=\mathbf{m}, \mathbf{x}_0=\mathbf{b})&=\frac{\left[\left(\frac{\alpha_t(\mathbf{b})}{\alpha_s(\mathbf{b})}\right)\mathbf{b}^{\top}\mathbf{m}+\left(1-\frac{\alpha_t(\mathbf{b})}{\alpha_s(\mathbf{b})}\right)\right]\left[\alpha_s(\mathbf{b})\mathbf{b}^{\top}\mathbf{b}+\left(1-\alpha_s(\mathbf{b})\right)\mathbf{m}^{\top}\mathbf{b}\right]}{\left(1-\alpha_t(\mathbf{b})\right)}\nonumber\\
    &=\frac{\left(1-\frac{1-t^w}{1-s^w}\right)(1-s^w)}{\left(1-(1-t^w)\right)}\nonumber\\
    &=\frac{\left(\frac{1-s^w-1+t^w}{1-s^w}\right)(1-s^w)}{t^w}\nonumber\\
    &=\frac{t^w-s^w}{t^w}\nonumber\\
    &=1-\frac{s^w}{t^w}\label{eq:case1pt1}
\end{align}
The probability of remaining in a masked state is
\begin{align}
    q(\mathbf{z}_s=\mathbf{m}|\mathbf{z}_t=\mathbf{m}, \mathbf{x}_0=\mathbf{b})&=\frac{\left[\left(\frac{\alpha_t(\mathbf{m})}{\alpha_s(\mathbf{m})}\right)\mathbf{m}^{\top}\mathbf{m}+\left(1-\frac{\alpha_t(\mathbf{m})}{\alpha_s(\mathbf{m})}\right)\right]\left[\alpha_s(\mathbf{b})\mathbf{b}^{\top}\mathbf{m}+\left(1-\alpha_s(\mathbf{b})\right)\mathbf{m}^{\top}\mathbf{m}\right]}{\left(1-\alpha_t(\mathbf{b})\right)}\nonumber\\
    &=\frac{\left[\left(\frac{\alpha_t(\mathbf{m})}{\alpha_s(\mathbf{m})}\right)+\left(1-\frac{\alpha_t(\mathbf{m})}{\alpha_s(\mathbf{m})}\right)\right]\left(1-\alpha_s(\mathbf{b})\right)}{\left(1-\alpha_t(\mathbf{b})\right)}\nonumber\\
    &=\frac{1-\alpha_s(\mathbf{b})}{1-\alpha_t(\mathbf{m})}\nonumber\\
    &=\frac{1-(1-s^w)}{1-(1-t^w)}\nonumber\\
    &=\frac{s^w}{t^w}\label{eq:case1pt2}
\end{align}
which aligns with the constraint that $\mathbf{z}_t\in \{\mathbf{m}, \mathbf{x}_0\}$ in the forward diffusion process.

\textbf{Case 2: }Consider the case when $\mathbf{x}_0\neq\mathbf{b}$ or the ground truth token $\mathbf{x}_0$ is not a peptide-bond token. From the baseline log-linear masking schedule, we have $\vec{\alpha}_t^{\top}\mathbf{x}_0=1-t$. Therefore, we can write the probability distribution for unmasking a peptide-bond token as
\begin{align}
    q(\mathbf{z}_s|\mathbf{z}_t=\mathbf{m}, \mathbf{x}_0\neq\mathbf{b})&=\frac{\left[\left(\frac{\alpha_t(\mathbf{z}_s)}{\alpha_s(\mathbf{z}_s)}\right)\mathbf{z}_s^{\top}\mathbf{m}+\left(1-\frac{\alpha_t(\mathbf{z}_s)}{\alpha_s(\mathbf{z}_s)}\right)\mathbf{m}^{\top}\mathbf{m}\right]\left[\alpha_s(\mathbf{x}_0)\mathbf{x}_0^{\top}\mathbf{m}+\left(1-\alpha_s(\mathbf{x}_0)\right)\mathbf{m}^{\top}\mathbf{m}\right]}{\left[\alpha_t(\mathbf{x}_0)\mathbf{x}_0^{\top}\mathbf{m}+\left(1-\alpha_t(\mathbf{x}_0)\right)\mathbf{m}^{\top}\mathbf{m}\right]}\nonumber\\
    &=\frac{\left[\left(\frac{\alpha_t(\mathbf{z}_s)}{\alpha_s(\mathbf{z}_s)}\right)\mathbf{z}_s^{\top}\mathbf{m}+\left(1-\frac{\alpha_t(\mathbf{z}_s)}{\alpha_s(\mathbf{z}_s)}\right)\right]\left[\alpha_s(\mathbf{x}_0)\mathbf{x}_0^{\top}\mathbf{m}+\left(1-\alpha_s(\mathbf{x}_0)\right)\right]}{\left[\alpha_t(\mathbf{x}_0)\mathbf{x}_0^{\top}\mathbf{m}+\left(1-\alpha_t(\mathbf{x}_0)\right)\right]}
\end{align}
With similar steps to Case 1, the probability of transitioning from a masked state to a non-peptide-bond token is given by
\begin{align}
    q(\mathbf{z}_s=\mathbf{x}_0|\mathbf{z}_t=\mathbf{m}, \mathbf{x}_0\neq \mathbf{b})&=\frac{\left(1-\frac{\alpha_t(\mathbf{x}_0)}{\alpha_s(\mathbf{x}_0)}\right)\left(1-\alpha_s(\mathbf{x}_0)\right)}{\left(1-\alpha_t(\mathbf{x}_0)\right)}\nonumber\\
    &=\frac{\left(1-\frac{1-t}{1-s}\right)\left(1-(1-s)\right)}{\left(1-(1-t)\right)}\nonumber\\
    &=\frac{t-s}{t}\nonumber\\
    &=1-\frac{s}{t}\label{eq:case2pt1}
\end{align}
It follows that the probability of remaining in a masked state in the reverse process is
\begin{align}
    q(\mathbf{z}_s=\mathbf{m}|\mathbf{z}_t=\mathbf{m}, \mathbf{x}_0\neq\mathbf{b})&=\frac{s}{t}\label{eq:case2pt2}
\end{align}
This demonstrates that the probability of remaining in a masked state when $\mathbf{x}_0=\mathbf{b}$ is smaller than when $\mathbf{x}_0\neq\mathbf{b}$, since taking the exponent of a fraction results in a smaller value. So we have $\frac{s^w}{t^w}< \frac{s}{t}$ for $w> 1$. 

Combining Equations (\ref{eq:case1pt2}) and (\ref{eq:case2pt2}) we get the following distribution for the case when $\mathbf{z}_t=\mathbf{m}$ and $\mathbf{z}_s=\mathbf{m}$
\begin{align}
    q(\mathbf{z}_s=\mathbf{m}|\mathbf{z}_t=\mathbf{m}, \mathbf{x}_0)&=\left(\frac{s^w}{t^w}-\frac{s}{t}\right)\mathbf{b}^{\top}\mathbf{x}_0+\frac{s}{t}\nonumber\\
    &=\left(\frac{s^w}{t^w}\mathbf{b}-\frac{s}{t}\mathbf{b}+\frac{s}{t}\mathbf{1}\right)^{\top}\mathbf{x}_0\nonumber\\
    &=\left(\left(\frac{s^w}{t^w}-\frac{s}{t}\right)\mathbf{b}+\frac{s}{t}\mathbf{1}\right)^{\top}\mathbf{x}_0
\end{align}
Similarly, combining (\ref{eq:case1pt1}) and (\ref{eq:case2pt1}) we get the following distribution for the case when $\mathbf{z}_t=\mathbf{m}$ and $\mathbf{z}_s\neq \mathbf{m}$ or equivalently $\mathbf{z}_s=\mathbf{x}_0$. 
\begin{align}
    q(\mathbf{z}_s= \mathbf{x}_0|\mathbf{z}_t=\mathbf{m}, \mathbf{x}_0)&=\left(\frac{s}{t}-\frac{s^w}{t^w}\right)\mathbf{b}^{\top}\mathbf{x}_0+\left(1-\frac{s}{t}\right)\nonumber\\
    &=\left(\frac{s}{t}\mathbf{b}-\frac{s^w}{t^w}\mathbf{b}+\mathbf{1}-\frac{s}{t}\mathbf{1}\right)^{\top}\mathbf{x}_0\nonumber\\
    &=\left(\left(\frac{s}{t}-\frac{s^w}{t^w}\right)\mathbf{b}+\frac{t-s}{t}\mathbf{1}\right)^{\top}\mathbf{x}_0
\end{align}
Now, we can write the true reverse posterior as
\begin{align}
    q(\mathbf{z}_s|\mathbf{z}_t, \mathbf{x}_0)=\begin{cases}
        \left\langle\left(\frac{s}{t}-\frac{s^w}{t^w}\right)\mathbf{b}+\frac{t-s}{t}\mathbf{1}, \mathbf{x}_0\right\rangle \mathbf{x}_0+\left\langle \left(\frac{s^w}{t^w}-\frac{s}{t}\right)\mathbf{b}+\frac{s}{t}\mathbf{1}, \mathbf{x}_0\right\rangle \mathbf{m}&\mathbf{z}_t=\mathbf{m}\\
        \mathbf{z}_t&\mathbf{z}_t\neq \mathbf{m}
    \end{cases}
\end{align}
Therefore, we get the following expression for the parameterized reverse posterior
\begin{align}
    p_{\theta}(\mathbf{z}_s|\mathbf{z}_t)=\begin{cases}
        \left\langle\left(\frac{s}{t}-\frac{s^w}{t^w}\right)\mathbf{b}+\frac{t-s}{t}\mathbf{1}, \mathbf{x}_{\theta}(\mathbf{z}_t,t)\right\rangle \mathbf{x}_{\theta}(\mathbf{z}_t,t)+\left\langle \left(\frac{s^w}{t^w}-\frac{s}{t}\right)\mathbf{b}+\frac{s}{t}\mathbf{1}, \mathbf{x}_{\theta}(\mathbf{z}_t,t)\right\rangle \mathbf{m}&\mathbf{z}_t=\mathbf{m}\\
        \mathbf{z}_t&\mathbf{z}_t\neq \mathbf{m}
    \end{cases}
\end{align}

\subsection{Derivation of Bond-Dependent NELBO Loss}
\label{appendix:G.3}
\paragraph{Proposition 2.2}
The bond-dependent continuous-time NELBO decomposes into the sum of the negative log-losses (NLLs) for all non-peptide bond tokens that follow a log-linear masking schedule and the sum of the NLLs for all peptide bond tokens that follow a log-polynomial schedule. 
\begin{align}
    \mathcal{L}^{\infty}_{\text{NELBO}}&=\mathbb{E}_{t, q(\mathbf{z}_t|\mathbf{x}_0)}\bigg[-\sum_{\ell:\mathbf{x}^{(\ell)}_0=\mathbf{b}}\frac{w}{t}\log\langle\mathbf{x}^{(\ell)}_0,\mathbf{x}^{(\ell)}_{\theta}(\mathbf{z}_t, t)\rangle-\sum_{\ell:\mathbf{x}^{(\ell)}_0\neq\mathbf{b}}\frac{1}{t}\log\langle\mathbf{x}^{(\ell)}_0,\mathbf{x}^{(\ell)}_{\theta}(\mathbf{z}_t, t)\rangle\bigg]
\end{align}

\textit{Proof.} The diffusion objective in its general form is given by
\begin{align}
    \mathcal{L}_{\text{NELBO}}&=\sum_{n=1}^{T-1}\mathbb{E}_{q(\mathbf{z}_{t(n)}|\mathbf{x}_0)}\bigg[\text{KL}\bigg(q(\mathbf{z}_{s(n)}|\mathbf{z}_{t(n)}, \mathbf{x}_0)||p_{\theta}(\mathbf{z}_{s(n)}|\mathbf{z}_{t(n)})\bigg)\bigg]\nonumber\\
    &=\mathbb{E}_{t\in\{\frac{1}{T}, \frac{2}{T}, \dots, 1\}}\mathbb{E}_{q(\mathbf{z}_t|\mathbf{x}_0)}\bigg[T\cdot \text{KL}\bigg(q(\mathbf{z}_s|\mathbf{z}_t, \mathbf{x}_0)\big|\big|p_{\theta}(\mathbf{z}_s|\mathbf{z}_t)\bigg)\bigg]
\end{align}
First, we will derive an expression for the bond-dependent KL-divergence, which measures the difference between the learned reverse posterior $q\big(\mathbf{z}_s|\mathbf{z}_t, \mathbf{x}_{\theta}(\mathbf{z}_t,t)\big)$ and the true reverse posterior $q(\mathbf{z}_s|\mathbf{z}_t,\mathbf{x}_0)$ conditioned on the training distribution $\mathbf{x}_0$. 
\begin{align}
    &\text{KL}(q(\mathbf{z}_s|\mathbf{z}_t, \mathbf{x}_0)||p_{\theta}(\mathbf{z}_s|\mathbf{z}_t))\nonumber\\
    &=\sum_{\mathbf{z}_s=\mathbf{e}_k}q(\mathbf{z}_s|\mathbf{z}_t=\mathbf{m}, \mathbf{x}_0)\log\frac{q(\mathbf{z}_s|\mathbf{z}_t=\mathbf{m},\mathbf{x}_0)}{p_{\theta}(\mathbf{z}_s|\mathbf{z}_t=\mathbf{m})}\nonumber\\
    &=\sum_{\mathbf{z}_s\in \{\mathbf{x}_0, \mathbf{m}\}}q(\mathbf{z}_s|\mathbf{z}_t=\mathbf{m}, \mathbf{x}_0)\log\frac{q(\mathbf{z}_s|\mathbf{z}_t=\mathbf{m},\mathbf{x}_0)}{p_{\theta}(\mathbf{z}_s|\mathbf{z}_t=\mathbf{m})}\nonumber\\
    &=q(\mathbf{z}_s=\mathbf{x}_0|\mathbf{z}_t=\mathbf{m}, \mathbf{x}_0)\log\frac{q(\mathbf{z}_s=\mathbf{x}_0|\mathbf{z}_t=\mathbf{m},\mathbf{x}_0)}{p_{\theta}(\mathbf{z}_s=\mathbf{x}_0|\mathbf{z}_t=\mathbf{m})}\nonumber\\
    &+q(\mathbf{z}_s=\mathbf{m}|\mathbf{z}_t=\mathbf{m}, \mathbf{x}_0)\log\frac{q(\mathbf{z}_s=\mathbf{m}|\mathbf{z}_t=\mathbf{m},\mathbf{x}_0)}{p_{\theta}(\mathbf{z}_s=\mathbf{m}|\mathbf{z}_t=\mathbf{m})}\nonumber\\
    &=\left(\left(\frac{s}{t}-\frac{s^w}{t^w}\right)\mathbf{b}+\frac{t-s}{t}\mathbf{1}\right)^{\top}\mathbf{x}_0\log\frac{\left(\left(\frac{s}{t}-\frac{s^w}{t^w}\right)\mathbf{b}+\frac{t-s}{t}\mathbf{1}\right)^{\top}\mathbf{x}_0}{\left(\left(\frac{s}{t}-\frac{s^w}{t^w}\right)\mathbf{b}+\frac{t-s}{t}\mathbf{1}\right)^{\top}\mathbf{x}_{\theta}(\mathbf{z}_t, t)}\nonumber\\
    &+\left(\left(\frac{s^w}{t^w}-\frac{s}{t}\right)\mathbf{b}+\frac{s}{t}\mathbf{1}\right)^{\top}\mathbf{x}_0\log\frac{\left(\left(\frac{s^w}{t^w}-\frac{s}{t}\right)\mathbf{b}+\frac{s}{t}\mathbf{1}\right)^{\top}\mathbf{x}_0}{\left(\left(\frac{s^w}{t^w}-\frac{s}{t}\right)\mathbf{b}+\frac{s}{t}\mathbf{1}\right)^{\top}\mathbf{x}_{\theta}(\mathbf{z}_t, t)}
\end{align}
In the case where the true token $\mathbf{x}_0=\mathbf{b}$, we can simplify to
\begin{align}
    \text{KL}(q(\mathbf{z}_s|\mathbf{z}_t, \mathbf{x}_0)||p_{\theta}(\mathbf{z}_s|\mathbf{z}_t))&=\left(\frac{s}{t}-\frac{s^w}{t^w}+1-\frac{s}{t}\right)\log\frac{\left(\frac{s^w}{t^w}-\frac{s}{t}+\frac{s}{t}\right)\mathbf{x}_0^{\top}\mathbf{x}_0}{\left(\frac{s^w}{t^w}-\frac{s}{t}+\frac{s}{t}\right)\mathbf{x}_0^{\top}\mathbf{x}_{\theta}(\mathbf{z}_t, t)}\nonumber\\
    &=-\left(1-\frac{s^w}{t^w}\right)\log\big(\mathbf{x}_0^{\top}\mathbf{x}_{\theta}(\mathbf{z}_t, t)\big)\nonumber\\
    &=-\left(\frac{t^w-s^w}{t^w}\right)\log\langle\mathbf{x}_0,\mathbf{x}_{\theta}(\mathbf{z}_t, t)\rangle
\end{align}
Substituting $s=t-\frac{1}{T}$, we can simplify $s^w$ to 
\begin{align}
    s^w&=\left(t-\frac{1}{T}\right)^w\nonumber\\
    &=\left[t\left(1-\frac{1}{tT}\right)\right]^w\nonumber\\
    &=t^w\left(1-\frac{1}{tT}\right)^w]\nonumber\\
    &=t^w\left(1-\frac{w}{tT}+o\left(\frac{1}{T^2}\right)\right)\tag{$(1+x)^w=1+wx+o(x^2)$}\nonumber\\
    &=t^w-\frac{wt^{w-1}}{T}+t^wo\left(\frac{1}{T^2}\right)
\end{align}
where $o\left(\frac{1}{T^2}\right)$ denotes higher order terms that grow slower than $\frac{1}{T^2}$. 

Now, we can write
\begin{align}
    \text{KL}(q(\mathbf{z}_s|\mathbf{z}_t, \mathbf{x}_0)||p_{\theta}(\mathbf{z}_s|\mathbf{z}_t))&=-\left(\frac{t^w-\left(t^w-\frac{wt^{w-1}}{T}+t^wo\left(\frac{1}{T^2}\right)\right)}{t^w}\right)\log\langle\mathbf{x}_0,\mathbf{x}_{\theta}(\mathbf{z}_t, t)\rangle\nonumber\\
    &=-\left(\frac{\frac{wt^{w-1}}{T}-t^wo\left(\frac{1}{T^2}\right)}{t^w}\right)\log\langle\mathbf{x}_0,\mathbf{x}_{\theta}(\mathbf{z}_t, t)\rangle\nonumber\\
    &=-\left(\frac{w}{tT}-o\left(\frac{1}{T^2}\right)\right)\log\langle\mathbf{x}_0,\mathbf{x}_{\theta}(\mathbf{z}_t, t)\rangle\nonumber\\
\end{align}
In the case where the true token $\mathbf{x}_0\neq\mathbf{b}$, we can simplify to 
\begin{align}
    \text{KL}(q(\mathbf{z}_s|\mathbf{z}_t, \mathbf{x}_0)||p_{\theta}(\mathbf{z}_s|\mathbf{z}_t))&=\left(1-\frac{s}{t}\right)\log\frac{\left(1-\frac{s}{t}\right)\mathbf{x}_0^{\top}\mathbf{x}_0}{\left(1-\frac{s}{t}\right)\mathbf{x}_0^{\top}\mathbf{x}_{\theta}(\mathbf{z}_t, t)}\nonumber\\
    &=-\left(1-\frac{s}{t}\right)\log\big(\mathbf{x}_0^{\top}\mathbf{x}_{\theta}(\mathbf{z}_t, t)\big)\nonumber\\
    &=-\left(\frac{t-s}{t}\right)\log\langle\mathbf{x}_0,\mathbf{x}_{\theta}(\mathbf{z}_t, t)\rangle
\end{align}
Similarly, substituting $s=t-\frac{1}{T}$, we have
\begin{align}
    \text{KL}(q(\mathbf{z}_s|\mathbf{z}_t, \mathbf{x}_0)||p_{\theta}(\mathbf{z}_s|\mathbf{z}_t))&=-\left(\frac{t-\left(t-\frac{1}{T}\right)}{t}\right)\log\langle\mathbf{x}_0,\mathbf{x}_{\theta}(\mathbf{z}_t, t)\rangle\nonumber\\
    &=-\frac{1}{tT}\log\langle\mathbf{x}_0,\mathbf{x}_{\theta}(\mathbf{z}_t, t)\rangle
\end{align}
Now, we can combine the two cases using the indicator functions $\mathbf{1}[\mathbf{x}_0=\mathbf{b}]$ that evaluates to 1 when $\mathbf{x}_0=\mathbf{b}$ and 0 otherwise and $\mathbf{1}[\mathbf{x}_0\neq\mathbf{b}]$ that evaluates to 1 when $\mathbf{x}_0\neq\mathbf{b}$ and 0 otherwise. Since this definition of KL divergence is only applicable when $\mathbf{z}_t=\mathbf{m}$, we have
\begin{align}
    &\text{KL}(q(\mathbf{z}_s|\mathbf{z}_t, \mathbf{x}_0)||p_{\theta}(\mathbf{z}_s|\mathbf{z}_t))\nonumber\\&=\bigg[-\mathbf{1}[\mathbf{x}_0=\mathbf{b}]\left(\frac{w}{tT}-o\left(\frac{1}{T^2}\right)\right)\log\langle\mathbf{x}_0,\mathbf{x}_{\theta}(\mathbf{z}_t, t)\rangle-\mathbf{1}[\mathbf{x}_0\neq\mathbf{b}]\frac{1}{tT}\log\langle\mathbf{x}_0,\mathbf{x}_{\theta}(\mathbf{z}_t, t)\rangle\bigg]
\end{align}
Substituting this back into the equation for the discrete-time diffusion loss, we get
\begin{align}
    &\mathcal{L}_{\text{NELBO}}=\mathbb{E}_{t\in\{\frac{1}{T}, \frac{2}{T}, \dots, 1\}}\mathbb{E}_{q(\mathbf{z}_t|\mathbf{x}_0)}\bigg[T\cdot \text{KL}\bigg(q(\mathbf{z}_s|\mathbf{z}_t, \mathbf{x}_0)\big|\big|p_{\theta}(\mathbf{z}_s|\mathbf{z}_t)\bigg)\bigg]\nonumber\\
    &=\mathbb{E}_{t\in\{\frac{1}{T}, \frac{2}{T}, \dots, 1\}}\mathbb{E}_{q(\mathbf{z}_t|\mathbf{x}_0)}T\cdot \bigg[-\mathbf{1}[\mathbf{x}_0=\mathbf{b}]\left(\frac{w}{tT}-o\left(\frac{1}{T^2}\right)\right)\log\langle\mathbf{x}_0,\mathbf{x}_{\theta}(\mathbf{z}_t, t)\rangle\nonumber\\
    &\hspace{30pt}-\mathbf{1}[\mathbf{x}_0\neq\mathbf{b}]\frac{1}{tT}\log\langle\mathbf{x}_0,\mathbf{x}_{\theta}(\mathbf{z}_t, t)\rangle\bigg]\nonumber\\
    &=\mathbb{E}_{t\in\{\frac{1}{T}, \frac{2}{T}, \dots, 1\}}\mathbb{E}_{q(\mathbf{z}_t|\mathbf{x}_0)}\bigg[-\mathbf{1}[\mathbf{x}_0=\mathbf{b}]\left(\frac{wT}{tT}-To\left(\frac{1}{T^2}\right)\right)\log\langle\mathbf{x}_0,\mathbf{x}_{\theta}(\mathbf{z}_t, t)\rangle\nonumber\\
    &\hspace{30pt}-\mathbf{1}[\mathbf{x}_0\neq\mathbf{b}]\frac{T}{tT}\log\langle\mathbf{x}_0,\mathbf{x}_{\theta}(\mathbf{z}_t, t)\rangle\bigg]\nonumber\\
    &=\mathbb{E}_{t\in\{\frac{1}{T}, \frac{2}{T}, \dots, 1\}}\mathbb{E}_{q(\mathbf{z}_t|\mathbf{x}_0)}\bigg[-\mathbf{1}[\mathbf{x}_0=\mathbf{b}]\left(\frac{w}{t}-To\left(\frac{1}{T^2}\right)\right)\log\langle\mathbf{x}_0,\mathbf{x}_{\theta}(\mathbf{z}_t, t)\rangle\nonumber\\
    &\hspace{30pt}-\mathbf{1}[\mathbf{x}_0\neq\mathbf{b}]\frac{1}{t}\log\langle\mathbf{x}_0,\mathbf{x}_{\theta}(\mathbf{z}_t, t)\rangle\bigg]
\end{align}
Finally, taking the limit as $T\to\infty$, the higher-order term $\lim_{T\to \infty}To\left(\frac{1}{T^2}\right)=0$ and we get 
\begin{align}
    &\mathcal{L}^{\infty}_{\text{NELBO}}=\lim_{T\to \infty}\mathcal{L}_{\text{NELBO}}\nonumber\\
    &=\lim_{T\to \infty}\mathbb{E}_{t\in\{\frac{1}{T}, \frac{2}{T}, \dots, 1\}}\mathbb{E}_{q(\mathbf{x}_t|\mathbf{x}_0)}\bigg[-\mathbf{1}[\mathbf{x}_0=\mathbf{b}]\left(\frac{w}{t}-To\left(\frac{1}{T^2}\right)\right)\log\langle\mathbf{x}_0,\mathbf{x}_{\theta}(\mathbf{z}_t, t)\rangle\nonumber\\
    &\hspace{30pt}-\mathbf{1}[\mathbf{x}_0\neq\mathbf{b}]\frac{1}{t}\log\langle\mathbf{x}_0,\mathbf{x}_{\theta}(\mathbf{z}_t, t)\rangle\bigg]\nonumber\\
    &=\mathbb{E}_{t\sim\mathcal{U}(0,1]}\mathbb{E}_{q(\mathbf{x}_t|\mathbf{x}_0)}\bigg[-\mathbf{1}[\mathbf{x}_0=\mathbf{b}]\frac{w}{t}\log\langle\mathbf{x}_0,\mathbf{x}_{\theta}(\mathbf{z}_t, t)\rangle\nonumber\\
    &\hspace{30pt}-\mathbf{1}[\mathbf{x}_0\neq\mathbf{b}]\frac{1}{t}\log\langle\mathbf{x}_0,\mathbf{x}_{\theta}(\mathbf{z}_t, t)\rangle\bigg]
\end{align}
which is the continuous-time NELBO loss for a single token. Therefore, the loss across a sequence of $L$ tokens denoted as $\mathbf{x}^{(\ell)}_0$, we have
\begin{small}\begin{align}
    \mathcal{L}^{\infty}_{\text{NELBO}}=\mathbb{E}_{t\sim\mathcal{U}(0,1]}\mathbb{E}_{q(\mathbf{x}_t|\mathbf{x}_0)}\bigg[-\sum_{\ell:\mathbf{x}^{(\ell)}_0=\mathbf{b}}\frac{w}{t}\log\langle\mathbf{x}_0,\mathbf{x}_{\theta}(\mathbf{z}_t, t)\rangle-\sum_{\ell:\mathbf{x}^{(\ell)}_0\neq\mathbf{b}}\frac{1}{t}\log\langle\mathbf{x}_0,\mathbf{x}_{\theta}(\mathbf{z}_t, t)\rangle\bigg]
\end{align}\end{small}
which proves the loss defined in (\ref{eq:13}).

\subsection{Gradient Flow of Invalid Loss}
\label{appendix:G.4}
\paragraph{Proposition 2.3}
By differentiating the invalid loss with respect to the probability vector $\mathbf{x}_{\theta}^{(\ell)}(\mathbf{z}_t, t)$ for position $\ell$, the gradient with respect to the predicted probability of the sampled token $j=k$ and all other tokens in the vocabulary $j\neq k$ is given by
\begin{align}
    \nabla\mathcal{L}_{\text{invalid}}=\begin{cases}
        \text{SM}(x^{(\ell)}_{\theta, k})\left(1-\text{SM}(x^{(\ell)}_{\theta. k})\right)&j=k\\
        -\text{SM}(x^{(\ell)}_{\theta, j})\text{SM}(x^{(\ell)}_{\theta, k})&j\neq k
    \end{cases}
\end{align}

\textit{Proof.} We aim to show that the penalty for invalid token samples through the \texttt{argmax} function on predicted logits can be effectively backpropagated through the model parameters via our \texttt{softmax} scaling strategy. Here, we will denote the predicted probability for the token $k=\arg\max_j\big(\mathbf{x}^{(\ell)}_{\theta}(\mathbf{z}_t,t)\big)$ with the highest probability as $x_{\theta, k}^{(\ell)}$ and all remaining token probabilities as $x_{\theta,j}^{(\ell)}$ for $j=[1\dots |\mathcal{V}|]$.

First, we define the softmax function as
\begin{align}
    \text{SM}\big(x^{(\ell)}_{\theta,k}\big)=\frac{\exp(x_{\theta,k}^{(\ell)})}{\sum_{j=1}^{|\mathcal{V}|}\exp(x_{\theta, j}^{(\ell)})}
\end{align}
The partial derivative of the softmax probability $x^j_{\theta}$ for every token $j$ is given by equation
\begin{align}
    \frac{\partial}{\partial x_{\theta,j}^{(\ell)}}\left(\frac{\exp(x_{\theta,k}^{(\ell)})}{\sum_{j=1}^{|\mathcal{V}|}\exp(x_{\theta,j}^{(\ell)})}\right)=\frac{\left(\frac{\partial}{\partial x_{\theta, j}^{(\ell)}}\exp(x_{\theta, k}^{(\ell)})\right)\left(\sum_{j=1}^{|\mathcal{V}|}\exp(x_{\theta,j}^{(\ell)})\right)-\left(\frac{\partial}{\partial x_{\theta,j}^{(\ell)}}\sum_{j=1}^{|\mathcal{V}|}\exp(x_{\theta,j}^{(\ell)})\right)\left(\exp(x_{\theta,k}^{(\ell)})\right)}{\left(\sum_{j=1}^{|\mathcal{V}|}\exp(x_{\theta,j}^{(\ell)})\right)^2}
\end{align}
Therefore, we have two cases for the derivative: first, the derivative with respect to $x^{(\ell)}_{\theta,k}$ which denotes the predicted probability for the token that was sampled, and second, the derivative with respect to $x^{(\ell)}_{\theta,j}$ for $j\neq k$ which denotes the predicted probabilities for all remaining tokens.

For the first case, when $j=k$, the partial derivative simplifies to
\begin{align}
    \frac{\partial}{\partial x_{\theta,k}^{(\ell)}}\left(\frac{\exp(x_{\theta,k}^{(\ell)})}{\sum_{j=1}^{|\mathcal{V}|}\exp(x_{\theta,j}^{(\ell)})}\right)&=\frac{\exp(x_{\theta,k}^{(\ell)})\sum_{j=1}^{|\mathcal{V}|}\exp(x_{\theta,j}^{(\ell)})-\exp(x_{\theta,k}^{(\ell)})\exp(x_{\theta,k}^{(\ell)})}{\left(\sum_{j=1}^{|\mathcal{V}|}\exp(x_{\theta,j}^{(\ell)})\right)^2}\nonumber\\
    &=\left(\frac{\exp(x_{\theta,k}^{(\ell)})}{\sum_{j=1}^{|\mathcal{V}|}\exp(x_{\theta,j}^{(\ell)})}\right)\left(\frac{\sum_{j=1}^{|\mathcal{V}|}\exp(x_{\theta,j}^{(\ell)})-\exp(x_{\theta,k}^{(\ell)})}{\sum_{j=1}^{|\mathcal{V}|}\exp(x_{\theta,j}^{(\ell)})}\right)\nonumber\\
    &=\text{SM}(x^{(\ell)}_{\theta,k})\left(1-\text{SM}(x^{(\ell)}_{\theta,k})\right)
\end{align}
For all $j\neq k$, the derivative simplifies to
\begin{align}
    \frac{\partial}{\partial x_{\theta,j}^{(\ell)}}\left(\frac{\exp(x_{\theta,k}^{(\ell)})}{\sum_{j=1}^K\exp(x_{\theta,j}^{(\ell)})}\right)&=\frac{0-\exp(x_{\theta,j}^{(\ell)})\exp(x_{\theta,k}^{(\ell)})}{\left(\sum_{j=1}^K\exp(x_{\theta,j}^{(\ell)})\right)^2}\nonumber\\
    &=-\left(\frac{\exp(x_{\theta,j}^{(\ell)})}{\sum_{j=1}^{|\mathcal{V}|}\exp(x_{\theta,j}^{(\ell)})}\right)\left(\frac{\exp(x_{\theta,k}^{(\ell)})}{\sum_{j=1}^{|\mathcal{V}|}\exp(x_{\theta,j}^{(\ell)})}\right)\nonumber\\
    &=-\text{SM}(x^{(\ell)}_{\theta,j})\text{SM}(x^{(\ell)}_{\theta,k})
\end{align}
The parameters $\theta$ are updated such that the predicted probability of sampling the token $\ell$ with \texttt{argmax} probability $x^{(\ell)}_{\theta,k}$, which resulted in an invalid peptide SMILES sample, is reduced. The gradient update is minimized for predicted probabilities near 0 and 1, suggesting that the loss function pushes the model towards higher confidence predictions from uncertain predictions to minimize invalid sampling. 
\begin{align}
    x'^{(\ell)}_{\theta,k} \gets x^{(\ell)}_{\theta,k}  -\eta\cdot\text{SM}(x^{(\ell)}_{\theta,k})\left(1-\text{SM}(x^{(\ell)}_{\theta,k})\right)
\end{align}
where $\eta$ is the learning rate. 

In contrast, the parameters of the remaining tokens $x^{(\ell)}_{\theta,j}$ are updated so that the predicted probability of sampling the other tokens increases proportionally to their original softmax probabilities. This prevents extreme changes in the predicted probabilities of the remaining tokens and ensures that the token distribution remains relatively consistent with the previous iteration. 
\begin{align}
    x'^{(\ell)}_{\theta,j} \gets x^{(\ell)}_{\theta,j}  +\eta\cdot\text{SM}(x^{(\ell)}_{\theta,j})\text{SM}(x^{(\ell)}_{\theta,k})
\end{align}
Here, we show that our invalid loss effectively updates parameters to reduce the position-specific token probabilities that result in invalid sequence samplings and push the model predictions toward other high-likelihood tokens.

\section{Hyperparameter Selection}
\label{appendix:H}
In this section, we discuss the choice of hyperparameters for PepTune. Specifically, we discuss the effects of changing the number of expanded children nodes, the number of iterations, and the tokenization scheme.

\paragraph{Number of Children} The number of children $M$ is the hyperparameter that determines the batch size during the expansion step of MCTS. A small number of expanded nodes would limit the degree of exploration and the number of generated sequences for evaluation at each iteration. If the initial iterations resulted in sub-optimal unmasking steps for all children, this could prevent the algorithm from discovering a local or global optimum across objectives. Suppose the number of children is too large. In that case, this can result in a lack of diversity if several sequences from the same expansion step are added to the Pareto-optimal set given their sequence similarity, leading to similar property scores. A large $M$ also slows down runtime significantly. To determine a value for $M$ within the two extremes, we evaluated the performance of the MCTS search for $M=10, 50, 70, 100$. Overall, we found that $M=50$ yields consistently increasing scores across all properties, which we use for the remainder of the study. 

\paragraph{Number of Iterations} The number of iterations $N_{\text{iter}}$ determines the number of selection, expansion, rollout, and backpropagation loops executed in a single MCTS run. In addition, $N_{\text{iter}}$ is the maximum value of $t$ or the number of unmasking steps that can be executed before the rollout begins, which corresponds to the maximum tree depth. We found that equating the number of diffusion steps $T$ to the number of MCTS iterations $N_{\text{iter}}$ results in convergence on the property prediction scores as the selection process becomes biased towards a single unmasking scheme. As shown in Figure \ref{fig:tfr}, all property scores converge for $N_{\text{iter}}=T=128$, which we use for the remainder of the study.

\paragraph{Tokenization Scheme} \label{appendix:Tokenization Scheme.} To evaluate the effect of different tokenization methods on the generation quality, we experimented with three different tokenization schemes: SMILES Pair Encoding (SPE) tokenization with the trained vocabulary used by PeptideCLM and Atom Pair Encoding (APE) tokenization for SMILES and SELFIES \cite{krenn2020self} representations. Overall, we found that the SPE tokenization scheme decreased perplexity and maintained precision while capturing common peptide motifs like bonds and recurring side chains.

\begin{table}[H]
  \centering
  \caption{\textbf{Effect of Tokenization on Sequence Length, Training, and Validation Loss after Convergence}}
  \label{table:tokenization}
  \vskip 0.05in
  \adjustbox{max width=\textwidth}{%
    \begin{tabular}{lccccc}
      \toprule
      \textbf{Tokenization Scheme} & \textbf{Vocab Size} & \textbf{Avg Sequence Length in Data} & \textbf{Train Loss ($\downarrow$)} & \textbf{Val Loss ($\downarrow$)} & \textbf{Val PPL ($\downarrow$)} \\
      \midrule
      SMILES SPE Tokenizer & \textbf{581} &  84 & \textbf{0.65}& \textbf{0.75}& \textbf{2.12}\\
      SMILES APE Tokenizer & 605 &  19& 1.33& 2.32 & 10.18 \\
      SELFIES APE Tokenizer & 605 & 21 & 1.56& 2.50 & 12.12\\
      \bottomrule
    \end{tabular}%
  }
\end{table}

\section{Additional Experiments}
\label{appendix:I}
\subsection{Case Study for Time-Dependent Multi-Objective Guidance} 
\label{appendix:I.1}

Some properties of peptides require more intense guidance towards specific structural or sub-structural features, while others may only require small changes in the side chain composition or non-natural modifications. To enable the prioritization of properties during guidance, we introduce a time-dependent multi-objective guidance strategy that guides the generation based on only a subset of properties, depending on the current iteration number of the MCTS search. To achieve this, we define a $K$-dimensional vector $\mathbf{i}=[i_1, i_2, \dots, i_K]$ where each $i_k$ is the iteration number to begin guidance for the $k$th objective. Properties where $i_k=0$ are used to guide all iterations, whereas properties where $i_k> 1$ are used to guide only the iterations from $i_k\to N_{\text{iter}}$. 

Our time-dependent guidance operates as follows. During the expansion and rollout steps on iteration $i$, the rolled-out child sequences $\mathbf{x}_{s,i}$ that are non-dominated across the sub-vector of property scores $\mathbf{s}_i=[s_k\;|\;i_k\leq i \leq N_{\text{iter}}]$ dependent on the iteration $i$ are added to the Pareto-optimal set $\mathcal{P}^*$. Therefore, $\mathbf{x}_{s}$ does not need to be non-dominated in the properties $k$ where $i_k> i$. Similarly, only the sequences $\mathbf{x}^*\in \mathcal{P}^*$ that become dominated when adding $\mathbf{x}_s$ in the subset of properties represented in $\mathbf{s}_i(\mathbf{x}^*)$ are removed from $\mathcal{P}^*$.
\begin{align}
    \mathcal{P}'^*&=\mathcal{P}^*\cup \big\{(\mathbf{z}_{s}, \mathbf{s}(\mathbf{x}_{s}))\;|\;\forall \mathbf{x}^*\in \mathcal{P}^* \;\; \mathbf{s}_i(\mathbf{x}_{s})\succeq \mathbf{s}_i(\mathbf{x}^*)\big\}\\
    \mathcal{P}'^*&=\mathcal{P}^*\setminus \big\{\mathbf{x}^*\;|\;\exists \mathbf{x}_{s}\in \text{children}(\mathbf{z}_t)\;\text{s.t.}\;\mathbf{s}_i(\mathbf{x}_{s})\succ \mathbf{s}_i(\mathbf{x}^*)\big\}
\end{align}
Then, during selection, we only consider the cumulative rewards $\mathbf{W}_i=[W_k\;|\; i_k\leq i \leq N_{\text{iter}}]$ for the properties where $i_k$ s.t. $i_k\leq i \leq N_{\text{iter}}$ when computing the selection score vector $\mathbf{U}_i$ to form the Pareto-optimal selection set. 
\begin{align}
    \mathcal{P}'^*_{\text{select}}&=\mathcal{P}^*_{\text{select}}\cup \big\{\mathbf{z}_{s}\;|\;\forall \mathbf{x}^*\in \mathcal{P}^*_{\text{select}} \;\; \mathbf{U}_i(\mathbf{z}_t, \mathbf{z}_{s})\succeq \mathbf{U}_i(\mathbf{z}_t, \mathbf{z}^*)\big\}\\
    \mathcal{P}'^*_{\text{select}}&=\mathcal{P}^*_{\text{select}}\setminus \big\{\mathbf{z}^*\;|\;\exists \mathbf{z}_{s}\in \text{children}(\mathbf{z}_t)\;\text{s.t.}\;\mathbf{U}_i(\mathbf{z}_t, \mathbf{z}_{s})\succ \mathbf{U}_i(\mathbf{z}_t, \mathbf{z}^*)\big\}
\end{align}
Finally, we select the next node $\mathbf{z}_s\sim \mathcal{P}'^*_{\text{select}}$ uniformly at random from the Pareto-optimal selection set.

\begin{figure}[h]
    \centering
    \includegraphics[width=\linewidth]{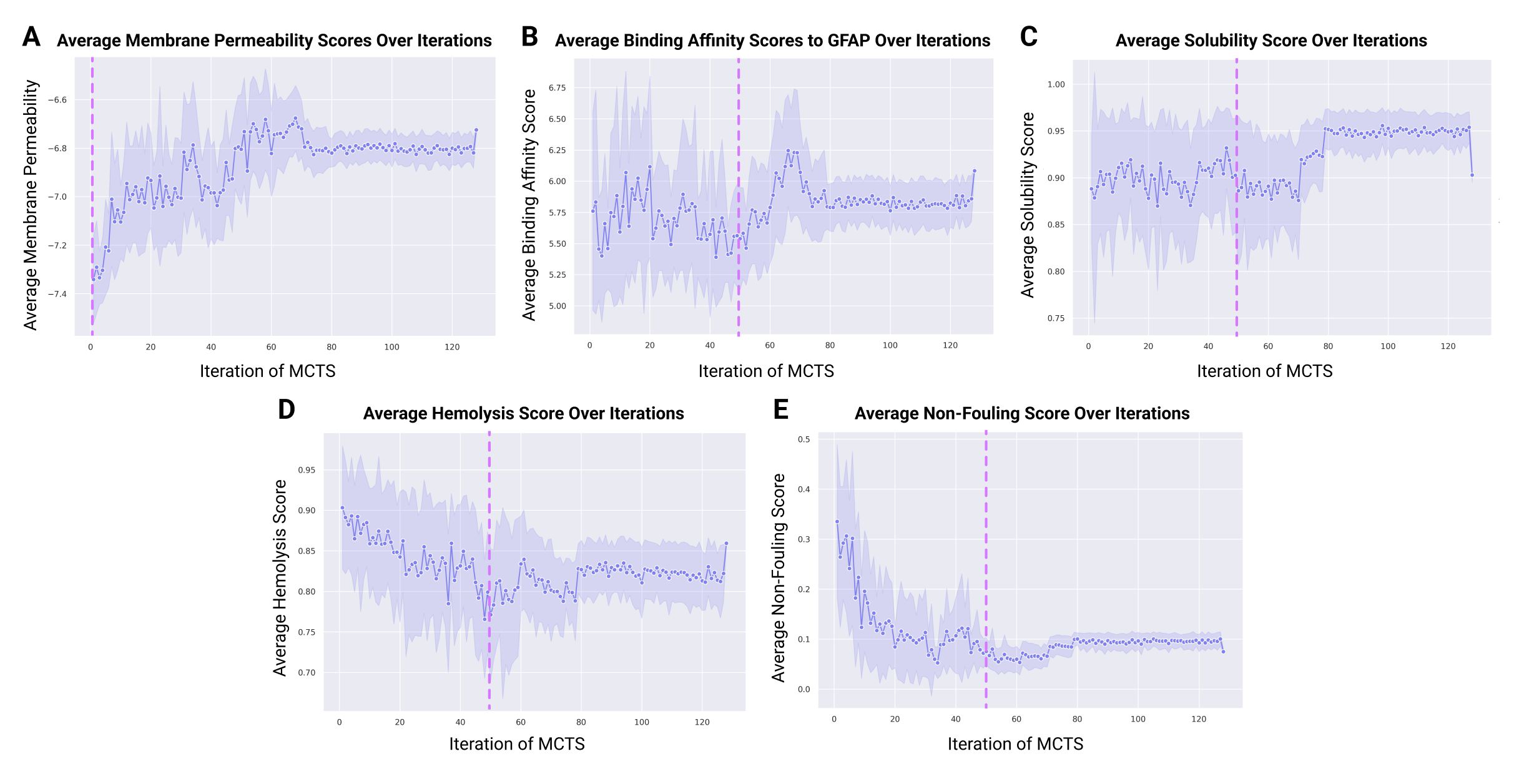}
    \caption{\textbf{Time-Dependent Multi-Objective Guidance.} (\textbf{A}) Plot of average membrane permeability score for 50 sampled sequences in the expansion and rollout step over iterations where the MCTS search is conditioned on permeability for all iterations. (\textbf{B}) Plot of average predicted binding affinity score to GFAP over iterations when conditioned starting from epoch 50. (\textbf{C}, \textbf{D}, \textbf{E}) Plot of average predicted solubility, hemolysis, and non-fouling scores over iterations when conditioned starting from epoch 50. Pink dotted lines denote the iteration where the MCTS search began conditioning on the property}
    \label{fig:time-dependent}
\end{figure}

To test this strategy, we generated 100 peptides conditioned only on membrane permeability for the first 50 iterations since we found it as the most challenging property to optimize. Then we conditioned all properties, including membrane permeability, binding affinity to GFAP, solubility, hemolysis, and non-fouling. We show that during the first 50 iterations, all properties except membrane permeability show relatively constant average scores, whereas the permeability score increased (Figure \ref{fig:time-dependent}). Then, after the 50 iteration mark, GFAP binding affinity and solubility curves increased significantly while the hemolysis and non-fouling curves increased slightly for the remainder of the iterations (Figure \ref{fig:time-dependent}). Although all the results in this paper leverage peptides without time-dependent guidance, this serves as a proof of concept for future experiments varying the start times across properties to refine certain properties at later time steps, where the generated sequences are already constrained to specific predefined substructures. 

\subsection{Ablation Studies}
\label{appendix:I.2}
We conduct an ablation study to evaluate the effect of our bond-dependent masking schedule and invalid loss on the fraction of unconditionally generated SMILES sequences that are classified as valid peptides by our SMILES2PEPTIDE decoder. We demonstrate that both components of the model are critical for the generation of sequences containing the necessary components that define a valid peptide (Table \ref{table:ablation}). Furthermore, our bond-dependent masking schedule and invalid loss can be applied for diverse sequence generation tasks, where preservation of fundamental structural components of the sequence is necessary (e.g., sentence structure in natural language, protein motifs, etc.).

\begin{table}[H]
  \centering
  \caption{\textbf{Ablation study on the effect of bond-dependent masking and invalid loss.} We unconditionally sampled 100 sequences from each model and evaluated validity with our SMILES2PEPTIDE decoder.}
  \label{table:ablation}
  \vskip 0.05in
  \adjustbox{max width=0.6\textwidth}{%
    \begin{tabular}{lc}
      \toprule
      \textbf{Model} & \textbf{Fraction of Valid Peptides} \\
      \midrule
      PepMDLM & 0.40 \\
      PepMDLM + No Bond Dependent Masking & 0.16 \\
      PepMDLM + No Invalidity Loss & 0.21 \\
      \bottomrule
    \end{tabular}%
  }
\end{table}

\newpage
\section{Algorithms}
\label{appendix:J}
Algorithm \ref{alg:1} outlines the training algorithm for PepMDLM, our bond-dependent masked discrete diffusion model for unconditional peptide SMILES generation. Algorithms \ref{alg:reverse step}, \ref{alg:2}, \ref{alg:3}, \ref{alg:4} and \ref{alg:5} describe PepTune, our MCTS-guided peptide SMILES generator. Algorithms \ref{alg:6} and \ref{alg:7} describe the bond mask function and peptide sequence decoder, which can also act as a validity filter.  
\begin{algorithm}
\caption{PepMDLM Training}\label{alg:1}
    \begin{algorithmic}[1]
        \State \textbf{Inputs:} Batched training examples $\mathbf{x}_0$
        \State \textbf{Output:} Trained unconditional MDLM for peptide SMILES generation $p_{\theta}(\mathbf{z}_s|\mathbf{z}_t)$
        \Procedure{Train}{}
            \State Sample $t\sim \text{Uniform}(0,1)$\Comment{sample continuous times}
            \LComment{bond-dependent masking schedule}
            \State $\alpha_t(\mathbf{x}_0)\gets \big(\mathbf{1}-$\textsc{BondMask}$(\mathbf{x}_0)\big)(1-t)+$\textsc{BondMask}$(\mathbf{x}_0)(1-t^w)$
            \LComment{mask each sequence in batch at varying degrees}
            \State Sample $\mathbf{z}_t\sim \text{Cat}(\mathbf{z}_t;
            \alpha_t(\mathbf{x}_0)\mathbf{x}_0+(1-\alpha_t(\mathbf{x}_0))\mathbf{m})$
            \State $\mathbf{x}_{\theta}(\mathbf{z}_t, t)\gets \text{RoFormer}_{\theta}(\mathbf{z}_t, t)$\Comment{predict token logits with RoFormer backbone}
            \If {$\mathbf{z}_t\neq \mathbf{m}$}
                \State $\mathbf{x}_{\theta}(\mathbf{z}_t, t)\gets\mathbf{z}_t$\Comment{carry-over unmasking}
            \ElsIf{$\mathbf{z}_t= \mathbf{m}$}
                \State $\mathbf{x}_{\theta}(\mathbf{z}_t, t)\gets\mathbf{x}_{\theta}(\mathbf{z}_t, t)-\infty \mathbf{m}$\Comment{zero-masking probability}
            \EndIf
            \State $\mathcal{L}^{\infty}_{\text{NELBO}}\gets \frac{1}{|B|}\sum_{\mathbf{x}_0\in B}\bigg(-\sum_{\ell:\mathbf{x}^{(\ell)}_0=\mathbf{b}}\frac{w}{t}\log\langle\mathbf{x}^{(\ell)}_0,\mathbf{x}^{(\ell)}_{\theta}(\mathbf{z}_t, t)\rangle-\sum_{\ell:\mathbf{x}^{(\ell)}_0\neq\mathbf{b}}\frac{1}{t}\log\langle\mathbf{x}^{(\ell)}_0,\mathbf{x}^{(\ell)}_{\theta}(\mathbf{z}_t, t)\rangle\bigg)$
            \State $\tilde{\mathbf{x}}_0^{(\ell)}\gets \arg\max \mathbf{x}^{(\ell)}_{\theta}(\mathbf{z}_t,t)$
            \State $\mathcal{L}_{\text{invalid}}\gets \frac{1}{|B|}\sum_{\mathbf{z}_t\in B}\bigg(\sum_{\ell=1}^L \tilde{\mathbf{x}}_{0}^{(\ell)\top}\text{SM}\big(\mathbf{x}_{\theta}^{(\ell)}(\mathbf{z}_t, t)\big)\cdot \mathbf{1}[\tilde{\mathbf{x}}_0\text{ is Invalid}]\bigg)$
            \State $\mathcal{L}\gets \mathcal{L}^{\infty}_{\text{NELBO}}+\mathcal{L}_{\text{invalid}}$\Comment{total loss}
            \State $\theta'\gets \theta -\eta \nabla_{\theta} \mathcal{L}$\Comment{update backbone parameters to minimize loss}
        \EndProcedure
    \end{algorithmic}
\end{algorithm}

\begin{algorithm}
\caption{Reverse Diffusion Step}\label{alg:reverse step}
    \begin{algorithmic}[1]
        \State \textbf{Inputs:} Partially unmasked sequence at time $t$ $\mathbf{z}_t$
        \State \textbf{Output:} Token probability distribution $p_{\theta}(\mathbf{z}_s|\mathbf{z}_t)$ for all positions in the sequence with the bond-dependent reverse posterior and SUBS parametrization
        \Procedure{ReverseDiffusionStep}{}
            \State $\mathbf{x}_{\theta}(\mathbf{z}_t,t)\gets \text{RoFormer}_{\theta}(\mathbf{z}_t, t)$
            \State $s\gets t-\frac{1}{T}$
            \If {$\mathbf{z}_t=\mathbf{m}$}
                \State $p_{\theta}(\mathbf{z}_s|\mathbf{z}_t)\gets \left\langle\left(\frac{s}{t}-\frac{s^w}{t^w}\right)\mathbf{b}+\frac{t-s}{t}\mathbf{1}, \mathbf{x}_{\theta}(\mathbf{z}_t,t)\right\rangle\mathbf{z}_s+
        \left\langle\left(\frac{s^w}{t^w}-\frac{s}{t}\right)\mathbf{b}+\frac{s}{t}\mathbf{1}, \mathbf{x}_{\theta}(\mathbf{z}_t,t)\right\rangle \mathbf{m}$
                \State $p_{\theta}(\mathbf{z}_s=\mathbf{m}|\mathbf{z}_t)\gets 0$\Comment{zero-masking probability}
            \ElsIf{$\mathbf{z}_t\neq \mathbf{m}$}
                \State $\mathbf{z}_s\gets \mathbf{z}_t$\Comment{carry-over unmasking}
            \EndIf
            \State \textbf{return} $p_{\theta}(\mathbf{z}_s|\mathbf{z}_t)$
        \EndProcedure
    \end{algorithmic}
\end{algorithm}

\begin{algorithm}
\caption{PepTune: Multi-Objective Guided Discrete Diffusion with Monte Carlo Tree Guidance (MCTG)}\label{alg:2}
    \begin{algorithmic}[1]
        \State \textbf{Inputs:} Trained MDLM denoiser $p_{\theta}(\mathbf{z}_s|\mathbf{z}_t)$, score function $\mathbf{s}(\mathbf{x}):\mathcal{V}^L\to \mathbb{R}^K$ containing classifiers for $K$ objectives $s_1, s_2, \dots, s_K$, number of time steps $T$, number of iterations $N_{\text{iter}}$
        \State \textbf{Output:} Set of Pareto-optimal sequences for the objectives and their $K$-dimensional classifier score vectors $\mathcal{P}^*=\{(\mathbf{x}^*_i, \mathbf{s}^*_i) \}$
        \Procedure{SamplePepTune}{}
            \State $\mathbf{z}_T\gets \text{[MASK]}^L$\Comment{initialize fully masked sequence} 
            \State $\mathcal{P}^*\gets \{\}$\Comment{initialize Pareto-front}
            \For{$i=1, \dots, N_{\text{iter}}$}
                \State $\mathbf{z}_{t(n)}, t(n)\gets$ \textsc{Select}$(\mathbf{z}_T)$\Comment{select expandable leaf node unmasked to time $t(n)$}
                \State $\mathbf{r}\gets 0$\Comment{initialize vector that will store the sum of all rewards from expanded children}
                \State $\text{children}(\mathbf{z}_{t(n)})\gets$  \textsc{BatchedReverseStep}$(\mathbf{z}_t)$
                \State $N_{\text{rollout}}\gets T-t(n)$
                \State $\vec{t}\gets [\frac{n}{T},\frac{n-1}{T}, \dots \frac{1}{T}]$
                \State $dt\gets \frac{1}{T}$
                \For{$i=1, \dots, M$}
                    \State $\mathbf{z}_{t(n)}\gets \mathbf{z}_{s,i}$
                    \For{$n=1$ to $N_{\text{rollout}}$}\Comment{rollout to fully unmasked sequence}
                        \State ${p}_{\theta,i}(\mathbf{z}_{s(n)}|\mathbf{z}_{t(n)}) \gets $\textsc{ReverseDiffusionStep}$(\mathbf{z}_{t(n)})$
                        \State $\mathbf{z}_{t(n)}\gets \arg\max {p}_{\theta,i}(\mathbf{z}_{s(n)}|\mathbf{z}_{t(n)})$
                    \EndFor
                    \State $\mathbf{x}_{s,i}\gets \arg\max {p}_{\theta,i}(\mathbf{x}_{s,i}|\mathbf{z}_{t(1)})$\Comment{get clean sequence}
                    \State $\mathbf{s}(\mathbf{x}_{s,i})\gets \mathbf{s}(\mathbf{x}_{s,i})$\Comment{compute score vector}
                    \LComment{add sequence if non-dominated}
                    \State $\mathbf{r}(\mathbf{z}_{s,i}), \mathcal{P}^*\gets $\textsc{UpdateParetoFront}$\big(\mathcal{P}^*, (\mathbf{z}_{s,i}, \mathbf{s}(\mathbf{z}_{s,i}))\big)$
                    \State children$(\mathbf{z}_t)$.append$\big(\mathbf{z}_{s,i}, \mathbf{s}(\mathbf{z}_{s,i})\big)$\Comment{add child node}
                    \State $\mathbf{r}\gets \mathbf{r}+\mathbf{r}(\mathbf{z}_{s,i})$\Comment{add child reward to total reward for node $\mathbf{z}_t$}
                    
                \EndFor
                \State $\mathbf{z}\gets \text{parent}(\mathbf{z}_{s,i})$
                \While{$\mathbf{z}$ not None}\Comment{backpropagate scores}
                    \State $\mathbf{W}(\mathbf{z})\gets \mathbf{W}(\mathbf{z})+ \mathbf{r}$
                    \State $N_{\text{visits}}(\mathbf{z})\gets N_{\text{visits}}(\mathbf{z})+1$
                    \State $\mathbf{z}\gets \text{parent}(\mathbf{z})$\Comment{repeat for parent node until root node}
                \EndWhile
            \EndFor
            \State \textbf{return} $\mathcal{P}^*$\Comment{return Pareto-optimal sequences}
        \EndProcedure
    \end{algorithmic}
\end{algorithm}

\begin{algorithm}
\caption{Batched Reverse Step}\label{alg:3}
    \begin{algorithmic}[1]
        \State \textbf{Inputs:} Partially unmasked sequence $\mathbf{z}_t$ at time $t$ (representing the selected node in MCTS search), value of $k$ for top $k$ sampling ($k=0$ for batched Gumbel-max sampling), total time steps $T$
        \State \textbf{Output:} Set of $M$ slightly unmasked sequences $\text{children}(\mathbf{z}_t)=\{\mathbf{z}_{s,1},\dots \mathbf{z}_{s, M}\}$ at time $s$ that become the child nodes of $\mathbf{z}_t$
        \Procedure{BatchedReverseStep}{}
            \State $\text{children}(\mathbf{z}_t)\gets \{\}$
            \State $\mathbf{x}_{\theta}(\mathbf{z}_t,t)\gets \text{RoFormer}_{\theta}(\mathbf{z}_t, t)$
            \State $s\gets t-\frac{1}{T}$
            \If {$\mathbf{z}_t=\mathbf{m}$}
                \State $p_{\theta}(\mathbf{z}_s|\mathbf{z}_t)\gets \left\langle\left(\frac{s}{t}-\frac{s^w}{t^w}\right)\mathbf{b}+\frac{t-s}{t}\mathbf{1}, \mathbf{x}_{\theta}(\mathbf{z}_t,t)\right\rangle\mathbf{z}_s+
        \left\langle\left(\frac{s^w}{t^w}-\frac{s}{t}\right)\mathbf{b}+\frac{s}{t}\mathbf{1}, \mathbf{x}_{\theta}(\mathbf{z}_t,t)\right\rangle \mathbf{m}$
                \State $p_{\theta}(\mathbf{z}_s=\mathbf{m}|\mathbf{z}_t)\gets 0$\Comment{zero-masking probability}
            \ElsIf{$\mathbf{z}_t\neq \mathbf{m}$}
                \State $\mathbf{z}_s\gets \mathbf{z}_t$\Comment{carry-over unmasking}
            \EndIf
            \For {$i = 1\dots M$}\Comment{define slightly different distribution for each sample in batch}
            \State $u_{i, j}\sim  \text{Uniform}(0, 1)$
            \State $G_{i,j}\gets-\log\left(-\log (u_{i, j}+\epsilon)+\epsilon\right)$
            \If {$k=0$}
                \State $\tilde{p}_{\theta, i}\big(\mathbf{z}_{s,i}|\mathbf{z}_t) \gets \log p_{\theta}\big(\mathbf{z}_{s,i}|\mathbf{z}_t\big)+\mathbf{G}_i$\Comment{batched Gumbel-max distributions}
                \State $\mathbf{z}_{s, i}\sim \tilde{p}_{\theta, i}\big(\mathbf{z}_{s,i}|\mathbf{z}_t)$
            \ElsIf{$k>0$}
                \State $\tilde{p}_{\theta, i}\big(\mathbf{z}_{s,i}|\mathbf{z}_t) \gets \text{SM}\bigg(\text{top}k\big\{\log p_{\theta}\big(\mathbf{z}_{s,i}|\mathbf{z}_t\big)+\mathbf{G}_i\big\}\bigg)$\Comment{batched top $k$ sampling}
            \EndIf
            \State $\mathbf{z}_{s, i}\sim \tilde{p}_{\theta, i}\big(\mathbf{z}_{s,i}|\mathbf{z}_t)$
            \State $\text{children}(\mathbf{z}_t).\text{append}(\mathbf{z}_{s, i})$
            \EndFor
            \State \textbf{return }$\text{children}(\mathbf{z}_t)$
        \EndProcedure
    \end{algorithmic}
\end{algorithm}

\begin{algorithm}
\caption{Selection}\label{alg:4}
    \begin{algorithmic}[1]
        \State \textbf{Inputs:} Masked root node $\mathbf{z}_{t(T)}$
        \State \textbf{Output:} Expandable leaf node $\mathbf{z}_t$
        \Procedure{Select}{}
            \While{True}
                \If{$\mathbf{z}_{t}$ is non-leaf node and $t\neq 0$}
                    \State $\mathcal{P}^*_{\text{select}}\gets$ \{\}\Comment{initialize Pareto front of select scores}
                    \For{$\mathbf{z}_{s, i}$ in children$(\mathbf{z}_t$)}
                        \If{$\mathbf{z}_{s, i}$ is non-leaf or expandable leaf node}
                            \State $\mathbf{U}(\mathbf{z}_t, \mathbf{z}_{s, i})\gets \frac{\mathbf{W}(\mathbf{z}_{s,i})}{N_{\text{visit}}(\mathbf{z}_{s,i})}+c\cdot p_{\theta}(\mathbf{z}_{s, i}|\mathbf{z}_t)\frac{\sqrt{N_{\text{visit}}(\mathbf{z}_t)}}{1+N_{\text{visit}}(\mathbf{z}_{s,i})}$
                            \State $\mathcal{P}^*_{\text{select}}\gets$\textsc{UpdateParetoFront}$\bigg(\mathcal{P}^*_{\text{select}}, (\mathbf{z}_{s, i},\mathbf{U}(\mathbf{z}_t, \mathbf{z}_{s, i}))\bigg)$
                        \EndIf
                    \EndFor
                    \LComment{set parent node for next iteration as a child node selected uniformly at random from Pareto-optimal set}
                    \State $\mathbf{z}_t\sim  \mathcal{P}^*_{\text{select}}$
                    \State \textsc{Select}$(\mathbf{z}_t)$\Comment{recursively call select until leaf node is reached}
                \ElsIf{$t=0$}\Comment{node is already fully unmasked}
                    \State \textsc{Select}$(\mathbf{z}_T)$\Comment{restart selection process from root node}
                \Else\Comment{return leaf node for expansion}
                    \State \textbf{return} $\mathbf{z}_{t}$
                \EndIf
            \EndWhile 
        \EndProcedure
    \end{algorithmic}
\end{algorithm}

\begin{algorithm}
\caption{Update Pareto Front}\label{alg:5}
    \begin{algorithmic}[1]
        \State \textbf{Inputs:} Current Pareto-front sequences and score vectors $\mathcal{P}^*=\{(\mathbf{x}^*_i, \mathbf{s}^*_i) \}$, newly sampled sequence and score vector $(\mathbf{z}_{s}, \mathbf{s}(\mathbf{x}_{s}))$
        \State \textbf{Output:} Reward vector $\mathbf{r}(\mathbf{z}_{s})$ and updated Pareto-optimal set $\mathcal{P}^*$
        \Procedure{UpdateParetoFront}{}
            \If{$\mathcal{P}^*$ is empty}
                \State $\mathcal{P}^*$.append($(\mathbf{z}_{s}, \mathbf{s}(\mathbf{x}_{s}))$)
                \State $\mathbf{r}(\mathbf{z}_{s})\gets \mathbf{1}^{K}$\Comment{set reward vector to ones}
            \Else
                \LComment{vector of boolean flags indicating which sequences are nondominant to $\mathbf{x}$}
                \State \text{nondominateFlag} $\gets $ new bool$[|\mathcal{P}^*|]$
                
                \State toDelete $\gets$ \{\}
                \State $\mathbf{r}(\mathbf{z}_{s})\gets \mathbf{0}^{K}$\Comment{set reward vector to zeroes}
                \For{$(\mathbf{x}_i^*, \mathbf{s}_i^*)$ in $\mathcal{P}^*$}
                    \LComment{define vector with 1 where $\mathbf{x}_{s}$ is non-dominated in the property}
                    \State $\mathbf{n}\gets [n_k=1 \text{ if }s_k(\mathbf{x}_{s})\succeq s^*_{k, i}]$
                    \LComment{define vector with 1 where $\mathbf{x}_{s,i}$ is dominant in the property}
                    \State $\mathbf{d}\gets [d_k=1 \text{ if }s_k(\mathbf{x}_{s})\succeq s^*_k]$

                    \State $\mathbf{r}(\mathbf{z}_{s, i})\gets \mathbf{r}(\mathbf{z}_{s})+\mathbf{n}$\Comment{update reward vector}
                    
                    \If {$(\forall n_k\in \mathbf{n}\;\text{s.t. } n_k=1)\land (\exists d_k\in \mathbf{d}\;\text{s.t. } d_k=1) $} \Comment{$\mathbf{x}$ dominates $\mathbf{x}^*$}
                        \State toDelete.append$(\mathbf{x}^*)$
                        \State \text{nondominateFlag}[$i$] $\gets $ True
                    \ElsIf{$\forall n_k\in \mathbf{n}\;\text{s.t. } n_k=1$}\Comment{$\mathbf{x}$ is not dominated by $\mathbf{x}^*$}
                        \State \text{nondominateFlag}[$i$] $\gets $ True
                    \Else \Comment{$\mathbf{x}^*$ dominates $\mathbf{x}$}
                        \State \text{nondominateFlag}[$i$] $\gets $ False
                    \EndIf
                \EndFor
                \LComment{if $\mathbf{x}_{s}$ is either dominant or non-dominated by all $\mathbf{x}^*$ in Pareto-optimal set $\mathcal{P}^*$, then add to $\mathcal{P}^*$}
                \If{$\forall i$ nondominateFlag$[i]$ = True}
                    \State $\mathcal{P}^*$.append$(\mathbf{z}_{s}, \mathbf{s}(\mathbf{x}_{s}))$
                \EndIf
                \For{$\mathbf{x}$ in toDelete}
                    \State $\mathcal{P}^*$.delete$\left(\mathbf{x}^*, \mathbf{s}^*\right)$
                \EndFor
            \EndIf
            \Comment{return reward vector and updated Pareto-optimal set}
            \State \textbf{return} $\mathbf{r}(\mathbf{z}_{s}), \mathcal{P}^*$
        \EndProcedure
    \end{algorithmic}
\end{algorithm}

\begin{algorithm}
\caption{Bond Mask}\label{alg:6}
    \begin{algorithmic}[1]
        \State \textbf{Inputs:} List of peptide SMILES strings \texttt{smiles\_list}
        \State \textbf{Output:} Position-wise bond mask for each sequence with 1 in positions of peptide bonds and 0 otherwise \texttt{mask}
        \Procedure{BondMask}{}
            \State \texttt{bond\_patterns} $\gets$ [(\texttt{r`OC(=O)'}, \texttt{`ester'}),
            \Statex \hspace{3cm}    (\texttt{r`N(C)C(=O)'}, \texttt{`n\_methyl'}),
            \Statex \hspace{3cm}    (\texttt{r`C(=O)N(C)'}, \texttt{`n\_methyl'}),
            \Statex \hspace{3cm}    (\texttt{r`N[12]C(=O)'}, \texttt{`peptide'}),
            \Statex \hspace{3cm}    (\texttt{r`C(=O)N[12]?'}, \texttt{`peptide'})
            ]\\
            \For{\texttt{batch\_idx,smiles} \textbf{in} \texttt{enumerate}(\texttt{smiles\_list})}
                \State \texttt{positions} $\gets$ \texttt{empty\_list()} \Comment{list to store bond positions}
                \State \texttt{used} $\gets$ \texttt{empty\_set()} \Comment{set to track used positions}
                \For{\texttt{pattern,bond\_type} \textbf{in} \texttt{bond\_patterns}}\Comment{identify bonds using patterns}
                    \For{\texttt{match} \textbf{in} \texttt{re.finditer}(\texttt{pattern,smiles})}\\
                        \If{ \textbf{not any}($p \in$ range(\texttt{match.start()}, \texttt{match.end()}) \textbf{for} $p$ \textbf{in} \texttt{used})}
                            \State \texttt{positions.append}\big(\{\texttt{start: match.start()}, \texttt{end: match.end(),}
                            \Statex \hspace{3cm} \texttt{type: bond\_type}, \texttt{pattern: match.group()}\}\big)
                            \State \texttt{used.update}\big(range(\texttt{match.start()}, \texttt{match.end()})\big)
                        \EndIf\\
                    \EndFor
                \EndFor
                \For{\texttt{pos} \textbf{in} \texttt{positions}} \Comment{update the mask for the current SMILES}
                    \State \texttt{mask[batch\_idx, pos[start]:pos[end]]} $\gets$ 1
                \EndFor
            \EndFor
            \State \textbf{return} \texttt{mask}
        \EndProcedure
    \end{algorithmic}
    
\end{algorithm}

\begin{algorithm}
\caption{SMILES2PEPTIDE}\label{alg:7}
    \begin{algorithmic}[1]
        \State \textbf{Inputs:} SMILES String $s$
        \State \textbf{Output:} Batch of $M$ sequences at time $s$. 
        \Procedure{Analyzer}{}
            \If{$s$ \text{is correct SMILES format}},
                \If{$s$ \text{contains peptide bond [NC(=O)]} or {N-methylated peptide bond [N(C)C(=O)]}},
                    \State \texttt{IS\_PEPTIDE} $\gets$ \texttt{TRUE}
                    \State \texttt{positions} $\gets$ \texttt{empty\_list()}
                    \For{\texttt{pattern,bond\_type} \textbf{in} \texttt{bond\_patterns}}
                        \For{\texttt{match} \textbf{in} \texttt{re.finditer}(\texttt{pattern,smiles})}
                            \State \texttt{positions} $\gets$ \textsc{BondMask}
                            \State \texttt{segments} $\gets$ \texttt{empty\_list()}
                            \State \texttt{positions.sort()}
                            \If{\texttt{positions[0]['start']} > 0},  \Comment{first segment}
                                \State \texttt{segments.append}\big({
                                            \texttt{content}: \texttt{smiles[0:positions[0]['start']]},
                                            \Statex \hspace{3cm}\texttt{bond\_after}: \texttt{positions[0]['pattern']}
                                            })
                            \EndIf
                            \For{\texttt{i} \textbf{in} \texttt{len(positions)-1}} \Comment{other segments}
                                \State \texttt{current} = \texttt{positions[i]}
                                \State \texttt{next\_pos} = \texttt{positions[i+1]}
                            \State \texttt{segments.append}\big({
                                        \texttt{content}: \texttt{smiles[current['end']:next\_pos['start']]},
                                        \Statex \hspace{3cm} \texttt{bond\_before}: \texttt{current['pattern']},
                                        \Statex \hspace{3cm}\texttt{bond\_after}: \texttt{next\_pos['pattern']}
                                        })
                            \EndFor
                            \If{\texttt{positions[-1]['end']} < \texttt{len(smiles)}}, \Comment{last segment}
                            \State \texttt{segments.append}\big({
                                            \texttt{content}: \texttt{smiles[positions[-1]['end']:]},
                                            \Statex \hspace{3cm}\texttt{bond\_after}: \texttt{positions[-1]['pattern']}
                                            }) 
                            \EndIf
                        \EndFor
                    \EndFor
                    \For{\texttt{residue} \textbf{in} \texttt{segments}} 
                        \State \texttt{residue} $\gets$ \texttt{Regex pattern} \Comment{Empirical Amino Acid Regex Pattern}
                    \EndFor
                \EndIf      
            \EndIf                
        \EndProcedure
    \end{algorithmic}
\end{algorithm}

\end{document}